\documentclass[preprint,flushrt]{aastex}
\usepackage{natbib}
\usepackage{lineno}
\newcommand\pflux{photons cm$^{-2}$ s$^{-1}$}

\def\mathbi#1{\textbf{\em #1}}

\slugcomment{Submitted to {\it The Astrophysical Journal}}

\shorttitle{First {\it Fermi}-LAT AGN Catalog}
\shortauthors{Abdo et al.}

\title{The First Catalog of Active Galactic Nuclei\\Detected by the {\it Fermi} Large Area Telescope}

\author{
A.~A.~Abdo\altaffilmark{2,3}, 
M.~Ackermann\altaffilmark{4}, 
M.~Ajello\altaffilmark{4}, 
A.~Allafort\altaffilmark{4}, 
E.~Antolini\altaffilmark{5,6}, 
W.~B.~Atwood\altaffilmark{7}, 
M.~Axelsson\altaffilmark{8,9}, 
L.~Baldini\altaffilmark{10}, 
J.~Ballet\altaffilmark{11}, 
G.~Barbiellini\altaffilmark{12,13}
% ...
}
% ...
\altaffiltext{2}{Space Science Division, Naval Research Laboratory, Washington, DC 20375, USA}
\altaffiltext{3}{National Research Council Research Associate, National Academy of Sciences, Washington, DC 20001, USA}
\altaffiltext{4}{W. W. Hansen Experimental Physics Laboratory, Kavli Institute for Particle Astrophysics and Cosmology, Department of Physics and SLAC National Accelerator Laboratory, Stanford University, Stanford, CA 94305, USA}
\altaffiltext{5}{Istituto Nazionale di Fisica Nucleare, Sezione di Perugia, I-06123 Perugia, Italy}
\altaffiltext{6}{Dipartimento di Fisica, Universit\`a degli Studi di Perugia, I-06123 Perugia, Italy}
\altaffiltext{7}{Santa Cruz Institute for Particle Physics, Department of Physics and Department of Astronomy and Astrophysics, University of California at Santa Cruz, Santa Cruz, CA 95064, USA}
\altaffiltext{8}{Department of Astronomy, Stockholm University, SE-106 91 Stockholm, Sweden}
\altaffiltext{9}{The Oskar Klein Centre for Cosmoparticle Physics, AlbaNova, SE-106 91 Stockholm, Sweden}
\altaffiltext{10}{Istituto Nazionale di Fisica Nucleare, Sezione di Pisa, I-56127 Pisa, Italy}
\altaffiltext{11}{Laboratoire AIM, CEA-IRFU/CNRS/Universit\'e Paris Diderot, Service d'Astrophysique, CEA Saclay, 91191 Gif sur Yvette, France}
\altaffiltext{12}{Istituto Nazionale di Fisica Nucleare, Sezione di Trieste, I-34127 Trieste, Italy}
\altaffiltext{13}{Dipartimento di Fisica, Universit\`a di Trieste, I-34127 Trieste, Italy}
% ...

\begin{abstract}
We present the first catalog of active galactic nuclei (AGN) detected by the
LAT, corresponding to 11~months of data collected in scientific operation mode.
The First LAT AGN Catalog (1LAC) includes 671 \mbox{$\gamma$-ray} sources
located at high Galactic latitudes ($|b|>10\arcdeg$) that are detected with a
test statistic ($TS$) greater than 25 and associated statistically with AGNs.
Some LAT sources are associated with multiple AGNs, and consequently, the
catalog includes 709 AGNs, comprising 300 BL~Lacertae objects (BL~Lacs), 296
flat-spectrum radio quasars (FSRQs), 41 AGNs of other types, and 72 AGNs of
unknown type.  We also classify the blazars based on their spectral energy
distributions (SEDs) as archival radio, optical, and \mbox{X-ray} data permit.
In addition to the formal 1LAC sample, we provide AGN associations for 51
low-latitude LAT sources and AGN ``affiliations'' (unquantified counterpart
candidates) for 104 high-latitude LAT sources without AGN associations.  The
overlap of the 1LAC with existing \mbox{$\gamma$-ray} AGN catalogs (LBAS,
EGRET, {\it AGILE}, {\it Swift}, {\it INTEGRAL}, TeVCat) is briefly discussed.
Various properties---such as \mbox{$\gamma$-ray} fluxes and photon power law
spectral indices, redshifts, \mbox{$\gamma$-ray} luminosities, variability, and
archival radio luminosities---and their correlations are presented and
discussed for the different blazar classes.  We compare the 1LAC results with
predictions regarding the $\gamma$-ray AGN populations, and we comment on the
power of the sample to address the question of the blazar sequence.

\end{abstract}


\begin{document}


\section{Introduction}

The {\it Gamma-ray Large Area Space Telescope} was launched on 2008 June 11.
It began its scientific operations two months later, and shortly thereafter,
it was renamed the {\it Fermi Gamma-ray Space Telescope}.  Its primary
instrument is the Large Area Telescope \citep[LAT;][]{atwood}, the successor
to the Energetic Gamma-Ray Experiment Telescope (EGRET) on board the
{\it Compton Gamma-Ray Observatory} \citep{egret}.  The LAT offers a major
increase in sensitivity over EGRET and the Italian Space Agency's
{\it Astro-rivelatore Gamma a Immagini Leggero} \citep[{\it AGILE};][]{agile},
allowing it to study the \mbox{$\gamma$-ray} sky in unprecedented detail.  In sky
survey mode ({\it Fermi}'s main observing mode), the LAT observes the entire
sky every 3~hours, providing effectively uniform exposure on the timescale of
days.

One of the major scientific goals of the {\it Fermi} mission is to investigate
high-energy emission in active galactic nuclei (AGNs).  Although it is
generally accepted that the \mbox{$\gamma$-rays} detected from blazars are emitted
from collimated jets of charged particles moving at relativistic speeds
\citep{blandrees,maraschi}, open questions remain.  The mechanisms by which
the particles are accelerated, the precise site of the \mbox{$\gamma$-ray} emission,
and the origin of AGN variability and the \mbox{$\gamma$-ray} duty cycle of blazars
are not well understood.  The physical reasons for the observational
differences between radio-loud and radio-quiet AGNs and between FSRQs and
BL~Lacs are also unclear.  LAT observations of blazars and other AGNs are
already helping to address these and other issues.  Several in-depth spectral
and/or multiwavelength studies of specific blazars
\citep[e.g.,][]{3c454, pks1454, pks2155} and of non-blazar radio galaxies
\citep{ngc1275,M87,pmnj0948} have been performed.

The high sensitivity and nearly uniform sky coverage of the LAT make it a
powerful tool for investigating the properties of large populations.  The first
list of bright AGNs detected by the LAT, the LAT Bright AGN Sample
\citep[LBAS;][]{LBAS} included bright AGNs at high Galactic latitude
($|b|>10\arcdeg$) detected with high significance ($TS>100$, or
$\ga$$10\sigma$) during the first three months of scientific operation.  This
list comprised 58 FSRQs, 42 BL~Lacs, two radio galaxies, and four AGNs of
unknown type.  Following the models used to describe the \mbox{$\gamma$-ray} spectra
obtained with previous \mbox{$\gamma$-ray} observatories \citep[e.g.,][]{mat96}, the
early analysis reported in the LBAS was carried out by fitting the \mbox{$\gamma$-ray}
spectra at energies above 200 MeV using a simple power-law model.  This
analysis revealed a fairly distinct spectral separation between FSRQs and
BL~Lacs, with FSRQs having significantly softer spectra.  The division between
the two classes was found to be at power law index $\Gamma \approx 2.2$.  It
has been suggested \citep{Ghi09} that this separation results from different
radiative cooling of the electrons due to distinct accretion regimes in the two
blazar classes.  The \mbox{$\gamma$-ray} spectral properties and time-resolved
multifrequency spectral energy distributions of LBAS sources were further
investigated in \citet{Spe09,SEDpaper}.

Here, we report on a larger AGN sample detected after 11 months of scientific
operations.  The LAT first-year catalog \citep[1FGL;][]{Catalog} contains a
total of 1451 sources detected with $TS > 25$, and 1043 of these are
at high Galactic latitudes ($|b| > 10\arcdeg$).  We present a catalog of the
high-latitude 1FGL sources that are associated with blazars and other AGNs.  We
refer to this as the First LAT AGN Catalog (1LAC).  In addition to the 1LAC, we
also provide, where possible, AGN associations for low-latitude LAT sources and
AGN ``affiliations''---candidate counterparts for which a quantitative
association probability could not yet be computed---for unassociated
high-latitude sources.

In Section 2, we describe the observations by the LAT and the analysis that led
to the first-year catalog.  In Section 3, we explain the method for associating
\mbox{$\gamma$-ray} sources with AGN counterparts in a statistically meaningful
way, present the results of this method, and describe the two schemes for
classifying 1LAC AGNs.  Section 4 provides a brief census of the 1LAC sample.
Section 5 summarizes some of the properties of the 1LAC, including the
\mbox{$\gamma$-ray} flux distribution, the \mbox{$\gamma$-ray} photon spectral
index distribution, the \mbox{$\gamma$-ray} variability properties, the
redshift distribution, and the \mbox{$\gamma$-ray} luminosity distribution.  In
Section 6, we describe the multiwavelength properties, from radio to TeV, of
the 1LAC AGNs.  We discuss the implications of the 1LAC results in Section 7
and conclude in Section 8.

In the following, we use a $\Lambda$CDM cosmology with values within $1\sigma$ of
the {\it Wilkinson Microwave Anisotropy Probe} ({\it WMAP}) results
\citep{komatsu}; in particular, we use $h = 0.71$, $\Omega_m = 0.27$, and
$\Omega_\Lambda = 0.73$, where the Hubble constant
$H_0=100h$~km~s$^{-1}$~Mpc$^{-1}$.  We also define the radio spectral indices
such that $S(\nu) \propto \nu^{-\alpha}$.

\section{\label{obs}Observations with the Large Area Telescope --- Analysis Procedures}

The \mbox{$\gamma$-ray} sources in the 1LAC are a subset of those in the 1FGL catalog;
we summarize here the procedures used in producing the 1FGL catalog.  The data
were collected from 2008 August 4 to 2009 July 4, primarily with standard
sky-survey observations.  Only photons in the ``Diffuse'' event
class\footnote{See http://fermi.gsfc.nasa.gov/ssc/data/analysis/documentation/Cicerone/Cicerone\_Data/LAT\_DP.html .}
with energies in the range 0.1-100~GeV were considered in this analysis in
order to minimize contamination from misclassified cosmic rays \citep{atwood}.
This photon class is described in more detail (comparison to the class used in
the Bright Source List \citep[BSL;][]{BSL} paper, systematic uncertainties) in
\citet{Catalog}.  To minimize contamination from \mbox{$\gamma$-rays} from the
Earth's limb, photons with incident directions greater than $105\arcdeg$ from
the local zenith were removed.  In addition, time ranges during which the
rocking angle of the LAT was greater than $43\arcdeg$ were excluded from the
data set because the bright limb of the Earth entered the field of view.  This
rocking angle limit removed only a small fraction of the data, with the
excluded time intervals occurring during occasional 5-hour pointed observations
at larger rocking angles (\mbox{$\gamma$-ray} burst [GRB] afterglow searches)
and during even briefer intervals related to Sun avoidance during survey mode
observations.  A few minutes were excised around two bright GRBs (GRB~080916C
and GRB~090510).  The few time intervals with poor data quality, flagged as
anything other than ``Good'' in the pointing/live time history (FT2) files,
were also excluded.  The resulting data set includes 245.6~d of live time.  The
standard {\it Fermi}-LAT {\it ScienceTools} software
package\footnote{http://fermi.gsfc.nasa.gov/ssc/data/analysis/documentation/Cicerone/}
(version v9r15p2) was used with the ``P6\_V3\_DIFFUSE'' set of instrument
response functions.

The source detection step made use of two wavelet algorithms, {\it mr\_filter}
\citep{sp98} and {\it PGWAVE} \citep{ciprini07}, as well as tools that maximize
a simplified likelihood function \citep[{\it pointfind};][]{Catalog} and that
implement a minimum spanning tree algorithm \citep{cmg08}.  The intention in
using a variety of algorithms to generate a list of ``seed'' positions for
sources was to keep the source-detection step from being a limiting factor in
the analysis.  As described in the 1FGL paper, the algorithms were run
independently for different energy ranges to find both soft- and hard-spectrum
sources.  Yet more seeds were introduced from the Roma-BZCAT \citep{bzcat}
and {\it WMAP} \citep{Hin07,Gio09} catalogs if no nearby LAT seeds were
present.  The seeds were essentially candidate sources, and each was evaluated
in detail in the subsequent steps of the 1FGL catalog analysis using the
standard {\it gtlike} tool to arrive at the final list.

The Galactic diffuse background model consistently employed throughout the
analysis is the currently recommended version (gll\_iem\_v02), publicly
released through the {\it Fermi} Science Support
Center\footnote{http://fermi.gsfc.nasa.gov/ssc/data/access/lat/BackgroundModels.html}.
The isotropic background (including the \mbox{$\gamma$-ray} diffuse and residual
instrumental backgrounds) model was derived from an overall fit of the diffuse
component over the $|b|>30\arcdeg$ sky.  The Galactic diffuse model and
corresponding isotropic spectrum are described in more detail in documentation
available from the {\it Fermi} Science Support Center.

\begin{figure}
\centering
\resizebox{13cm}{!}{\rotatebox[]{0}{\includegraphics{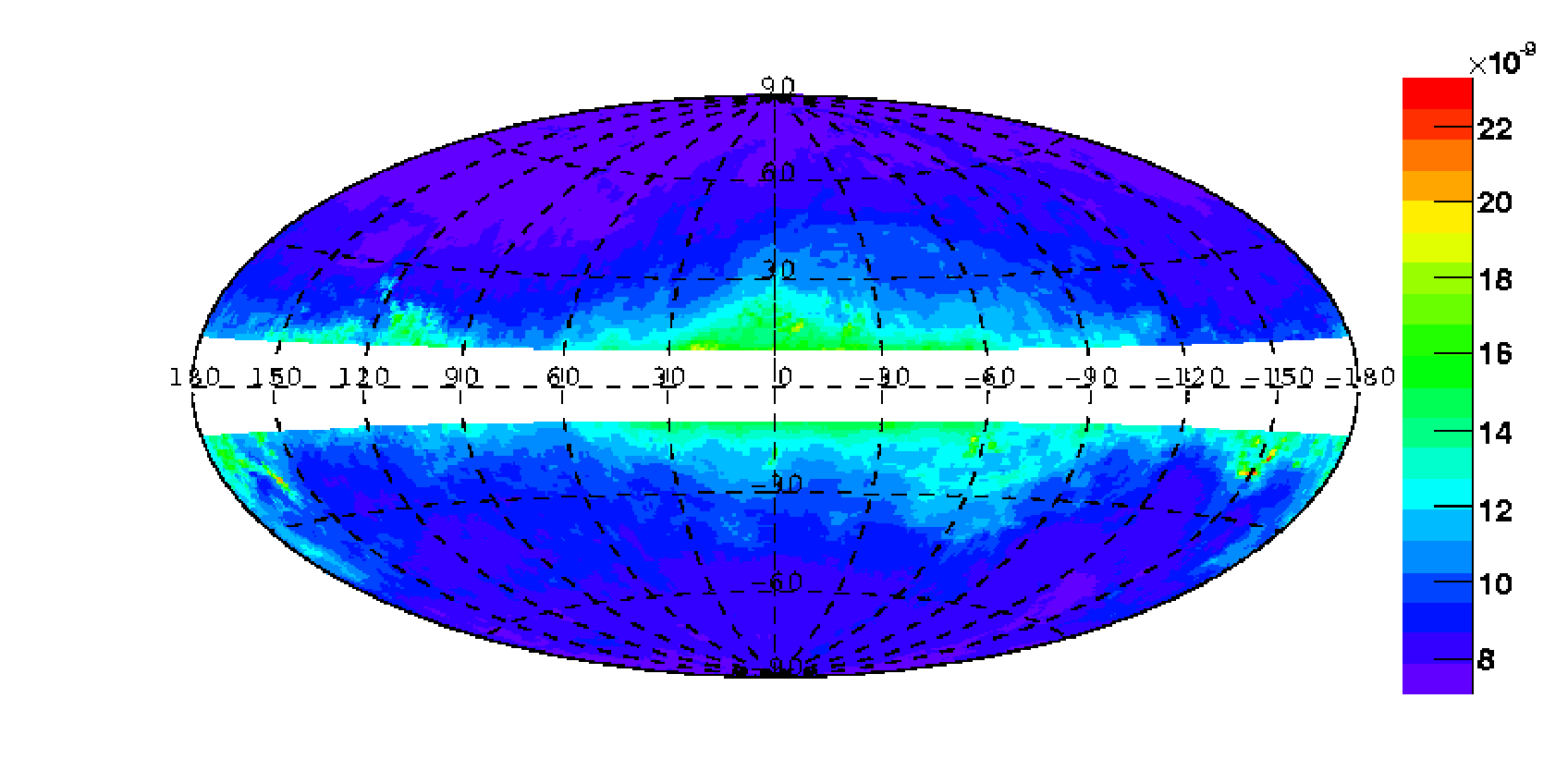}}}
\caption{Point-source flux limit in units of \pflux{} for $E > 100$~MeV and photon spectral index
$\Gamma = 2.2$ as a function of sky location (in Galactic coordinates).}
\label{fig:sens}
\end{figure}

To evaluate the source significance, we used the maximum-likelihood algorithm
implemented in {\it gtlike}.  For the 1FGL catalog, a threshold of 25 was
adopted for the test statistic ($TS$) from the {\it gtlike} likelihood
analysis.  Sources found to have $TS > 25$ were included in the 1FGL catalog.
This corresponds approximately to a minimum significance of $4.1\sigma$.
Figure~\ref{fig:sens} displays a sky map, in Galactic coordinates, of the flux
limit for photon spectral index $\Gamma=2.2$ and $TS=25$.  The anisotropy
(about a factor of 2) is due to the non-uniform Galactic diffuse background and
non-uniform exposure (mostly arising from the passage of the {\it Fermi}
satellite through the South Atlantic Anomaly).  For soft sources, source
confusion decreases the sensitivity to some extent \citep{Catalog}.

The flux, photon spectral index ($\Gamma$), and test statistic of each source
in the energy range 0.1-100~GeV were determined by analyzing regions of
interest (ROI) typically $12\arcdeg$ in radius.  The model of the ROI used to
fit the data was built taking into account all the sources detected within a
given ROI.  The fluxes in five bands (0.1-0.3, 0.3-1, 1-3, 3-10, and
10-100~GeV) were also evaluated, with the photon spectral index held fixed to
the best fit over the whole interval.  The energy flux was determined from
these fluxes, resulting in better accuracy than would be obtained from the
power-law fitted function.  For hard sources, the test statistics provided by
the 10-100~GeV band can be appreciable, which represents a notable difference
with respect to EGRET, for which the acceptance dropped sharply in this range.
The departure of the spectrum from a power-law shape can be estimated via a
simple $\chi^2$ test on the five fluxes, referred to as the curvature index
\citep{Catalog}.

\begin{figure}
\centering
\resizebox{10cm}{!}{\rotatebox[]{0}{\includegraphics{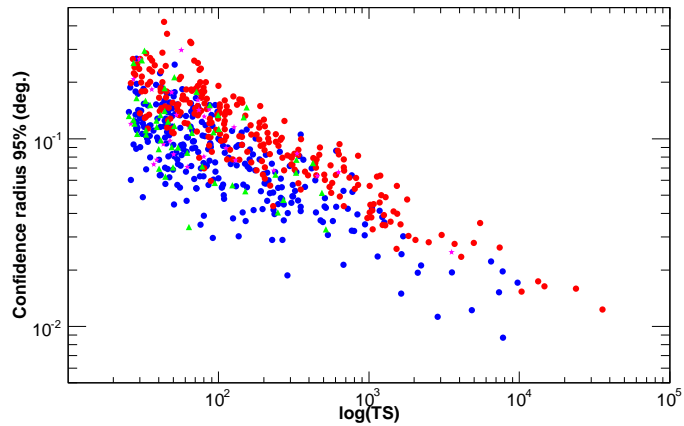}}}
\caption{95\% error radius vs.\ TS for the 1LAC clean sample.  Red circles:\ FSRQs, blue circles:\ BL~Lacs,
magenta stars:\ radio galaxies, green triangles:\ AGNs of unknown type.}
\label{fig:r95}
\end{figure}

A TS map---a 2-dimensional array of likelihood TS values evaluated at a finely
spaced grid centered on the direction of a \mbox{$\gamma$-ray} source---was generated
for each source using {\it pointfit}.  TS values are determined for each grid
position independently by maximum-likelihood fitting of a test point source.
From the TS maps, elliptical fits to the 95\% confidence source location
contours were derived; for this, the decrease of the TS away from the
maximum-likelihood position of a source is interpreted in terms of the $\chi^2$
distribution with two degrees of freedom (the coordinates of the test source).
The semimajor and semiminor axes of these ellipses were multiplied by 1.1 to
account for systematic errors, which were evaluated by comparing the measured
positions of bright sources to the known positions.  The fiducial 95\% error
radius (the geometric mean of the semimajor and semiminor axes) is plotted
against TS (derived as described above) in Figure~\ref{fig:r95}.  For each
source, a simple variability index was derived from the $\chi^2$ value of the
monthly flux distribution with a 3\% systematic uncertainty included.

\section{Source Associations}

Any procedure for associating a \mbox{$\gamma$-ray} source with a lower-energy
counterpart necessarily relies on a spatial coincidence between the two.  In
the EGRET era, the problem of counterpart associations was made very difficult
by the large \mbox{$\gamma$-ray} localization contours:\ the mean 95\% error radius
for sources in the 18-month EGRET sky survey \citep{1eg} was $0\fdg62$.  The
LAT offers a great improvement in source localization, resulting in a mean 95\%
error radius of $0\fdg15$ for the high-latitude 1FGL sources.  (For comparison
to LBAS, which included only sources with $TS > 100$ and had a mean 95\% error
radius of $0\fdg14$, the corresponding value for high-latitude 1FGL sources
with $TS > 100$ is $0\fdg09$.)

However, the LAT localization accuracy is not good enough to permit the
determination of a lower-energy counterpart based only on positional
coincidence.  A firm counterpart identification is asserted only if the
variability detected by the LAT corresponds with variability at longer
wavelengths.  In practice, such identifications are made only for a few sources
(see Table~7 in \citealt{Catalog}).  For the rest, we use a method for finding
associations between LAT sources and AGNs based on the calculation of
association probabilities using a Bayesian approach implemented in the
{\it gtsrcid} tool included in the LAT {\it ScienceTools} package and described
in the BSL paper.

\subsection{\label{sec:gtsrcid}The Bayesian Association Method}

The Bayesian method \citep{deRuiter77,ss92}, implemented by the {\it gtsrcid}
tool in the LAT {\it ScienceTools}, is similar to that used by \citet{mhr01} to
associate EGRET sources with flat-spectrum radio sources.  An earlier version
of the method was used and described in the BSL and LBAS papers.  A more
complete description is given in \citet{Catalog}, but we provide a basic
summary here.  The method uses Bayes's theorem to calculate the posterior
probability that a source from a catalog of candidate counterparts is truly an
emitter of \mbox{$\gamma$-rays} detected by the LAT.  The significance of a
spatial coincidence between a candidate counterpart from a catalog $C$ and a
LAT-detected \mbox{$\gamma$-ray} source is evaluated by examining the local
density of counterparts from $C$ in the vicinity of the LAT source.  We can
then estimate the likelihood that such a coincidence is due to random chance
and establish whether the association is likely to be real.  To each catalog
$C$, we assign a prior probability, assumed for simplicity to be the same for
all sources in $C$, for detection by the LAT.  The prior probability for each
catalog can be tuned to give the desired number of false positive associations
for a given threshold on the posterior probability, above which the
associations are considered reliable (see Section~\ref{sec:cat}).  We use a
slightly different configuration of {\it gtsrcid} from that used for the 1FGL
catalog.  This allowed us to assign multiple associations to a single LAT
source and to find associations with probabilities above 50\% (compared with a
threshold of 80\% for the 1FGL associations).

Candidate counterparts were drawn from a number of source catalogs.  The most
important ones are the Combined Radio All-sky Targeted Eight GHz Survey
\citep[CRATES;][]{crates}, the Candidate Gamma-Ray Blazar Survey
\citep[CGRaBS;][]{cgrabs}, and the Roma-BZCAT \citep{bzcat}.  The CRATES
catalog contains precise positions, 8.4~GHz flux densities, and radio spectral
indices for over 11,000 flat-spectrum sources over the entire $|b| > 10\arcdeg$
sky.  CGRaBS, a sample of the 1,625 CRATES sources with radio and \mbox{X-ray}
properties most similar to the blazars in the Third EGRET Catalog
\citep[3EG;][]{3EGcatalog}, provides optical magnitudes, optical
classifications, and spectroscopic redshifts.  Roma-BZCAT is a list of blazars
compiled based on an accurate examination of data from the literature and
currently includes over 2800 sources, all observed at radio and optical
frequencies and showing the observational characteristics of blazars.  A
complete list of the source catalogs used by {\it gtsrcid} can be found in
\citet{Catalog}.

The same association method can be used at low latitudes; most of the candidate
counterparts in this region are drawn from the VLBA Calibrator Survey
\citep{vcs1,vcs2,vcs3,vcs4,vcs5,vcs6}.  These associations are discussed in
Section~\ref{sec:lowlat}.

\subsection{Association Results}

\subsubsection{\label{sec:cat}The First LAT AGN Catalog (1LAC)}

The 1LAC comprises all high-latitude ($|b| > 10\arcdeg$) sources with an
association from {\it gtsrcid}; the full catalog includes 709 AGN associations
for 671 distinct 1FGL sources and is shown in Table~1.  An AGN is in the
``high-confidence'' sample if and only if its association probability $P$ is at
least 80\%; this sample contains 663 AGNs.  An AGN is in the ``clean'' sample
if and only if it has $P \ge 80$\%, it is the sole AGN associated with the
corresponding 1FGL \mbox{$\gamma$-ray} source (as indicated by an ``S'' in the
last column of Table~1), and it is not ``flagged'' in the 1FGL catalog as
exhibiting some problem or anomaly that casts doubt on its detection.  This
last criterion eliminates 12 sources from the clean sample:\ 
1FGL~J0217.8$+$7353, 1FGL~J0258.0$+$2033, 1FGL~J0407.5$+$0749,
1FGL~J0433.5$+$3230, 1FGL~J0539.4$-$0400, 1FGL~J0540.9$-$0547, 1FGL~J1424.5$-$7847, 1FGL~J1702.7$-$6217,
\linebreak
1FGL~J1727.9$+$5010,
1FGL~J1938.2$-$3957, 1FGL~J2212.9$+$0654, and 1FGL~J2343.6$+$3437.  All figures
presented here are for the clean sample, which contains 599 AGNs.

The LBAS associations included one LAT source that was associated with two
radio counterparts.  \citet{LBAS} noted that, as the number of LAT detections
increased, source confusion was likely to be more of a problem, and indeed, the
1LAC includes 35 LAT sources that are associated with more than one AGN (for a
total of 73 such associations).  In cases of multiple associations, we list
each counterpart separately in Table~1 and indicate them in the last column of
the table.

The prefix ``FRBA'' in the column of AGN names refers to sources observed at
8.4~GHz as part of VLA program AH996 (``Finding and Rejecting Blazar
Associations for {\it Fermi}-LAT \mbox{$\gamma$-ray} Sources'').  The prefix
``CLASS'' refers to sources from the Cosmic Lens All-Sky Survey \citep{class1,
class2}.

\begin{figure}
\centering
\resizebox{10cm}{!}{\rotatebox[]{0}{\includegraphics{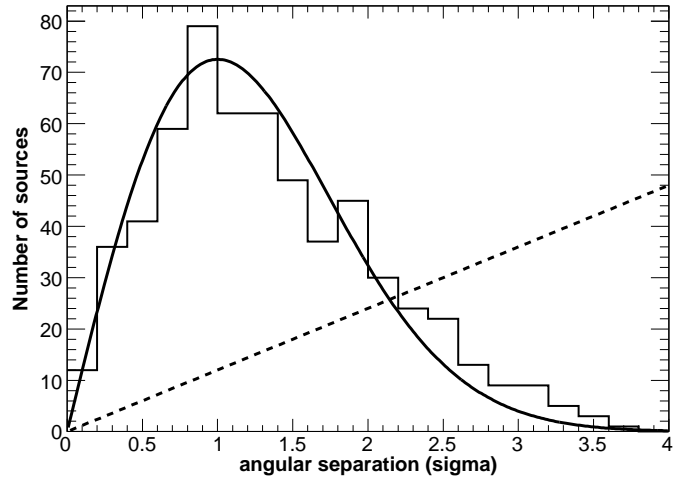}}}
\caption{Normalized angular separation between the {\it Fermi}-LAT source position and that of the AGN counterpart.  The solid curve corresponds
to the expected distribution ($\chi^2$ with two degrees of freedom) for real associations, the dashed one for purely accidental associations.}
\label{fig:separation}
\end{figure}

Figure~\ref{fig:separation} shows the distribution of normalized angular
separations between the \mbox{$\gamma$-ray} sources and their AGN counterparts. The
solid curve corresponds to the expected distribution for true associations
while the dashed curve represents the expected distribution for purely random
associations.  These results provide confidence that most of the associations
found are real.

The association probabilities can be used to estimate the number of false
positive associations.  In a sample of $k$ sources with association
probabilities $P_i$, the number of false positives is
\mbox{$N_\mathrm{false} \approx \sum_{i=1}^k (1-P_i)$}.  Among the 709
associations in the entire 1LAC, $\sim$30 are false, but of the 663 sources
in the high-confidence list, there are only $\sim$14 false positives, and only
$\sim$11 of the sources in the clean sample are falsely associated.
Additionally, there should be less than one false positive among the 363
most likely associations in the whole catalog.

\subsubsection{\label{sec:lowlat}Low-Latitude AGNs}

A simple extrapolation, based on the global density of 1LAC sources on the sky
and the solid angle subtended by the Galactic plane region ($|b| < 10\arcdeg$),
indicates that the LAT should be detecting $\sim$150 AGNs at low Galactic
latitudes.  Diffuse radio emission, interloping Galactic point sources, and
heavy optical extinction make the low-latitude sky a difficult region for AGN
studies, and catalogues of AGNs and AGN candidates often avoid it partially or
entirely.  However, we are able to make associations with 51 low-latitude AGNs;
these are presented in Table~\ref{tbl-lowlat}.  Although the associations are
considered valid, these sources have, in general, been studied much less
uniformly and much less thoroughly than the high-latitude sources at virtually
all wavelengths, so we do not include them as part of the 1LAC in order to keep
them from skewing any of our analyses of the overall \mbox{$\gamma$-ray} AGN
population.

\subsubsection{\label{sec:affil}AGN ``Affiliations''}

For many of the 1FGL sources that are not formally associated with AGNs in the
1LAC, it is still possible to find nearby AGNs or AGN candidates for which 
reliable association probabilities can not (yet) be computed but which show
some indication that they may be the correct counterpart.  For example, using
the ASDC multifrequency tools\footnote{These tools are available from the ASDC
web site at the following URLs:\ http://tools.asdc.asi.it/ and
http://www.asdc.asi.it/ .}, we performed a visual inspection of the vicinity of
each unassociated 1FGL source.  We considered objects inside the 95\% error
ellipse that showed hints of any blazar properties, such as coincident radio
and \mbox{X-ray} emission and indications in the literature of variability,
polarization, etc.  For some sources, an optical spectrum, often from the
Sloan Digital Sky Survey \citep[SDSS;][]{sdss}, was available, allowing us to
classify them as BL~Lacs or FSRQs.  These sources have been evaluated by a
method based on the known $\log N - \log S$ relationships of several types of
AGNs (FSRQs, BL~Lacs, flat-spectrum radio sources, etc.).  We also used the
figure of merit methodology developed by \citet{srm03} and employed in the
assembly of the LBAS source list; although we have discovered some problems
with the calibration of the probabilities calculated by this approach, most of
the resulting AGN associations seem likely to be legitimate.  Both of these
methods are being studied more carefully, but we list all of the unquantified
correspondences (which we call ``affiliations'') derived by them, along with
some of the properties of the affiliated AGNs, in Table~\ref{tbl-extras}.  In
all, we find 109 AGN affiliations for 104 high-latitude LAT sources.  We
expect that future refinements to our association methods will allow us to turn
many of these into true, quantitative associations.

\subsection{\label{sec:classif}Source Classification}

\subsubsection{\label{sec:optclass}Optical Classification}

We classify each AGN according to its optical spectrum where available.
Blazars are assigned optical classifications either as flat-spectrum radio
quasars (FSRQs) or BL~Lacertae objects (BL~Lacs) using the same scheme as for
CGRaBS.  In particular, following \citet{stocke91}, \citet{up95}, and
\citet{marcha96}, we classify an object as a BL~Lac if the equivalent width
(EW) of the strongest optical emission line is $< 5$~\AA, the optical spectrum
shows a \ion{Ca}{2} H/K break ratio $C < 0.4$, and the wavelength coverage of
the spectrum satisfies
$(\lambda_\mathrm{max} - \lambda_\mathrm{min})/\lambda_\mathrm{max} > 1.7$ in
order to ensure that at least one strong emission line would have been detected
if it were present.  Although other definitions of BL~Lac objects are sometimes
applied (e.g., using the EWs of the [\ion{O}{2}]~$\lambda$3727 and
[\ion{O}{3}]~$\lambda$5007 lines and/or different limits on $C$; see, e.g.,
\citealt{landt04}), the definition used here can be applied over a large
redshift range, with the caveat that high-redshift blazars may be classified as
BL~Lacs or FSRQs using different emission lines from those used for
low-redshift objects.  The classification of higher-redshift sources will
preferentially use lines at shorter wavelengths (e.g., Ly$\alpha$~$\lambda$1216
and \ion{C}{4}~$\lambda$1549) than for low-redshift sources (e.g.,
\ion{Mg}{2}~$\lambda$2798 and H$\alpha$~$\lambda$6563).  A few of the 1LAC
sources (e.g., radio galaxies such as Centaurus~A and NGC~1275) that are not
considered blazars are listed in Table 1 simply as ``AGNs''; these objects are
discussed individually in greater detail in the following section.  Sources for
which no optical spectrum was available or for which the optical spectrum was
of insufficient quality to determine the optical classification are listed as
being of unknown type.  Some redshifts and source types given in Table~1 may
differ from previously published results (generally from the NASA Extragalactic
Database\footnote{http://nedwww.ipac.caltech.edu/} [NED], SDSS, and/or
\citealt{vcv}).  We thoroughly examined the data in the literature for accuracy
and compatibility with our classification scheme, and our reevaluations of
these results are reflected in Table~1.

A substantial number of the redshifts and optical classifications presented in
Table~1 are from our own optical follow-up campaigns.  Most of these come from
spectroscopic observations conducted with the Marcario Low-Resolution
Spectrograph \citep{lrs} on the 9.2~m Hobby-Eberly Telescope (HET) at McDonald
Observatory.  Other facilities that have contributed optical results include
the 3.6~m New Technology Telescope at La~Silla, the 5~m Hale Telescope at
Palomar, the 8.2~m Very Large Telescope at Paranal, and the 10~m Keck~I
Telescope at Mauna Kea.  These spectra will be examined in greater detail in a
subsequent paper \citep{Shaw10}.  A number of the HET redshifts were confirmed,
and new results were obtained, with the 3.6~m Telescopio Nazionale Galileo at
La~Palma.  This work will also be detailed in an upcoming publication
\citep{tng}.

\subsubsection{\label{sec:sedclass}SED Classification}

\begin{figure}
\centering
\resizebox{13cm}{!}{\rotatebox[]{0}{\includegraphics{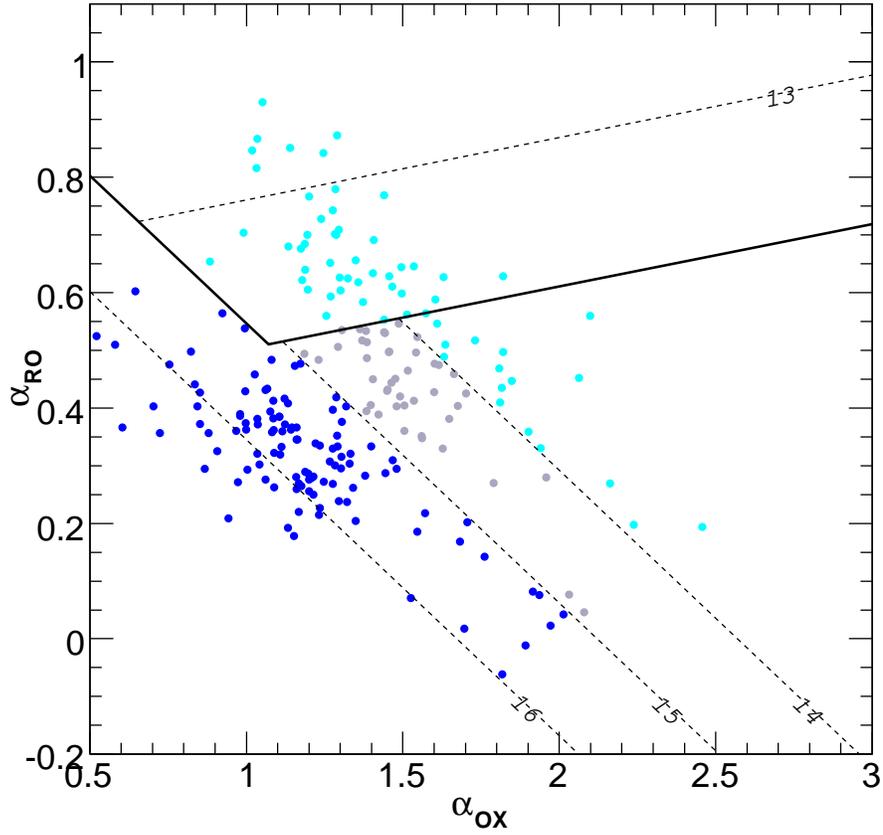}}}
\caption{Loci of the BL~Lacs in the clean sample in the ($\alpha_\mathrm{ox}$, $\alpha_\mathrm{ro}$) plane (cyan:\ LSP-BL~Lacs,
gray:\ ISP-BL~Lacs, blue:\ HSP-BL~Lacs).  The $\{X<0\} \cap \{Y<0.3\}$ region (see Section \ref{sec:sedclass}) is delineated in solid black.
The dashed lines correspond to $\log$~$(\nu_\mathrm{peak}^\mathrm{S}$ [Hz]$)=13$, 14, 15, and 16.}
\label{fig:alpha}
\end{figure}

\begin{figure}
\centering
\resizebox{13cm}{!}{\rotatebox[]{0}{\includegraphics{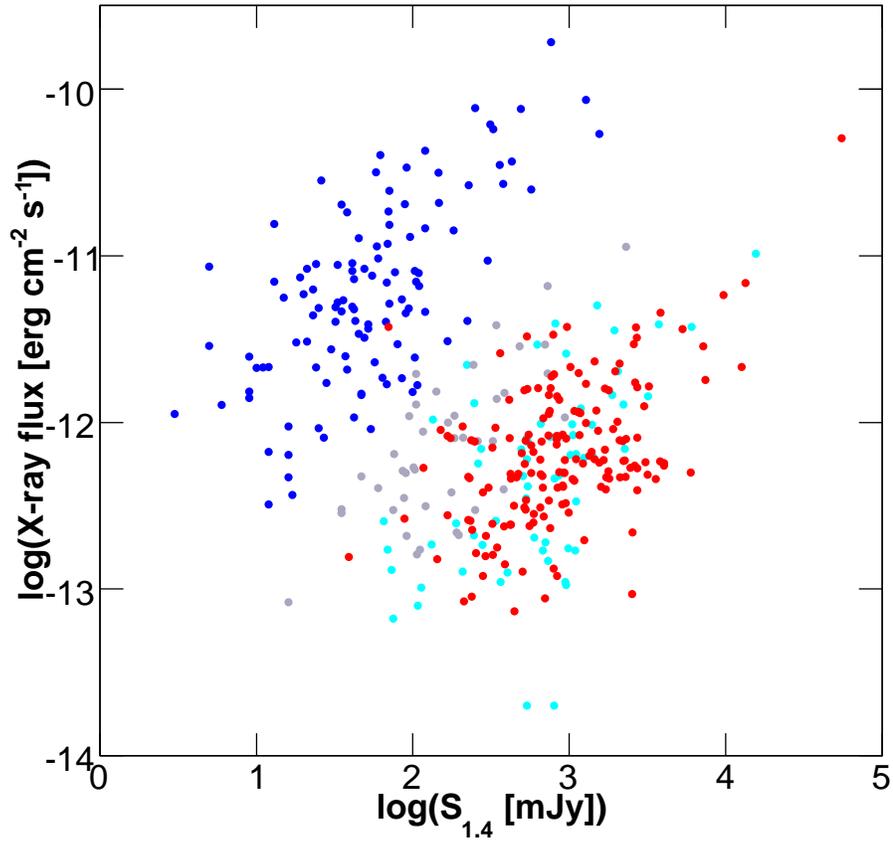}}}
\caption{Soft \mbox{X-ray} (0.1-2.4~keV) flux vs.\ radio flux density at 1.4~GHz for the clean sample (red:\ FSRQs, cyan:\ LSP-BL~Lacs,
gray:\ ISP-BL~Lacs, blue:\ HSP-BL~Lacs).}
\label{fig:Fr_Fx}
\end{figure}

The 1LAC blazars are also classified based on the peak of the synchrotron
component of the broadband SED.  In most cases, it is not possible to build a
complete SED with simultaneous data, so we follow the scheme outlined by
\citet{SEDpaper}.  This scheme is an extension to all blazars of a standard
classification system introduced by \citet{pg95} for BL~Lacs.  We estimate the
frequency of the synchrotron peak, $\nu_\mathrm{peak}^\mathrm{S}$, using the
broadband spectral indices $\alpha_\mathrm{ro}$ (between 5~GHz and 5000~\AA)
and $\alpha_\mathrm{ox}$ (between 5000~\AA\ and 1~keV).  This method uses an
analytic relationship calibrated on the $\nu_\mathrm{peak}^\mathrm{S}$ values
directly measured from the SEDs of the 48 sources studied by \citet{SEDpaper},
who confirm that these sources are representative of the global sample.  This
relationship is as follows:

$$ \log \nu_\mathrm{peak}^\mathrm{S} = \left\{ \begin{array}{ll}
 13.85 + 2.30X & \mbox{ if $X<0$ and $Y<0.3$,} \\
 13.15 + 6.58Y & \mbox{ otherwise,}
 \end{array} \right. $$
where $X = 0.565 - 1.433\alpha_\mathrm{ro} + 0.155\alpha_\mathrm{ox}$ and
$Y = 1.000 - 0.661\alpha_\mathrm{ro} - 0.339\alpha_\mathrm{ox}$.  We use the
estimated value of $\nu_\mathrm{peak}^\mathrm{S}$ to classify the source as a
low-synchrotron-peaked, or LSP, blazar (for sources with
$\nu_\mathrm{peak}^\mathrm{S} < 10^{14}$~Hz), an
intermediate-synchrotron-peaked, or ISP, blazar (for
$10^{14}$~Hz~$< \nu_\mathrm{peak}^\mathrm{S} < 10^{15}$~Hz), or a
high-synchrotron-peaked, or HSP, blazar (for sources with 
$\nu_\mathrm{peak}^\mathrm{S} > 10^{15}$~Hz).  Figure~\ref{fig:alpha} displays
the locations of BL~Lacs from the clean sample in the ($\alpha_\mathrm{ox}$,
$\alpha_\mathrm{ro}$) plane.  The data used for calculating the broadband
spectral indices were obtained mainly from {\it Swift}-XRT observations,
the NRAO VLA Sky Survey \citep[NVSS;][]{nvss}, and the USNO-B1.0 catalog
\citep{usno} and completed with data from Roma-BZCAT.  It is evident from
Figure~\ref{fig:alpha} that \mbox{$\gamma$-ray}--selected BL~Lacs cover the
region in this plane typically occupied by blazars selected at other
frequencies, such as radio and \mbox{X-rays}.  Figure~\ref{fig:Fr_Fx} shows the
soft \mbox{X-ray} flux vs.\ the radio flux density at 1.4~GHz for the blazars
in the clean sample.  As in LBAS, we find that the broadband properties of the
1LAC blazars are consistent with those of the parent populations of BL~Lacs and
FSRQs.

We note that the SED classification method assumes that the optical and \mbox{X-ray}
fluxes come from the non-thermal emission without contamination from the disk
or accretion.  For blazars in which the thermal components are non-negligible,
this method may lead to a significant overestimation of the position of
$\nu_\mathrm{peak}^\mathrm{S}$.

\section{\label{sec:census}1LAC Population Census}

Table~\ref{tbl-census} summarizes the breakdown of 1LAC sources by type for
the full catalog, the high-confidence sample, and the clean sample.
The fraction of blazars (BL~Lacs and FSRQs) that are BL~Lacs (51\% for the
high-confidence sample) is higher than in LBAS (39\%) and much higher than in
3EG (21\%).  As discussed in Section~\ref{sec:flux}, for a given significance,
BL~Lacs can typically be detected by the LAT at a lower flux than can FSRQs,
which have softer spectra.  The presence of one ISP-FSRQ and one HSP-FSRQ
warrants comment.  The existence of higher-peaked FSRQs has not been
definitively proven yet.  We stress that the SED classifications given here can
be improved with simultaneous data, and for blazars in which the thermal
components are non-negligible, it may lead to a significant overestimation of
$\nu_\mathrm{peak}^\mathrm{S}$.  Moreover, the average error on the
\mbox{$\gamma$-ray} spectral indices is 0.14, so it can not be ruled out that the
candidate ISP- and HSP-FSRQs actually have lower synchrotron peak frequencies
and are consistent with the rest of the sample.

\begin{figure}
\centering
\resizebox{13cm}{!}{\rotatebox[]{0}{\includegraphics{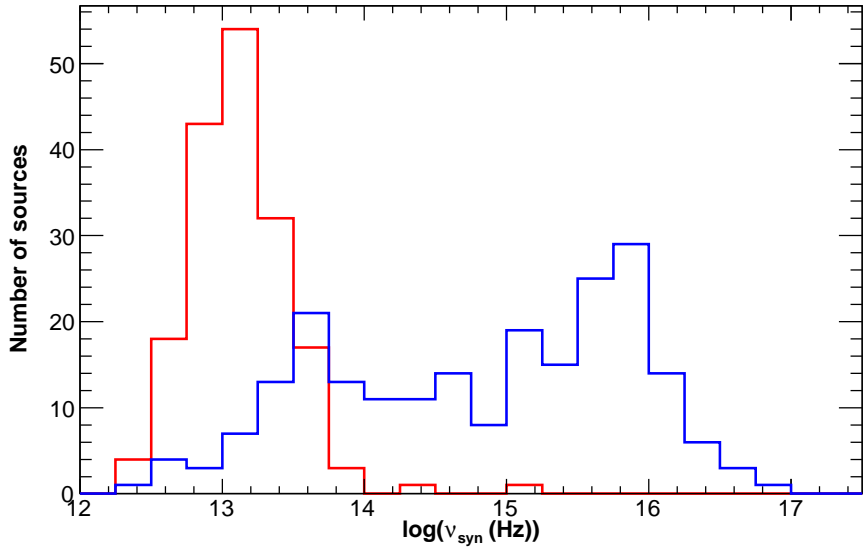}}}
\caption{Distributions of $\nu_\mathrm{peak}^\mathrm{S}$ for FSRQs (red) and BL~Lacs (blue) in the clean sample.}
\label{fig:syn_hist}
\end{figure}

The distributions of synchrotron peak positions $\nu_\mathrm{peak}^\mathrm{S}$
are shown in Figure~\ref{fig:syn_hist}.  The large fraction of HSPs in the 1LAC
sample contrasts sharply with the results from the 3EG sample, in which most
BL~Lacs ($>$80\%) were of the LSP type.  This is also a consequence of the
dependence of the limiting flux on the spectral index (see
Section~\ref{sec:flux} and Figure~\ref{fig:index_flux}).

\begin{figure}
\centering
\resizebox{13cm}{!}{\rotatebox[]{0}{\includegraphics{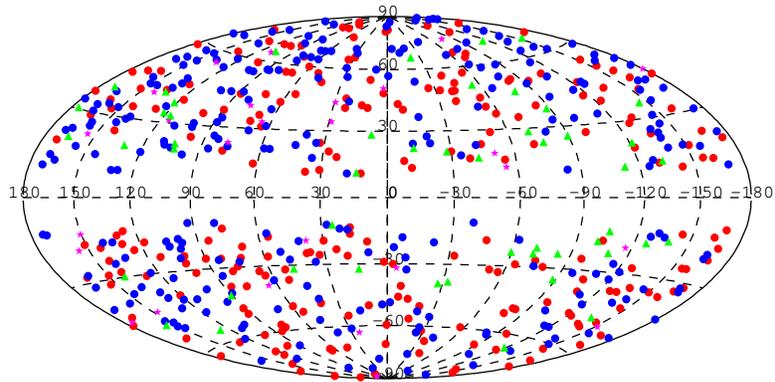}}}
\caption{Locations of the sources in the clean sample.  Red:\ FSRQs, blue:\ BL~Lacs, magenta:\ radio galaxies, green:\ AGNs of unknown type.}
\label{fig:sky_map}
\end{figure}

\begin{figure}
\centering
\resizebox{13cm}{!}{\rotatebox[]{0}{\includegraphics{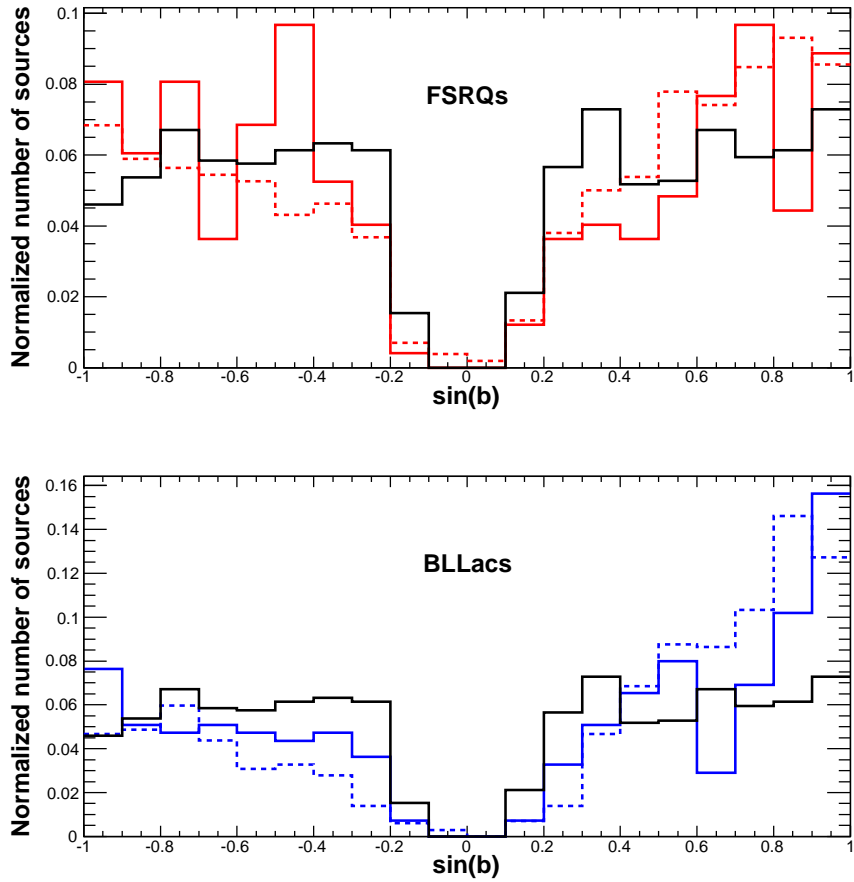}}}
\caption{Galactic latitude distributions of FSRQs (top) and BL~Lacs (bottom) in the clean sample (solid red).  The corresponding distributions
for the Roma-BZCAT blazars (dashed red) and the 1FGL sources with $|b| > 10\arcdeg$ (solid black) are shown for comparison.}
\label{fig:sky_lat}
\end{figure}

Figure~\ref{fig:sky_map} shows the sky locations of the sources in the clean
sample.  It is clear that the distribution is not isotropic; there are more
sources in the northern galactic hemisphere than in the southern one.  The
Galactic latitude distributions for FSRQs and BL~Lacs are shown in
Figure~\ref{fig:sky_lat} along with the corresponding distributions for the
Roma-BZCAT and 1FGL.  The anisotropy, defined as $(N_+ - N_-)/(N_+ + N_-)$,
where $N_+$ and $N_-$ are the number of sources at $b > 10\arcdeg$ and
$b < -10\arcdeg$ respectively, amounts to $-4$\% ($N_+ = 135$, $N_- = 146$) for
high-confidence FSRQs and 18\% ($N_+ = 171$, $N_- = 120$) for high-confidence
BL~Lacs.  For comparison, the sources in the 1FGL catalog are very evenly
distributed over the two hemispheres:\ 559 sources with $b > 10\arcdeg$ and 550
sources with $b < -10\arcdeg$.

\subsection{\label{sec:misaligned}Misaligned AGNs}

We have performed a search for possible associations between LAT detections and
non-blazar radio-loud sources (i.e., radio galaxies and steep-spectrum radio
quasars) using three main low-frequency surveys:\ the 3CR catalog \citep{3cr,
3cr2}; its revised version, the 3CRR catalog \citep{3crr}; and the Molonglo
Southern 4~Jy Sample \citep[MS4;][]{ms4}.  These catalogs are flux limited
(3CR:\ 9~Jy; 3CRR:\ 10.9~Jy; MS4:\ 4~Jy) and cover large portions of the
northern (3CR, 3CRR) and southern (MS4) sky.  In addition, because these
surveys were conducted at low frequencies (3CR and 3CRR:\ 178~MHz;
MS4:\ 408~MHz), they detect radio sources primarily in the relatively steep
part of the synchrotron emission spectrum, which is generally associated with
the extended lobes rather than the compact cores.

As a result, these catalogs are particularly appropriate for selecting AGNs
with jets misaligned with respect to the line of sight.  The sources in these
surveys exhibit radio maps with resolved and possibly symmetrical structures
and steep radio spectra ($\alpha_\mathrm{r}>0.5$).  At higher radio
frequencies, the distinction between blazars and misaligned radio sources can
be less sharp as the compact core emission (characterized by a flat or inverted
spectrum) can emerge and dominate the optically thin synchrotron emission from
extended regions.  In addition to Cen~A \citep{LBAS}, NGC~1275
\citep{LBAS,ngc1275}, and M~87 \citep{M87}, this search has resulted in the
classification of five LAT-detected sources as misaligned AGNs:\ three radio
galaxies (NGC~1218~=~3C~78, PKS~0625$-$35, and NGC~6251) and two steep-spectrum
radio quasars (3C~207 and 3C~380).  The properties of steep-spectrum radio
sources detected by the LAT will be examined further in a future publication
\citep{ssrs}.

Five other 1LAC sources have flat or nearly flat radio spectra between
$\sim$1~GHz and 8.4~GHz and strong emission lines in their optical spectra, but
these lines are narrow, in contrast to the broad emission features seen in
FSRQs.  Under the standard AGN unification paradigm \citep{antonucci,up95},
this is interpreted as an indication that the jet axis is at a larger angle to
the line of sight than for a typical FSRQ.  This causes the broad-line region
to be obscured by the dusty torus surrounding the central black hole, leaving
only narrow lines in the resulting spectrum, and such sources are commonly
described as narrow-line radio galaxies (NLRGs).  The NLRGs appearing in the
1LAC are 4C~+15.05, PKS~1106+023, CGRaBS~J1330+5202, 4C~+15.54, and
CGRaBS~J2250$-$2806.

\subsection{\label{sec:sy}Radio-Quiet AGNs}

Radio-loud narrow-line Seyfert~1 galaxies have previously been shown to be
emitters of \mbox{$\gamma$-rays} (PMN J0948$+$0022, \citealt{pmnj0948}; B2~0321$+$33B,
PKS~1502+036, and PKS~2004$-$447, \citealt{rlnlsy1}).  There is a tentative
indication that the LAT may have detected radio-quiet Seyfert galaxies or
radio-quiet quasars.  Indeed, the 1LAC includes 10 radio-quiet AGN associations
for LAT sources (RX~J0008.0$+$1450, 1WGA~0405.6$-$1313,
\mbox{CXOMP}~J084045.2$+$131617,
\linebreak
\mbox{CXOMP}~J084054.3$+$131456, SDSS~J112042.47$+$071311.5,
\mbox{CXOCY}~J113008.8$-$144737,
\linebreak
SDSS~J125257.95$+$525925.6, SDSS~J143847.94$+$371342.7,
QSO~J150255.20$-$415430.2, and
\linebreak
SDSS~J155140.52$+$085226.1).  In seven of these cases, however, the
LAT source is also associated with at least one radio-loud AGN (a blazar or a
radio galaxy) with higher probability, so the \mbox{$\gamma$-ray} emission is very
likely attributable to the radio-loud source.  1FGL~J1120.4$+$0710 is
associated both with SDSS~J112042.47$+$071311.5, a Seyfert~1 galaxy, and
CRATES~J1120$+$0704; although the latter is of unknown type, it is a radio-loud
source (probably a blazar) and is the likely source of the \mbox{$\gamma$-ray}
emission.  The remaining two radio-quiet 1LAC associations, RX~J0008.0$+$1450
and QSO~J150255.20$-$415430.2, have association probabilities under 70\%.  More
data and further study will be necessary to establish whether these objects are
indeed emitters of \mbox{$\gamma$-rays}.

\subsection{\label{sec:indiv}Notes on Individual Sources}

\noindent {\bf 1FGL~J0047.3$-$2512:} This source is associated with NGC~253,
a starburst galaxy previously detected by the LAT \citep{starbursts}.  It has
for some time been designated a Seyfert galaxy in the V{\'e}ron catalog (most
recently in \citealt{vcv}), but this classification is questionable
\citep[see, e.g.,][]{foerster}.  \citet{engelbracht} determine that the
LINER-like characteristics of NGC~253 can be attributed entirely to starburst
activity.



\noindent {\bf 1FGL~J0339.1$-$1734:} This source is associated with
PKS~0336-177.  The optical spectrum from the final release of 6dFGS
\citep{6dF04,6dF09} shows no strong emission lines, but the \ion{Ca}{2} H/K
break is too large to satisfy the criterion $C < 0.4$ for classification as a
BL~Lac.

\noindent {\bf 1FGL~J0405.6$-$1309:} This source is associated with
PKS~0403$-$13 (an FSRQ) and with
\linebreak
1WGA~J0405.6$-$1313.  \citet{rixos} classify the latter simply as an ``AGN''
but provide the redshift ($z = 0.226$) based on the detection of three emission
lines, including broad H$\beta$.  The source is radio-faint (undetected in
NVSS) and likely a Seyfert galaxy.

\noindent {\bf 1FGL~J0627.3$-$3530:} This source is associated with
PKS~0625$-$35, which shows an FR-I radio morphology.  However, \citet{wills2}
suggest a BL~Lac classification for this object.  They also note that, in
contrast with the other genuine FR-I sources in their sample, an adequate fit
to the optical continuum of PKS~0625$-$35 required a nonthermal power law in
addition to emission from the host galaxy.

\noindent {\bf 1FGL~J0645.5$+$6033:} This source is associated with
BZU~J0645+6024, a source with broad emission lines but with a steep radio
spectrum.

\noindent {\bf 1FGL~J0923.2$+$4121:} This source is associated with
B3~0920+416, which \citet{falco} characterize as a late-type galaxy.

\noindent {\bf 1FGL~J0956.5$+$6938:} This source is associated with M~82, also
known as NGC~3034 and 3C~231, a starburst galaxy previously detected by the LAT
\citep{starbursts}.  There are indications that it may host a weak AGN
\citep{wills}, though this is not confirmed.


\noindent {\bf 1FGL~J1202.9$+$6032:} This source is associated with
CRATES~J1203+6031.  \citet{falco} describes it as an early-type galaxy; it is
listed in NED as a LINER.

\noindent {\bf 1FGL~J1305.4$-$4928:} This source is associated with NGC~4945,
which is both a Seyfert~2 galaxy and a starburst galaxy.  It is the third
starburst galaxy seen by the LAT, and its detection is reported here for the
first time.  A more careful analysis will be required to determine whether
the \mbox{$\gamma$-ray} emission comes from the starburst activity or the AGN
or both.

\noindent {\bf 1FGL~J1307.0$-$4030:} This source is associated with
\mbox{ESO~323-G77}, a nearby ($z=0.015$) Seyfert 1.2 galaxy detected also by
{\it Swift}-BAT (see, e.g., \citealt{ajello09} and Section~\ref{sec:hardX}).
It exhibits a hard \mbox{X-ray} spectrum typical of radio-quiet AGN.  There is
little indication that the source is a starburst galaxy.  If the
\mbox{$\gamma$-ray} emission from 1FGL~J1307.0$-$4030 truly comes from
\mbox{ESO~323-G77}, then its origin might be connected to the central AGN; this
hypothesis is being examined more closely.



\noindent {\bf 1FGL~J1641.0$+$1143:} This source is associated with
CRATES~J1640+1144, described by \citet{mitton} only as a ``galaxy.''  Since
the authors do classify other sources as BL~Lacs, it seems likely that
CRATES~J1640+1144 is not a BL~Lac.

\noindent {\bf 1FGL~J1647.4$+$4948:} This source is associated with
CGRaBS~J1647+4950, which \citet{falco} characterize as a late-type galaxy.

\noindent {\bf 1FGL~J1724.0$+$4002:} This source is associated with B2~1722+40.
The optical spectrum from \citet{vermeulen} shows narrow forbidden emission
lines and absorption features from the host galaxy.

\noindent {\bf 1FGL~J1756.6$+$5524:} This source is associated with
BZB~J1756$+$5522 (a BL~Lac) and with CRATES~J1757+5523.  The optical spectrum
of the latter from \citet{caccianiga} shows strong absorption features from the
host galaxy and a large \ion{Ca}{2} H/K break.  The authors describe it as a
passive elliptical galaxy.

\noindent {\bf 1FGL~J2008.6$-$0419:} This source is associated with 3C~407, a
source with broad emission lines but with a fairly steep radio spectrum.

\noindent {\bf 1FGL~J2038.1$+$6552:} This source is associated with NGC~6951,
which has been classified as a Seyfert~2 galaxy and a LINER.  It also has a
circumnuclear ring of starburst activity.

\noindent {\bf 1FGL~J2204.6$+$0442:} This source is associated with
4C~$+$04.77, which is often called a BL~Lac in the literature.  A spectrum
from \citet{vcv1993}, however, shows that the object exhibits strong broad
emission lines characteristic of Seyfert~1 galaxies.


\section{Properties of the 1LAC Sources}

\subsection{\label{sec:flux}Flux and Photon Spectral Index Distributions}

As demonstrated in \cite{Spe09}, many bright LAT blazars (most notably FSRQs
and some LSP-BL~Lacs) exhibit breaks in their \mbox{$\gamma$-ray} spectra.  Despite
this caveat, determining a photon spectral index fitted over the whole band is
useful as it does reflect the source's spectral hardness and can be obtained
with reasonable accuracy even for fairly faint sources.

\begin{figure}
\centering
\plottwo{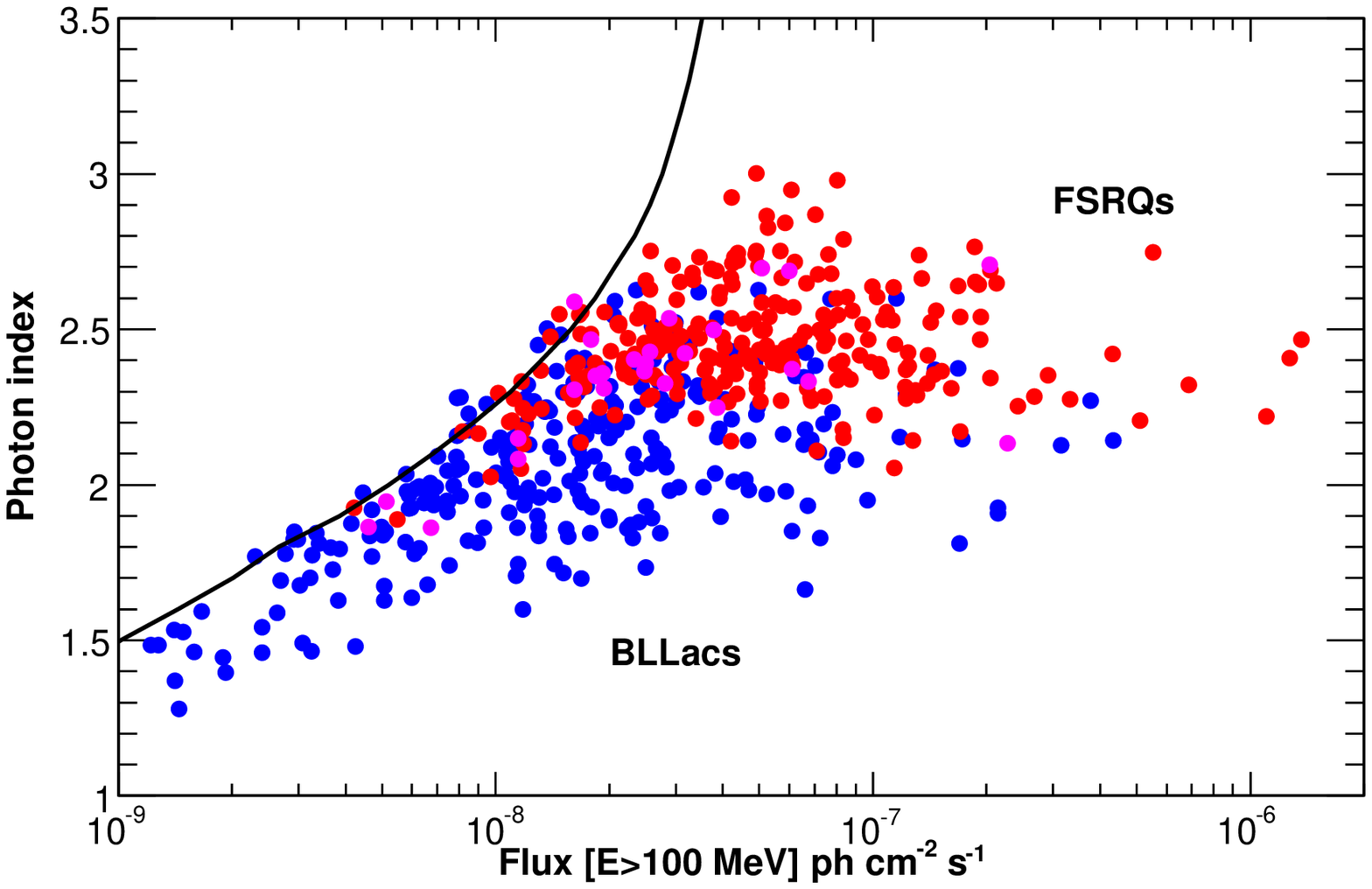}{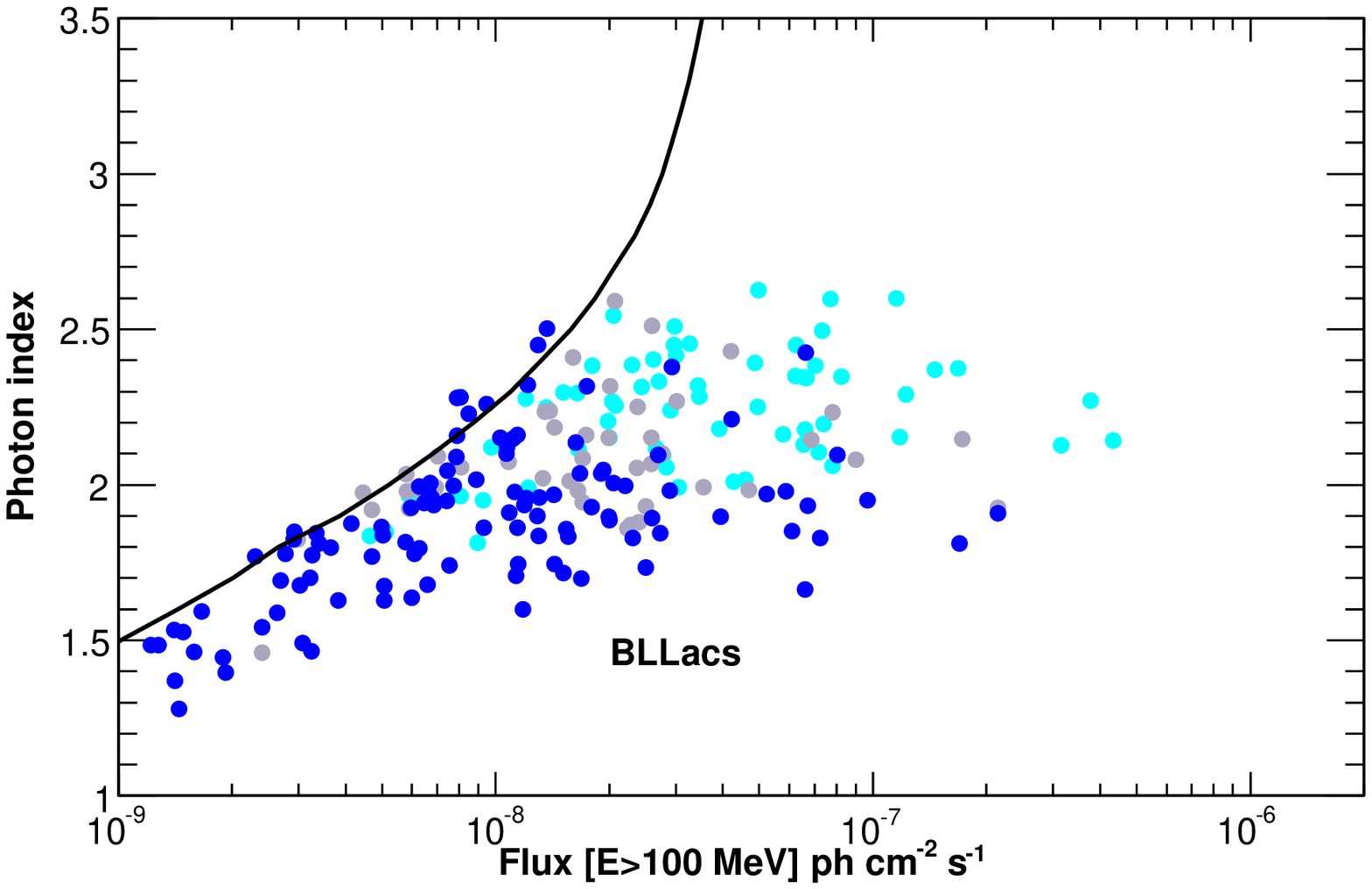}
\caption{Left:\ Flux ($E>100$~MeV) vs.\ photon spectral index for the sources in the clean sample.
The solid curves represents the $TS=25$ flux limit estimated for a Galactic position
$(l$, $b)=(40\arcdeg$, $40\arcdeg)$. Right:\ the same for the different BL~Lac subclasses.  Cyan:\ LSPs, gray:\ ISPs, blue:\ HSPs.}
\label{fig:index_flux}
\end{figure}

\begin{figure}
\centering
\resizebox{10cm}{!}{\rotatebox[]{0}{\includegraphics{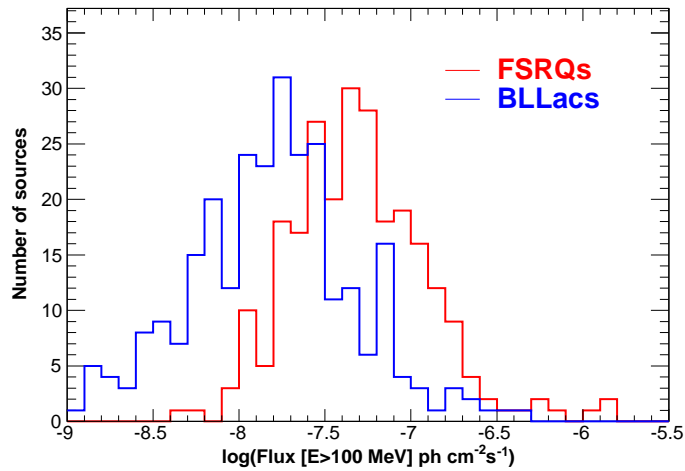}}}
\caption{Flux ($E>100$~MeV) distributions for the FSRQs (red) and BL~Lacs (blue) in the clean sample.}
\label{fig:flux}
\end{figure}

The 11-month average photon spectral index is plotted against the flux
($E>100$~MeV) estimated from a power law fit in Figure~\ref{fig:index_flux} for
blazars in the clean sample.  As the figure shows, the limiting flux
corresponding to $TS=25$ depends fairly strongly on the photon spectral index;
the solid curve corresponds to a simple analytical estimate \citep{Lot07}.
This effect is discussed in greater detail in \citet{Catalog}.  FSRQs (red
circles in Figure~\ref{fig:index_flux}) mostly cluster in the soft spectral
index region, while BL~Lacs (blue circles) primarily occupy the hard spectral
index region, confirming the trend seen for the LBAS sources \citep{LBAS}.
This implies that the limiting flux is different for FSRQs and BL~Lacs.  The
respective flux distributions are compared in Figure~\ref{fig:flux}.  The mean
fluxes are $8.5 \times 10^{-8}$ \pflux{} and $2.9 \times 10^{-8}$ \pflux{} for
FSRQs and BL~Lacs respectively.  The faintest BL~Lacs are about a factor of 3
fainter than the faintest FSRQs.

\begin{figure}
\centering
\resizebox{12cm}{!}{\rotatebox[]{0}{\includegraphics{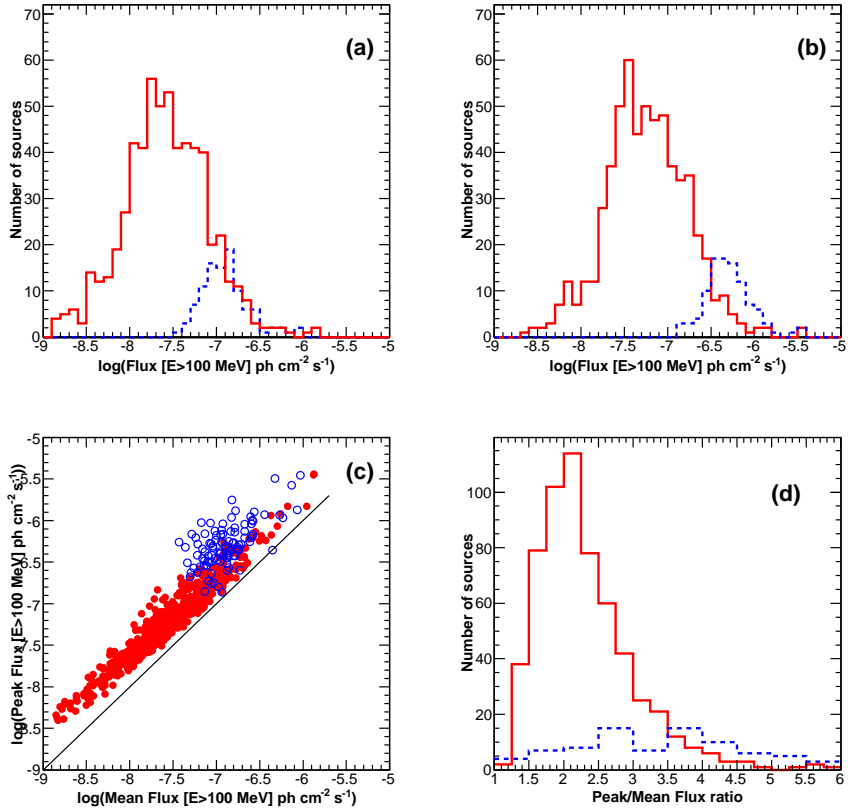}}}
\caption{(a) Comparison of mean flux distributions for blazars detected by the {\it Fermi}-LAT (solid) and by EGRET (dashed).
(b) Same as (a) but for the peak flux (i.e., maximum flux in a $\sim$15-day viewing period for EGRET, maximum monthly flux for the LAT) distributions.
(c) Peak flux vs.\ mean flux for 1LAC AGNs (filled circles) and for EGRET AGNs (open circles).
(d) Same as (a), but for the peak flux/mean flux ratio.}
\label{fig:flux_lat_egret}
\end{figure}

The overall flux distribution (of FSRQs and BL~Lacs combined) is compared to
that measured by EGRET in Figure~\ref{fig:flux_lat_egret}a. The high-flux ends
of the two distributions are in reasonable agreement.  The peak flux
distributions (maximum flux in a $\approx$15-day viewing period for EGRET vs.\ 
maximum monthly flux for the LAT) are compared in
Figure~\ref{fig:flux_lat_egret}b.  The peak fluxes observed by EGRET are
substantially higher than those observed for the brightest 1LAC sources; this
is illustrated in Figures~\ref{fig:flux_lat_egret}c and
\ref{fig:flux_lat_egret}d, which show the peak flux vs.\ mean flux and the peak
flux/mean flux ratio respectively.  The effect of using two different time
binnings to determine the peak flux (15~days for EGRET vs.\ 1~month for the
LAT) has been studied for the bright LBAS sources; the 15-day peak flux is only
13\% higher, on average, than the 1-month peak flux.  Hence, the main reason
for the different peak fluxes observed by EGRET and the LAT is probably the
different time spans over which the observations were conducted (4.5~years for
EGRET vs.\ 11~months for the LAT), enabling the sources to explore a wider
range of different states during the EGRET era.  Note also that the EGRET
observations were often triggered by flaring alerts provided by other
facilities, causing a bias that is not present for the LAT.

\begin{figure}
\centering
\resizebox{13cm}{!}{\rotatebox[]{0}{\includegraphics{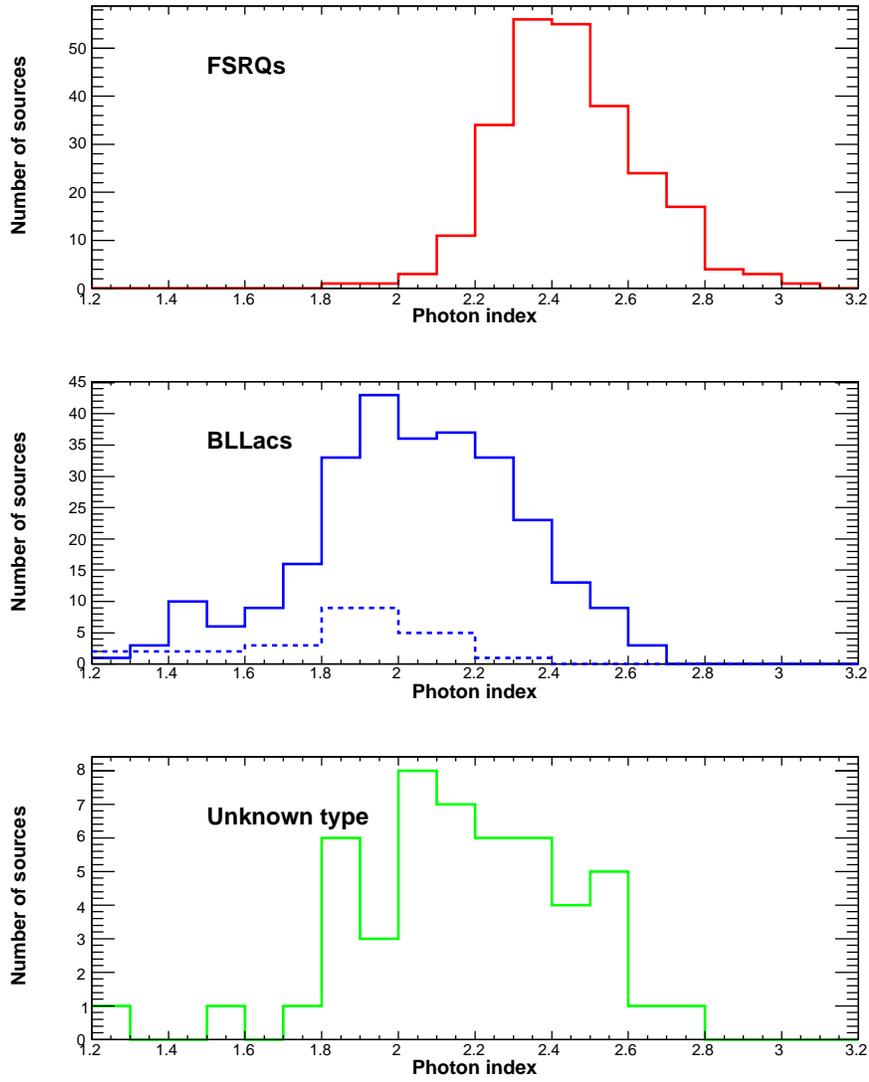}}}
\caption{Photon spectral index distributions for the FSRQs (top), BL~Lacs (middle), and AGNs of unknown type (bottom) in the clean sample.
For BL~Lacs, the dashed histogram corresponds to the TeV-detected sources.}
\label{fig:index}
\end{figure}

\begin{figure}
\centering
\resizebox{13cm}{!}{\rotatebox[]{0}{\includegraphics{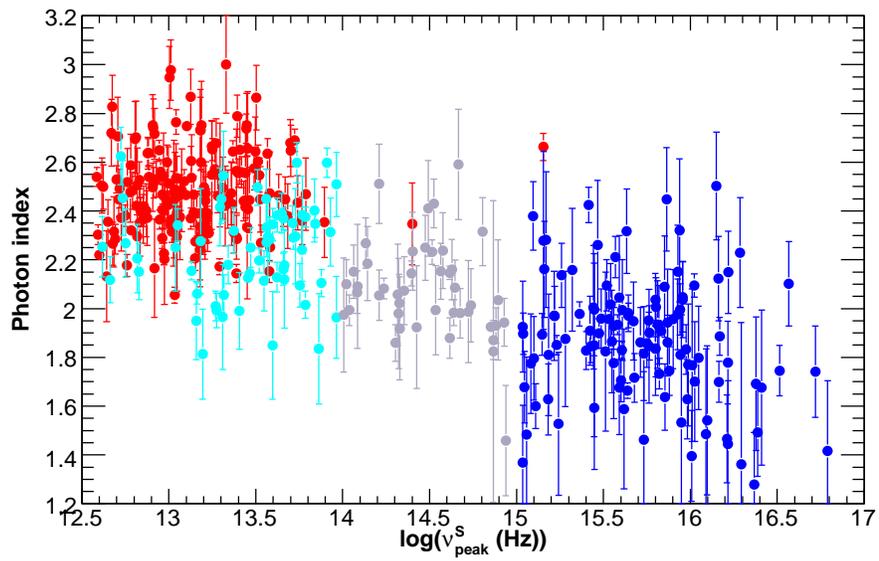}}}
\caption{Photon spectral index vs.\ peak frequency of the synchrotron component of the SED for FSRQs (red) and BL~Lacs
(cyan:\ LSPs, gray:\ ISPs, blue:\ HSPs) in the clean sample.}
\label{fig:index_log_nu_syn}
\end{figure}

\begin{figure}
\centering
\resizebox{13cm}{!}{\rotatebox[]{0}{\includegraphics{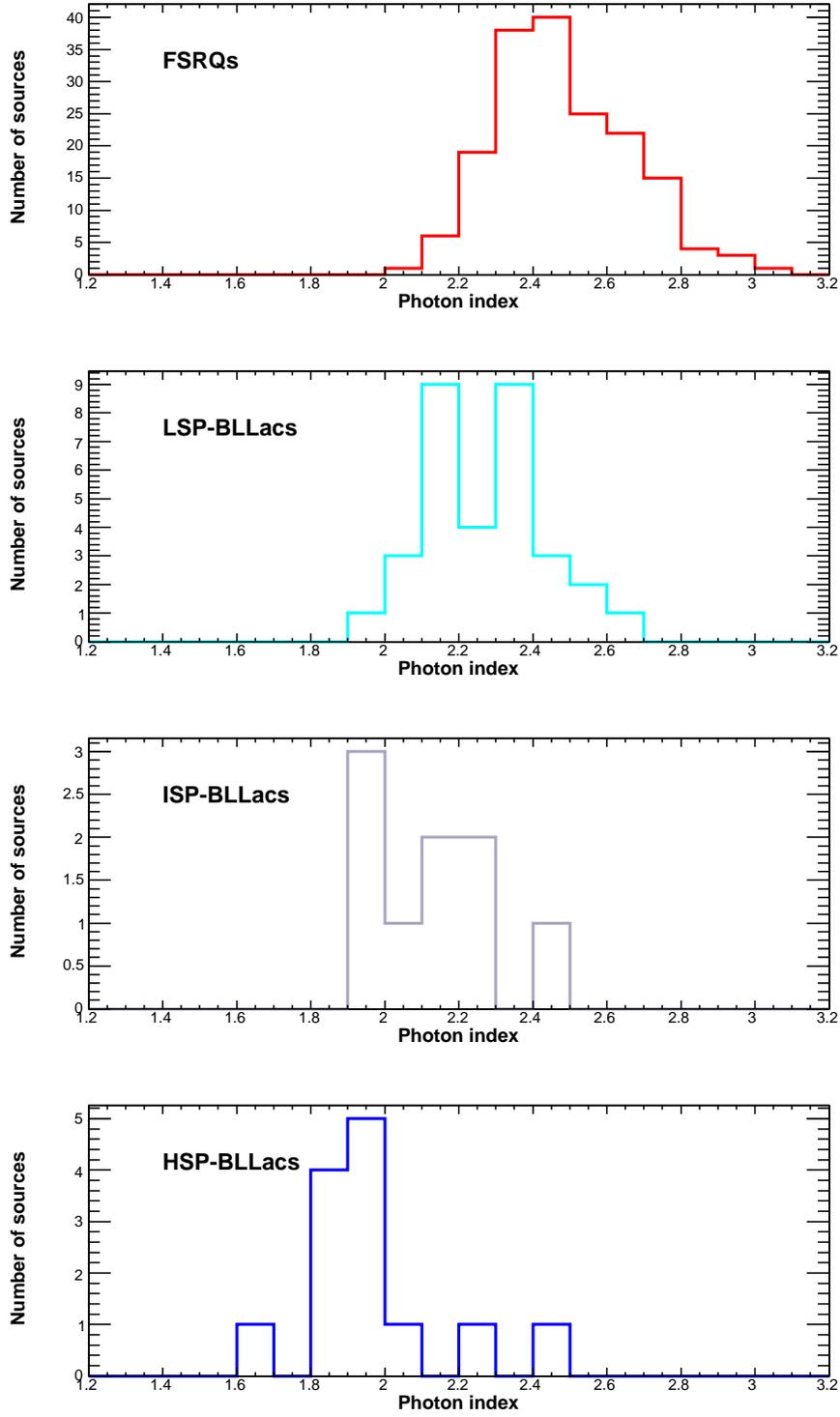}}}
\caption{Photon spectral index distributions for the different blazar classes for sources in the clean sample with $F$[${E>100}$~MeV]~$>3\times10^{-8}$ \pflux{}.}
\label{fig:index_c}
\end{figure}

The photon spectral index distributions for the clean sample, shown in
Figure~\ref{fig:index}, confirm the trend found for the LBAS sources
\citep{LBAS} and already mentioned above, with a clear difference between FSRQs
and BL~Lacs. Only 7\% of FSRQs (17/231) have a photon spectral index
$\Gamma < 2.2$, and only two FSRQs have $\Gamma < 2$ (with uncertainties of
$\sim$0.2).  Thus, it is quite clearly established that LAT-detected FSRQs
exhibit soft \mbox{$\gamma$-ray} spectra.  The 1LAC BL~Lac distribution is broader
than that observed for LBAS; it shows a larger overlap with the FSRQ
distribution.  In \cite{Spe09}, the BL~Lacs with $\Gamma > 2.2$ were found to
be mostly LSP sources.  This trend is also confirmed for the 1LAC sources, as
illustrated in Figure~\ref{fig:index_log_nu_syn}, which shows $\Gamma$ vs.
$\nu_\mathrm{peak}^\mathrm{S}$.  \cite{Spe09} also point out that the BL~Lac
photon spectral index distribution suffers from a bias since the
high-spectral-index (i.e., soft) end is cut off somewhat due to the TS
threshold, as seen in Figure~\ref{fig:index_flux}.  Figure~\ref{fig:index_c}
shows the photon spectral index distributions for the different blazar classes
for sources with $F$[$E>100$~MeV] $>3\times10^{-8}$ \pflux{}, above which the
sample is essentially complete (see Figure~\ref{fig:index_flux}).  The
completeness of the sample will be examined more carefully in an upcoming paper
on the LAT blazar population \citep{blazpop}.

\begin{figure}
\centering
\resizebox{10cm}{!}{\rotatebox[]{0}{\includegraphics{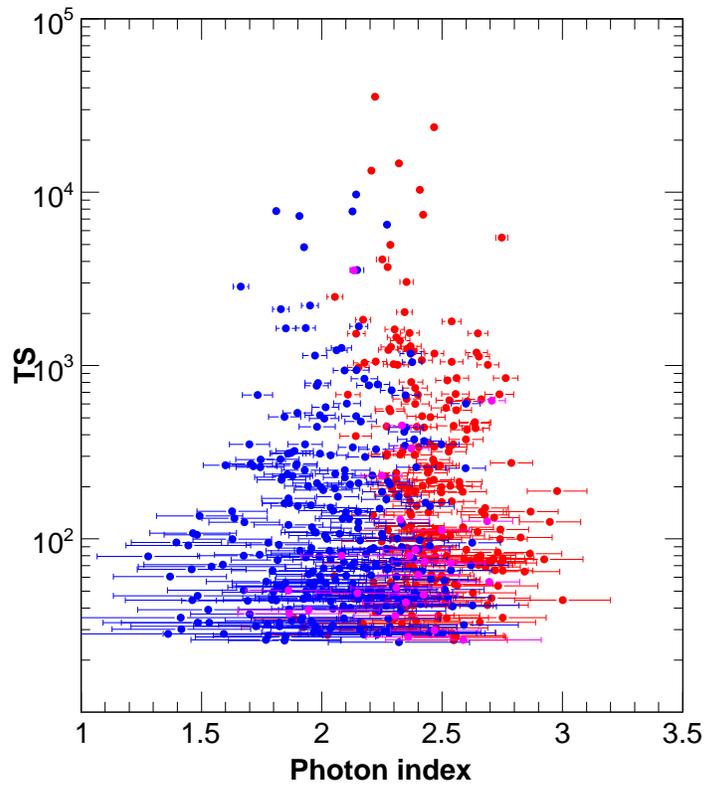}}}
\caption{TS vs.\ photon spectral index for FSRQs (red) and BL~Lacs (blue) in the clean sample.}
\label{fig:TS_index}
\end{figure}

The presence of a few very hard ($\Gamma < 1.7$) sources warrants comment.
These are fairly low-significance sources ($TS<150$), as is evident from
Figure~\ref{fig:TS_index}.  Note that typical statistical error bars reach
0.1--0.2 for all sources with TS in that range.  Figure~\ref{fig:index} also
displays the photon spectral index distribution for sources with unknown types.
This distribution overlaps with both those of FSRQs and BL~Lacs, which seems to
rule out that these sources belong preferentially to either class.

\begin{figure}
\centering
\resizebox{13cm}{!}{\rotatebox[]{0}{\includegraphics{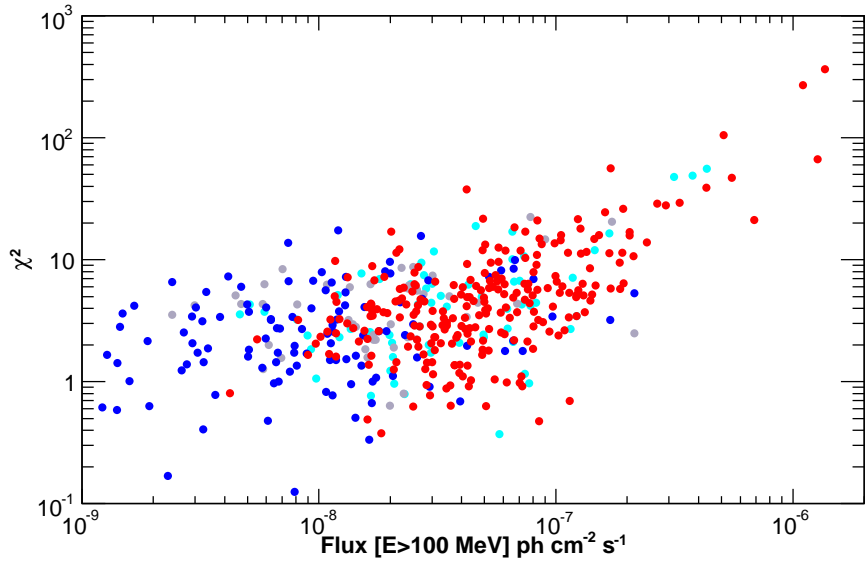}}}
\caption{Spectral curvature index vs.\ flux for FSRQs (red) and BL~Lacs (cyan:\ LSPs, gray:\ ISPs, blue:\ HSPs) in the clean sample.}
\label{fig:chi2_flux}
\end{figure}

Since the flux of each source is evaluated in five different bands, the
departure of the spectrum from a power-law (PL) shape can be estimated via a
simple $\chi^2$ test, referred to as the curvature index \citep{Catalog}.
Figure~\ref{fig:chi2_flux} displays the curvature index vs.\ total flux for
FSRQs and for the three BL~Lac subclasses.  The trend corroborates the findings
reported in \cite{Spe09}, namely, that the spectra of bright FSRQs and
LSP-BL~Lacs show strong departures from a PL shape while those of HSP-BL~Lacs
are essentially compatible with PLs.  Note that due to their harder spectra,
any deviation from a PL behavior would be more easily visible for HSP-BL~Lacs.
Because of larger statistical uncertainties, no conclusion can be drawn from
the $\chi^2$ test for fainter sources (a similar limitation affects the
variability index and is described below).

\subsection{Variability}

One of the defining characteristics of AGNs is their variability, measured at
all time scales and at all wavelengths.  In the LBAS sample, 46 sources (35
FSRQs, 10 BL~Lacs, and one blazar of unknown type) were flagged as variable
based on the results of a $\chi^2$ test applied to weekly light curves covering
the first three months of the LAT sky survey.  Recently, \citet{LBASvar}
detected variability in 64 LBAS sources by analyzing weekly light curves over a
period of 11 months.

The electronic version of the 1FGL catalog provides the light curves for all
sources over 11 time intervals of 30.37 days each in the energy range
100~MeV--100~GeV.  The procedure used to make these light curves is described
in \citet{Catalog}.  Our analysis of the light curves of the sources in the
clean sample shows that only six sources (3C~454.3, PKS~1510$-$08, 3C~279,
PKS~1502$+$106, 3C~273, and PKS~0235$+$164) have a maximum monthy flux
($E > 100$~MeV) greater than $10^{-6}$ \pflux{} (see also
Figure~\ref{fig:flux_lat_egret}b).  The number of sources with peak flux values
above $10^{-6}$ \pflux{} increases if we consider shorter time intervals, as
illustrated by the 41 Astronomer's Telegrams issued by the LAT collaboration
(see below), mostly related to sources that showed short flares that reached
this flux level on the time scale of days.

\begin{figure}
\plotone{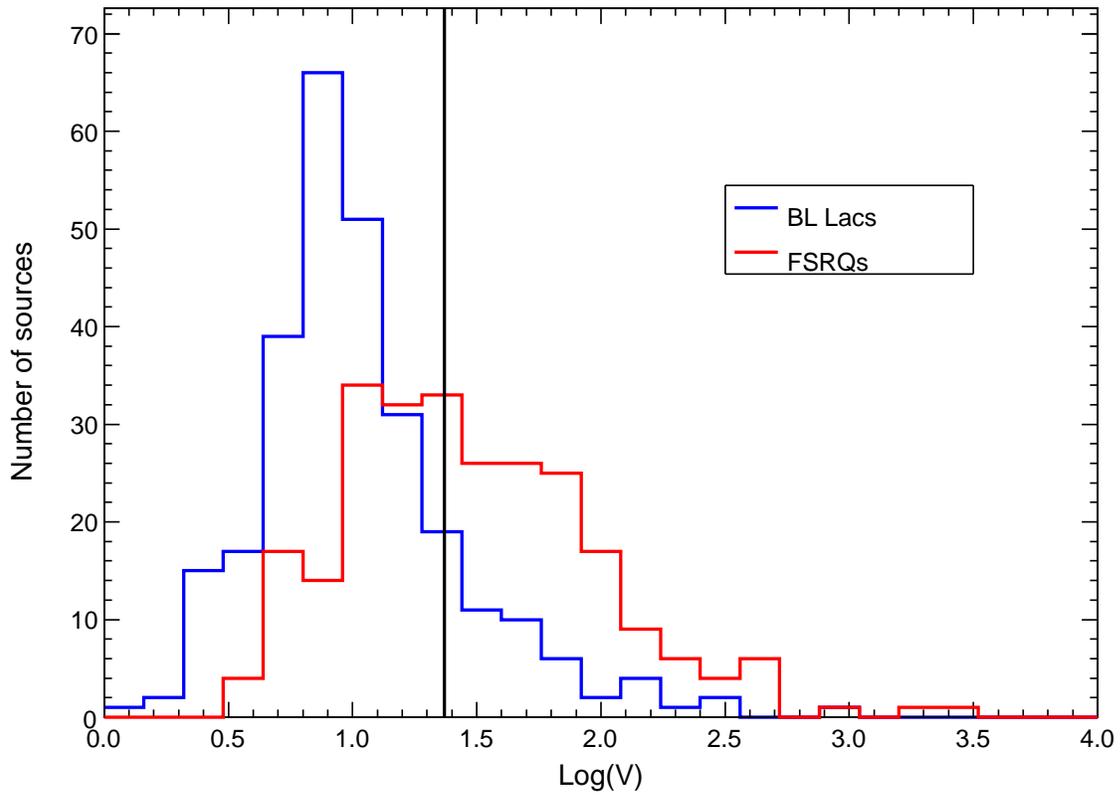}
\caption{Distribution of the variability index for all BL~Lacs and FSRQs in the clean sample.
The vertical line is drawn at $V = 23.21$, which corresponds to a 99\% probability that a source is variable.}
\label{fig:varind}
\end{figure}

\begin{figure}
\plotone{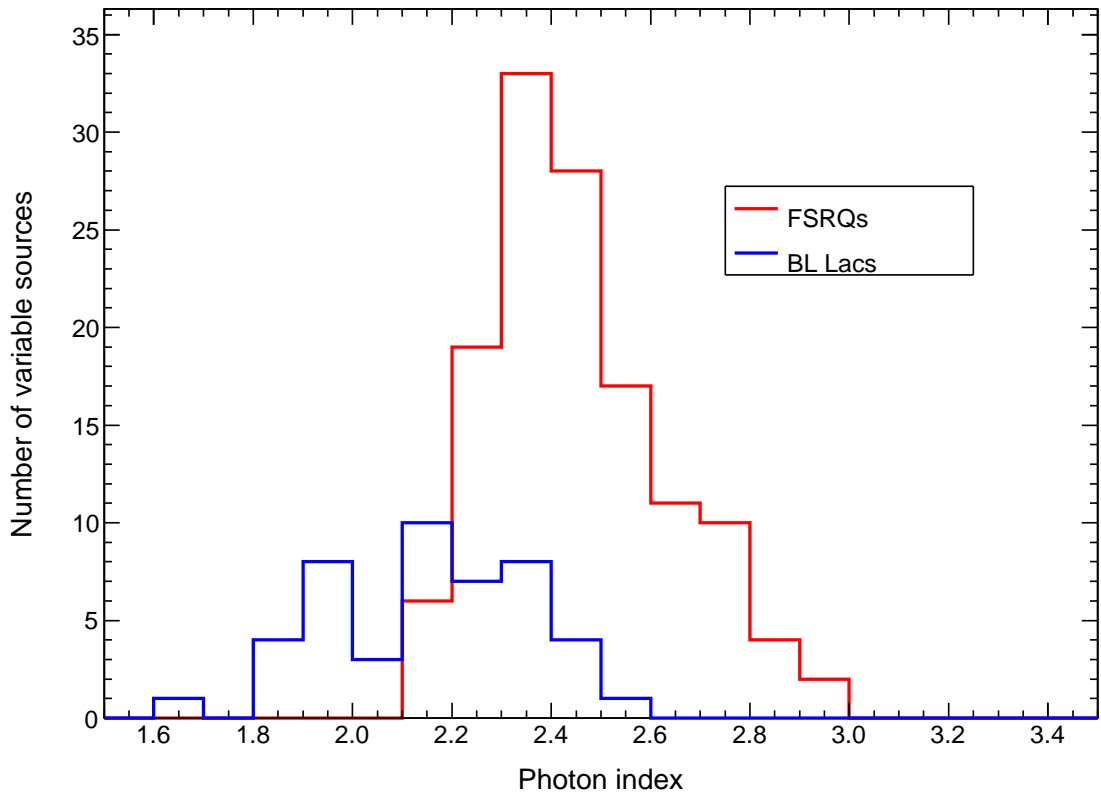}
\caption{Distribution of the photon spectral index for the variable BL~Lacs and FSRQs in the clean sample.}
\label{fig:varph}
\end{figure}

The 1FGL catalog also lists a variability index $V$, obtained using a simple
$\chi^2$ test, that can be used to determine the probability that a source is
variable.  Sources for which $V > 23.21$ have a 99\% probability (for 10
degrees of freedom) of being variable.  Figure~\ref{fig:varind} shows the
distribution of the variability index for blazars (BL~Lacs and FSRQs) in the
clean sample.  189 blazars in the clean sample are found to be variable (to the
right of the vertical line in Figure~\ref{fig:varind}); they comprise 129 FSRQs
(68\%), 46 BL~Lacs (24\%), five AGNs of other types (NGC~1275, B2~0321$+$33B,
4C~$+$15.54, B2~1722$+$40, and CGRaBS~J2250$-$2806), and nine blazars of
unknown type.  The 46 variable BL~Lacs include 24 LSPs, nine ISPs, 10 HSPs, and
three BL~Lacs (CRATES~J0058$+$3311, CGRaBS~J0211$+$1051, and
CRATES~J1303$+$2433) for which there are not enough multiwavelength data to
permit an SED classification.  Figure~\ref{fig:varph} shows the distribution of
the photon spectral index for the variable blazars.  This distribution is
dominated by sources with $\Gamma > 2.2$.  These are quite bright sources
observed at energies higher than the peak energies of their SEDs, where the
amplitude of the variability is generally larger.

\begin{figure}
\plotone{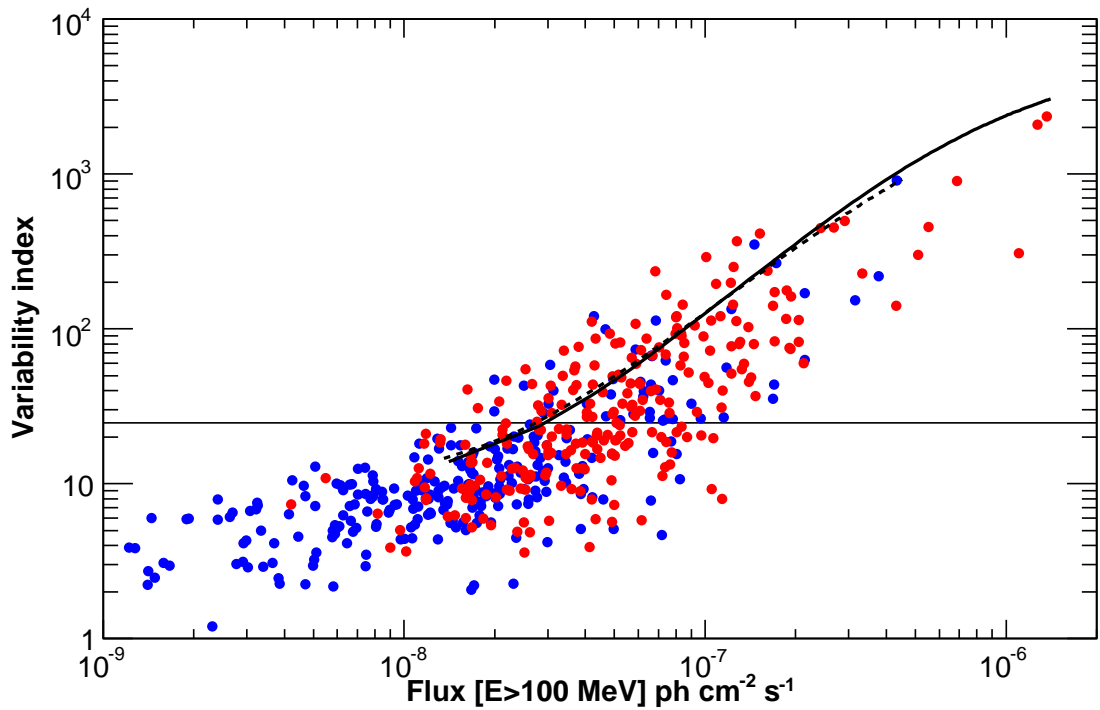}
\caption{Variability index versus flux for FSRQs (red) and BL~Lacs (blue) in the clean sample.  The curves display the evolution of the
variability index for the FSRQ 3C~454.3 (solid) and the BL~Lac object AO~0235+164 (dashed) that would be observed
for the same temporal variation but lower mean fluxes.}
\label{fig:varind_flux}
\end{figure}

It is important to bear in mind that for a source to be labeled as variable on
the basis of its variability index, it must be both intrinsically variable and
sufficiently bright.  Larger statistical flux uncertainties obtained for
fainter sources lead to a reduction of the variability index for a given
fractional flux variation.  To illustrate this effect, the variability index is
plotted against the flux in Figure~\ref{fig:varind_flux} for FSRQs and BL~Lacs.
The curves display the evolution of the variability index for two sample 
sources (the FSRQ 3C~454.3 and the BL~Lac object AO~0235$+$164) that would be
observed for the same temporal variation but lower mean fluxes.  As a result of
this effect, fainter sources appear less variable than brighter sources simply
because we can not measure their variability as well.  This may also explain
why only $\sim$17\% of the 1LAC BL~Lacs, which are generally fainter and harder
than FSRQs, are found to be variable by this criterion.  The histograms in
Figure~\ref{fig:varind} are artificially broadened by the flux dependence of
the variability index shown in Figure~\ref{fig:varind_flux}.  Allowing for the
behavior shown by the the trends for 3C~454.3 and AO~0235$+$164, there is no
strong evidence for a significant difference in the intrinsic variability of
FSRQs and BL~Lacs.

The LAT Collaboration routinely issues Astronomer's Telegrams (ATels) to alert
the community to transient sources or sources that are exhibiting flaring
states in order to encourage simultaneous multiwavelength follow-up.  Through
2010 Jan 20, 43 1LAC sources (and one at low latitude) have been the subjects
of ATels from the LAT team.  They comprise 32 FSRQs (with redshifts as high as
$z = 2.534$, for B3~1343+451), 10 BL~Lacs, and one AGN of another type
(CGRaBS~J2250$-$2806, a narrow-line radio galaxy).

\subsection{\label{sec:z}Redshift Distributions}

\begin{figure}
\centering
\resizebox{15cm}{!}{\rotatebox[]{0}{\includegraphics{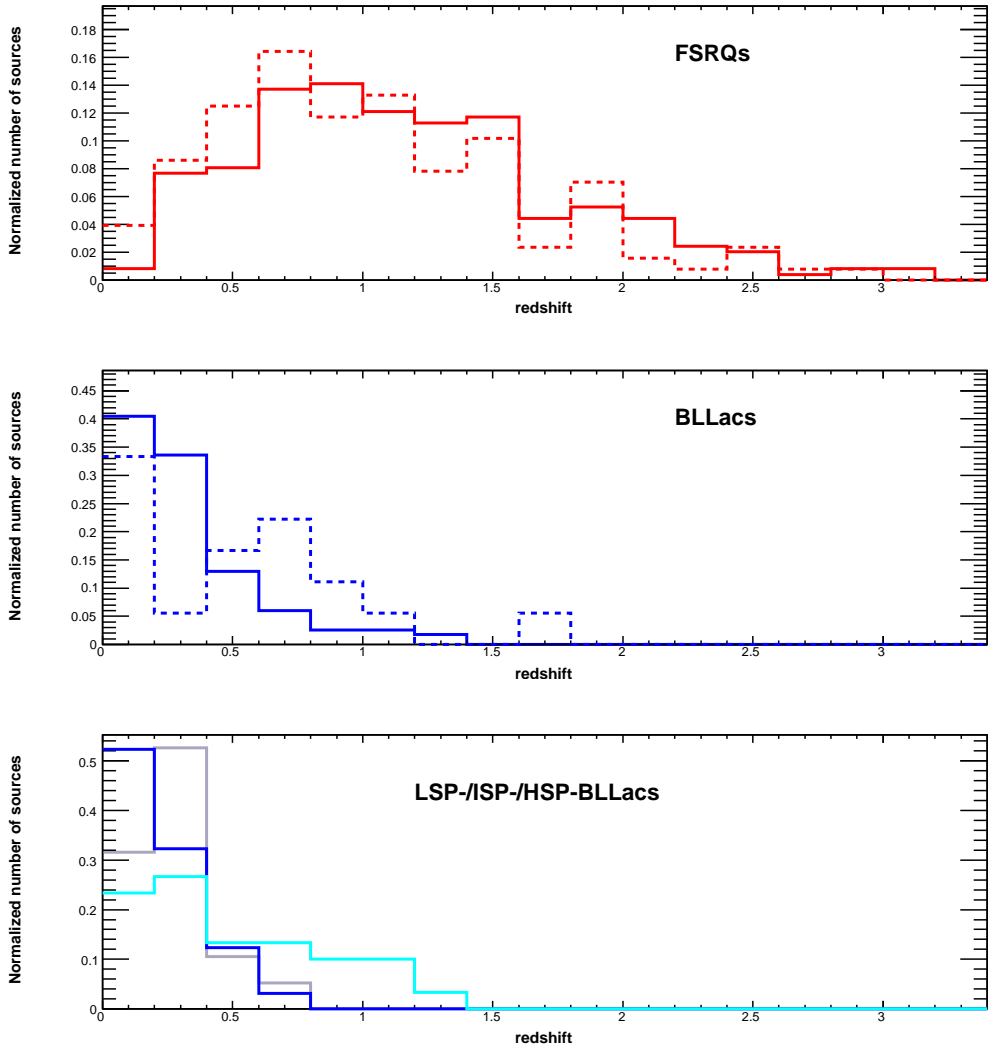}}}
\caption{Top:\ Normalized redshift distribution for the FSRQs in the clean sample (solid histogram). The redshift distribution for FSRQs in the
{\it WMAP} catalog is also shown for comparison (dashed histogram).  Middle:\ The same for BL~Lacs.  Bottom:\ Redshift distributions for
LSP-BL~Lacs (cyan), ISP-BL~Lacs (gray), HSP-BL~Lacs (blue) in the clean sample.}
\label{fig:redshift}
\end{figure}

The redshift distributions for FSRQs and BL~Lacs in the clean sample are
presented in Figure~\ref{fig:redshift} along with the corresponding ones for
the {\it WMAP} blazars \citep{Hin07,Gio09}, which constitute a flux-limited
all-sky sample above 1~Jy in the central {\it WMAP} observing band (Q band,
41~GHz).  Note that only 121 out of 291 (42\%) high-confidence 1LAC BL~Lacs
have measured redshifts.  This is notably worse than for LBAS, in which 29 out
of 42 (69\%) BL~Lacs had measured redshifts.  The redshift distributions of
\mbox{$\gamma$-ray}--detected blazars---both for FSRQs and for BL~Lacs---are
fairly similar to those for {\it WMAP}, peaking around $z=1$ for FSRQs and at a
lower redshift for BL~Lacs (the lowest-redshift 1LAC BL~Lac has $z = 0.030$).
The highest redshift for a high-confidence 1LAC FSRQ is $z=3.10$.

\begin{figure}
\centering
\resizebox{10cm}{!}{\rotatebox[]{0}{\includegraphics{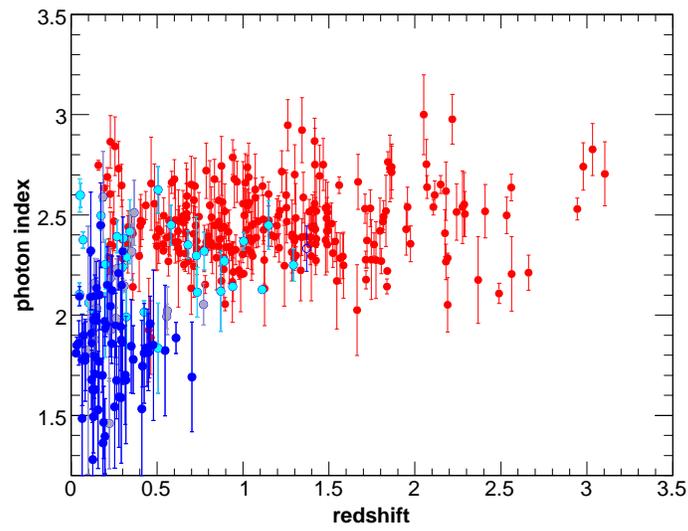}}}
\caption{Photon spectral index vs.\ redshift for the FSRQs (red) and BL~Lacs (cyan:\ LSPs, gray:\ ISPs, blue:\ HSPs) in the clean sample.}
\label{fig:index_redshift}
\end{figure}

\begin{figure}
\centering
\resizebox{13cm}{!}{\rotatebox[]{0}{\includegraphics{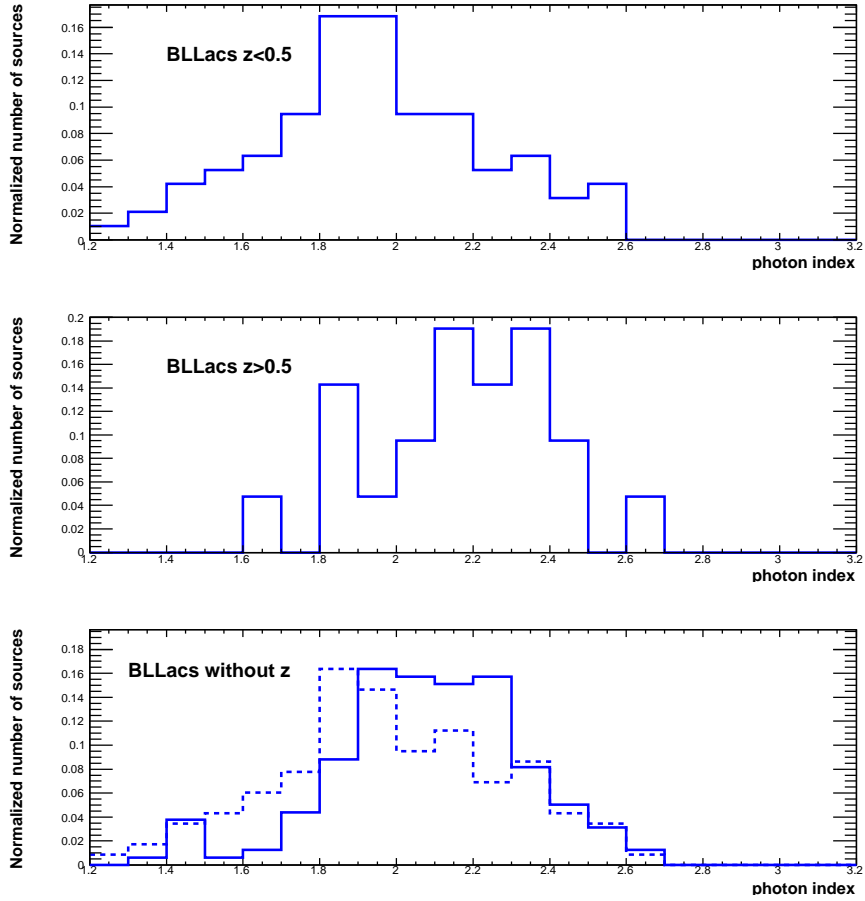}}}
\caption{Photon spectral index distributions for BL~Lacs in the clean sample with $z<0.5$ (top), with $z>0.5$ (middle), and with unknown redshifts
(bottom, solid histogram). The total distribution for BL~Lacs with known redshifts is shown for comparison as a
dashed histogram in the bottom panel.}
\label{fig:bll_z}
\end{figure}

The photon spectral index is plotted against redshift in
Figure~\ref{fig:index_redshift}.  For FSRQs, no significant evolution is
visible.  This behavior is compatible with what was previously observed for
LBAS.  The attenuation effect of the extragalactic background light (EBL) would
tend to introduce spurious evidence of evolution \citep{Chen04}, but the soft
spectra of FSRQs and the common presence of spectral breaks at a few GeV
\citep{Spe09} both minimize this effect.  A stronger evolution is seen for
BL~Lacs:\ hard sources are mostly located at low redshifts, while most
high-redshift sources are softer than average (though it is important to bear
in mind that most BL~Lacs do not have measured redshifts).  This trend was less
clear for the LBAS BL~Lacs \citep{LBAS} due to lower statistics.
Figure~\ref{fig:bll_z} displays the photon spectral index distributions for
sources with $z<0.5$ (top) and $z>0.5$ (middle), which are clearly different.
This provides some insight into the properties of BL~Lacs without measured
redshifts.  The photon spectral index distributions of BL~Lacs with measured
redshifts and with unknown redshifts are shown in the bottom panel of
Figure~\ref{fig:bll_z}.  The distribution of BL~Lacs with unknown redshifts
includes notably fewer hard sources than that for BL~Lacs with known redshifts.
A Kolmogorov-Smirnov (K-S) test yields a probability of $6 \times 10^{-3}$ that
the two distributions are drawn from the same underlying population.  The
similarity of the distribution of BL~Lacs with unknown redshifts to that of
BL~Lacs with $z > 0.5$ (K-S probability of 0.54 that the two distributions are
drawn from the same underlying population) supports the idea of some bias
toward higher redshifts for this class of objects.

\subsection{Luminosity Distributions}

\begin{figure}
\centering
\resizebox{10cm}{!}{\rotatebox[]{0}{\includegraphics{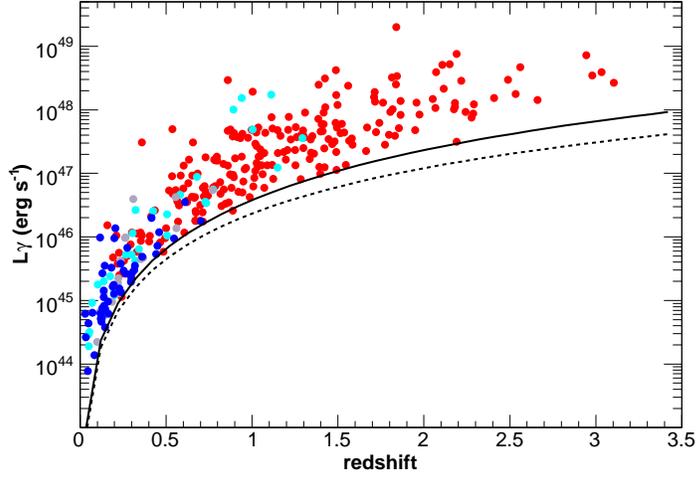}}}
\caption{\mbox{$\gamma$-ray} luminosity vs.\ redshift for the different AGN classes (red:\ FSRQs, cyan:\ LSP-BL~Lacs,
gray:\ ISP-BL~Lacs, blue:\ HSP-BL~Lacs, magenta:\ radio galaxies) in the clean sample.  The curves correspond to approximate
experimental limits calculated for a photon spectral index $\Gamma = 2.2$ (solid) and $\Gamma = 1.8$ (dashed).}
\label{fig:Lum_redshift}
\end{figure}

The \mbox{$\gamma$-ray} luminosity $L_\gamma$ is calculated as in \cite{Ghi09}:\
$$
L_\gamma = 4 \pi d_L^2 \frac{S (E_1, E_2)}{(1+z)^{2-\Gamma}}
$$
where $d_L$ is the luminosity distance, $\Gamma$ is the photon spectral index,
$S$ is the energy flux, and $E_1$ and $E_2$ are the lower and upper energy
bounds (taken here to be 100~MeV and 100~GeV) respectively.  Implicit in this
derivation is the assumption that $E_2\gg E_1$, which is satisfied for our
calculations.  Figure~\ref{fig:Lum_redshift} shows $L_\gamma$ plotted against
the source redshift.  The curves represent approximate instrumental limits
calculated for two photon indices, $\Gamma = 1.8$ and $\Gamma = 2.2$.  The
low-redshift FSRQs with $L_\gamma > 10^{48}$ erg s$^{-1}$ could potentially
still be detected at redshifts $z > 3.1$.

\begin{figure}
\centering
\resizebox{10cm}{!}{\rotatebox[]{0}{\includegraphics{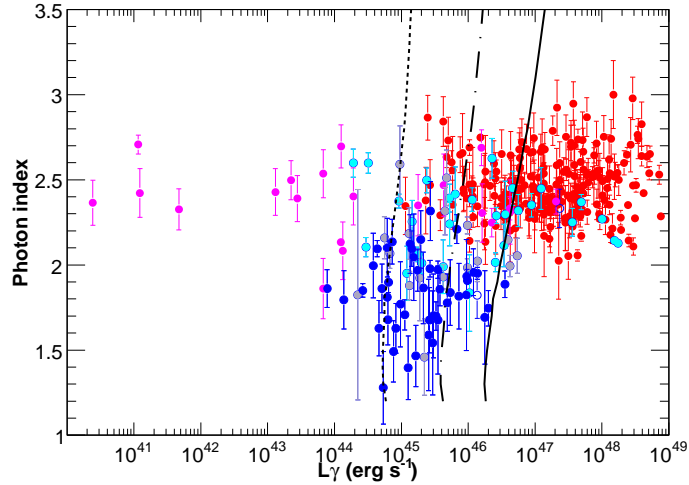}}}
\caption{Photon spectral index vs.\ \mbox{$\gamma$-ray} luminosity for the different AGN classes (red:\ FSRQs,
cyan:\ LSP-BL~Lacs, gray:\ ISP-BL~Lacs, blue:\ HSP-BL~Lacs, magenta:\ radio galaxies) in the clean sample.  The curves represent
approximate instrumental limits for $z=0.2$ (dashed), $z=0.5$ (dot-dashed), and $z=1$ (solid).}
\label{fig:index_lum}
\end{figure}

Figure~\ref{fig:index_lum} shows the photon spectral index plotted against the
\mbox{$\gamma$-ray} luminosity.  The Pearson correlation coefficient for the
two parameters is 0.17.  It is important to bear in mind two issues when
interpreting this correlation:\ the higher flux limit for soft sources (see
Figures~\ref{fig:index_flux} and \ref{fig:flux}) and the difference in redshift
distributions between FSRQs and BL~Lacs (see Figure~\ref{fig:redshift}).  Given
their relative softness (and thus, high flux limit), FSRQs are excluded from
the soft-faint region.  BL~Lacs are also partly excluded because of their
relatively low fluxes.  Luminosity limits calculated for different redshifts
(with the same approach as for the flux limit shown in
Figure~\ref{fig:index_flux}) are displayed for reference in
Figure~\ref{fig:index_lum} and illustrate somewhat the effect of the Malmquist
bias\footnote{In a flux-limited sample, sources located at larger distances
appear more luminous than closer ones since fainter sources are below the
detection threshold \citep{Mal20}.}.

\section{Multiwavelength Properties of 1LAC Sources}

\subsection{Sources Detected at TeV Energies}

Over the last two decades, ground-based \mbox{$\gamma$-ray} instruments operating in
the ``TeV'' or very-high-energy (VHE; $E\gtrsim100$~GeV) regime have detected
32 AGNs, with the pace of discovery increasing significantly\footnote{TeVCat
(\url{http://tevcat.uchicago.edu/}) presents an up-to-date catalog of TeV
sources.} as the latest generation of instruments---CANGAROO, H.~E.~S.~S.,
MAGIC and VERITAS---has been commissioned.  Of these 32 TeV AGNs, the majority
are BL~Lacs (23 HSPs, three ISPs, and two LSPs), with the remainder comprising
one FSRQ, two FR-I radio galaxies, and one AGN of unknown type.  A detailed
analysis of the TeV AGNs detected with {\it Fermi} during the first 5.5~months
of operation is given by \citet[][and see references to the TeV detections
therein]{FermiTeV}.  Five new TeV AGNs have since been detected
(1ES~0414$+$009, \citealt{ATelHofmann}; PKS~0447$-$437, \citealt{ATelRaue};
RBS~0413, \citealt{ATelOng1}; 1ES~0502$+$675, \citealt{ATelOng2};
VER~J0521$+$211, \citealt{ATelOng3}).  Two of these detections and one previous
detection \citep[PKS~1424$+$240;][]{pks1424} were motivated directly by the
detection of GeV emission with {\it Fermi}.

\begin{figure}
\centering
\plotone{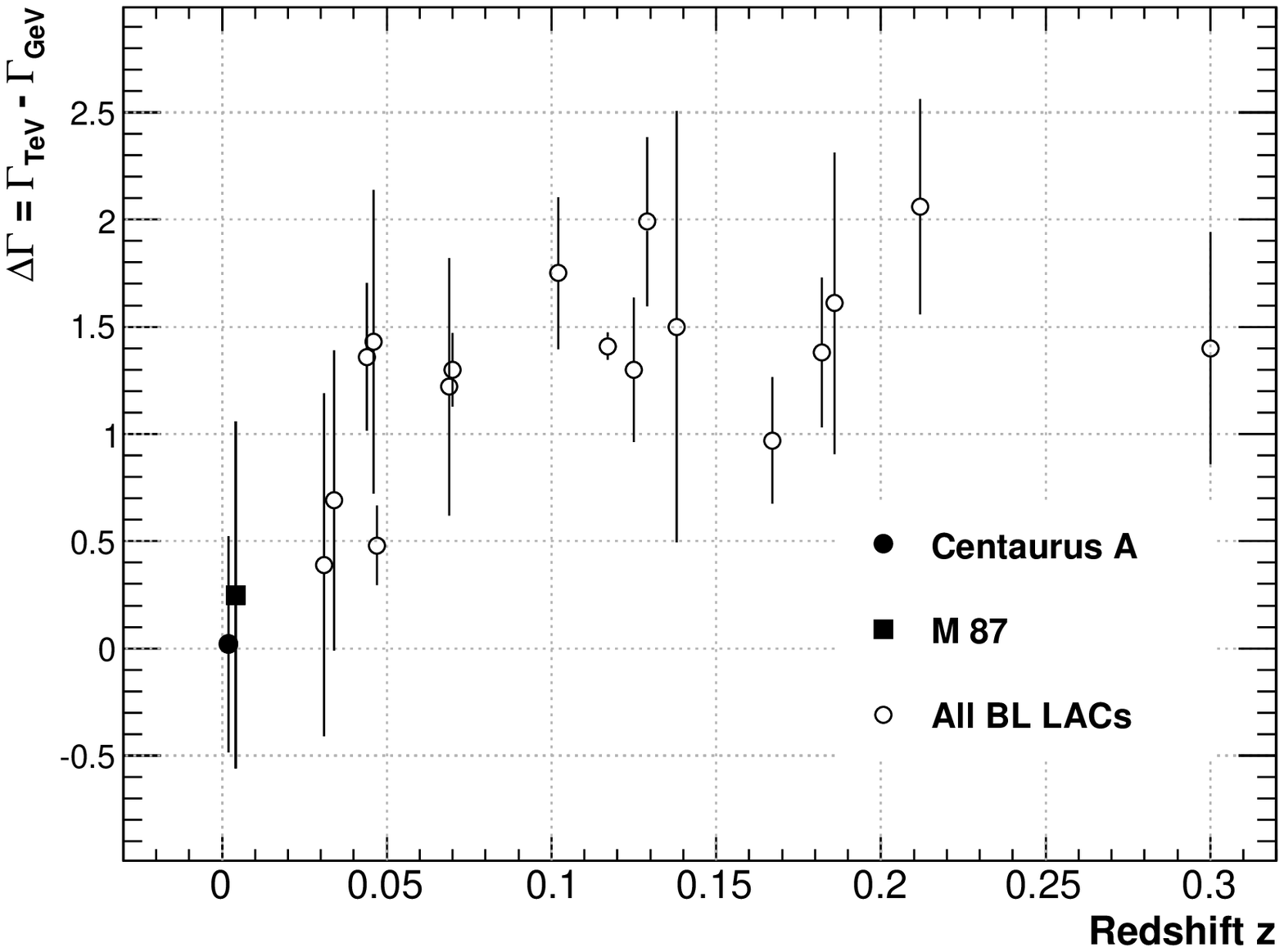}
\caption{GeV/TeV spectral break $\Delta\Gamma \equiv \Gamma_\mathrm{TeV} - \Gamma_\mathrm{GeV}$ vs.\ redshift.
Only sources with well measured redshifts and TeV photon spectral indices are shown.  The TeV measurements come
from the references given in \citet{FermiTeV}; new measurements for PKS~2005$-$489 \citep{tevindex:pks2005} and
S5~0716$+$714 \citep{tevindex:0716} are used.  3C~279, an FSRQ whose TeV spectrum was measured during an extreme
flaring episode, is also excluded.}
\label{fig:break}
\end{figure}

Of the TeV AGNs, the 28 listed in Table~\ref{tbl-tev} are detected by
\textit{Fermi} as 1FGL sources, with a mean photon spectral index of
$\langle\Gamma_\mathrm{GeV}\rangle=2.02\pm0.01$.  Taking only the subsample of
25 GeV--TeV BL~Lacs, the mean index is
$\langle\Gamma_\mathrm{GeV}\rangle=1.92\pm0.01$ ($\sigma_\Gamma=0.26$),
compared with $2.07\pm0.01$ ($\sigma_\Gamma=0.28$) for the larger sample of
{\it Fermi} BL~Lacs (see Figure~\ref{fig:index}b), illustrating that the TeV
sources are among the hardest BL~Lacs in the GeV regime.  The measured spectra
of the majority of the GeV--TeV BL~Lacs are well described by power laws in
both regimes.  For many of the sources, the photon spectral index of the GeV
emission differs significantly from that of the TeV emission, which may be an
indication of the presence of a break in the \mbox{$\gamma$-ray} spectrum between the
two regimes.  However, most of the TeV spectra have not been measured
simultaneously with the GeV spectra, and caution is advised when comparing the
spectra in detail.  The largest such break, consistent with
$\Delta\Gamma \equiv \Gamma_\mathrm{TeV}-\Gamma_\mathrm{GeV}\sim2$, is evident
in the spectra of 1ES~1101$+$496, H~1426$+$428 and PG~1553$+$113.  By contrast,
the spectra of the nearby radio galaxies M~87 and Cen~A show no evidence of a
spectral break.  The mean break index is $\langle\Delta\Gamma\rangle=1.3$. The
values of $\Delta\Gamma$ for the GeV--TeV sources are shown in
Figure~\ref{fig:break}, plotted against the redshift of the source.  Among the
possible explanations for the apparent deficit of sources with small
$\Delta\Gamma$ at high redshift is the effects of pair production with the EBL,
which is expected to introduce a redshift-dependent steepening into the TeV
spectra of extragalactic objects.

The four TeV AGNs not detected thus far by {\it Fermi} are RGB~J0152$+$017,
1ES~0229$+$200, 1ES~0347$-$121 and PKS~0548$-$322, all HSP-BL~Lacs.
Extrapolating the measured power-law TeV spectra from all the TeV AGNs down to
200~GeV, it is evident that these four have among the smallest fluxes of all
TeV AGN at this energy.

\subsection{Sources Detected Previously at GeV Energies}

Of the 709 sources in the 1LAC, 114 sources were included in the BSL and were
thus significantly detected (at $10\sigma$ or greater) during the first three
months of LAT observations.  All of those sources were previously associated
with AGNs as part of LBAS.  Three low-confidence LBAS sources
(0FGL~J0909.7$+$0145, 0FGL~J1248.7$+$5811, and 0FGL~J1641.4$+$3939) are not
confirmed in the 1LAC sample. 

Ten years after EGRET, it is interesting to look at the fraction of the AGNs
that were active in the EGRET era and are detected again by the LAT with a
comparable flux.  We consider two sources to be ``positionally coincident''
when the separation between their positions is less than the quadratic sum of
their 95\% error radii.  In the 1LAC sample, 63 AGNs are positionally
coincident with 3EG sources.  Of these, the 3EG catalog lists 51 sources as AGN
identifications and four as AGN associations.  The 3EG catalog listed a total
of 62 sources as AGN identifications, so there are 11 identified EGRET AGN that
are not positionally coincident with 1LAC sources.  45 sources from the 1LAC
have positions compatible with the revised EGRET catalog \citep[EGR;][]{EGR},
while 22 sources are positionally coincident with sources from the high-energy
EGRET catalog \citep[GEV;][]{Lamb1997}.  In all, 75 1LAC sources have
coincident detections from one or more of these three EGRET catalogs.

11 of the sources in this catalog are included in the first year {\it AGILE}
catalog \citep[1AGL;][]{AGILEcatalog}.  All 11 are identified as blazars by
{\it AGILE}, and indeed, these 11 comprise the total sample of blazars in the
1AGL catalog.  Only two of these {\it AGILE}-detected sources do not have 
similar EGRET detections.  In all, 77 1LAC sources have been cataloged by other
GeV instruments. 

These 77 sources are listed in Table \ref{tbl-fmrdetect} along with the mean
fluxes and photon indices measured by the LAT and by EGRET.  These AGNs are
composed of 49 FSRQs, 21 BL~Lacs, four AGNs of other types, and three AGNs of
unknown types.

During its 4.5-year mission, EGRET found few sources with flux
(${E > 100}$~MeV) less than ${10 \times 10^{-8}}$~\pflux{}. A large number
of the 1LAC sources have fluxes well below this value; such sources would not
have been visible to EGRET.  Some sources, such as 1FGL~J0428.6$-$3756
(associated with PKS 0426$-$380) and 1FGL~J2229.7$-$0832
(associated with PKS 2227$-$08), have flux values well above the EGRET
threshold but were not seen by EGRET and yet are not noted as being variable in
the 1FGL data.  These examples continue to demonstrate that long-duration
variability is evident in AGN \mbox{$\gamma$-ray} flux.

All of the EGRET sources seen at $10\sigma$ significance and associated with
flaring blazars have been detected by the LAT and appear with blazar
associations.

\subsection{\label{sec:hardX}Sources Detected in the Hard X-ray Band}

In recent years, a new generation of hard \mbox{X-ray} telescopes has drastically
improved our knowledge of the source populations in this energy band.  The
{\it Swift}-BAT and {\it INTEGRAL}-IBIS instruments have been conducting
surveys at hard \mbox{X-ray} energies; the most recent catalogs are the fourth IBIS
catalog \citep{bird} of 723 sources detected at 17-100 keV and the 54-month
Palermo BAT catalog \citep{swiftCat} of 1049 sources\footnote{\citet{swiftCat}
describes the 39-month catalog; the 54-month catalog is available at:\ 
http://www.ifc.inaf.it/cgi-bin/INAF/pub.cgi?href=activities/bat/index.html .}
detected at 14-150 keV.  These two telescopes perform in a complementary
fashion; while IBIS is the more sensitive instrument with better angular
resolution, it has a smaller field of view and has concentrated exposure in the
Galactic plane.  Conversely, the BAT instrument has lower instantaneous
sensitivity but has a much larger field of view and a much more uniform
exposure across the whole sky.  An initial comparison between the LBAS sources
and the fourth IBIS catalog by \citet{ubertini09} showed that only a small
subset of the $>$250 {\it INTEGRAL} AGNs, of which 19 are blazars, was detected
by the LAT.

Restricting the catalogs to $|b|>10\arcdeg$, there are 291 IBIS sources and
736 BAT sources.  If we consider all of the 1LAC sources with associated
counterparts that fall within the error circles of the hard \mbox{X-ray} catalogs, we
find that 50 of the 1LAC sources can be associated with known hard \mbox{X-ray}
sources.  Of these 50 sources, one appears only in the fourth IBIS catalog,
16 appear in both the fourth IBIS catalog and the 54-month Palermo BAT catalog,
and the remaining 35 are detected only in the 54-month Palermo BAT catalog.
This is still a very small subset of AGNs despite the depth of the LAT
catalog and the sky coverage of the BAT catalog.

Table~\ref{tab:hardXray} lists the high-confidence 1LAC AGNs detected in hard
\mbox{X-rays}.  This sample contains 27 FSRQs, 16 BL~Lacs, and seven AGNs of
other types. Most of the AGNs (54\%) detected in hard \mbox{X-rays} are FSRQs;
this is expected since the external Compton peak of these high-luminosity
objects reaches a maximum at MeV energies and declines in the hard
\mbox{X-rays}, maintaining a substantial fraction of the emitting power.  The
difference in the spectral shapes of BL~Lacs and FSRQs observed in hard
\mbox{X-rays} indicates that this energy range probes the high-energy tail of
the synchrotron peak in BL~Lacs and the ascending part of the Compton peak in
FSRQs.

Finally, there is evidence that the hard \mbox{X-ray} emission from most bright LAT
blazars is, apparently, missed in spite of the sub-milliCrab sensitivity
reached with {\it INTEGRAL} and {\it Swift} in the deepest fields.

\subsection{Radio Properties}

The high-latitude 1FGL sources associated with AGNs and presented here in the
1LAC are all radio sources at some level.  The large number of sources found in
CRATES can, by definition, be characterized as relatively bright, flat-spectrum
radio sources.  However, since a significant number of associations are found
from other catalogs, it is interesting to contemplate the distributions of the
flux densities, luminosities, and spectral indices.

\begin{figure}
\centering
\resizebox{12cm}{!}{\rotatebox[]{0}{\includegraphics{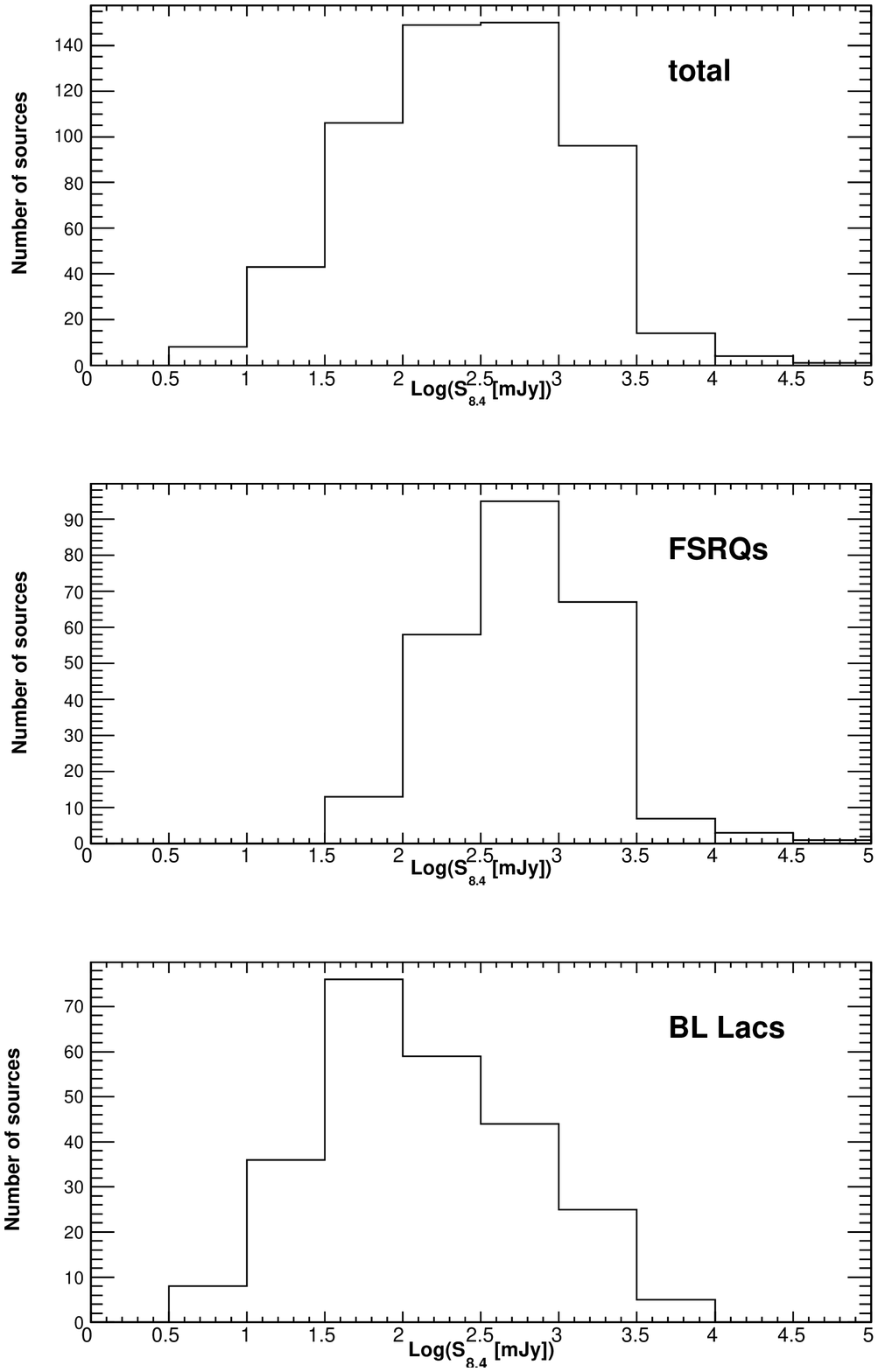}}}
\caption{Distributions of radio flux densities at 8.4~GHz for all sources in the clean sample (top), FSRQs (middle), and BL~Lacs (bottom).}
\label{fig:radioflux}
\end{figure}

In Figure~\ref{fig:radioflux}, we plot the distribution of the radio flux
density at 8.4~GHz for the clean sample and separately for the FSRQs and the
BL~Lacs in this sample.  The overall distribution, including also the blazars
of unknown type and the other AGNs, has one main peak (corresponding to
$\langle S_\mathrm{8\,GHz,\,1LAC}\rangle \sim 900$~mJy).  However, the FSRQs
are on average brighter than the BL~Lacs, with
$\langle S_\mathrm{8\,GHz,\,FSRQ}\rangle \sim 1200$~mJy and
$\langle S_\mathrm{8\,GHz,\,BLL}\rangle \sim 400$~mJy.

\begin{figure}
\centering
\resizebox{12cm}{!}{\rotatebox[]{0}{\includegraphics{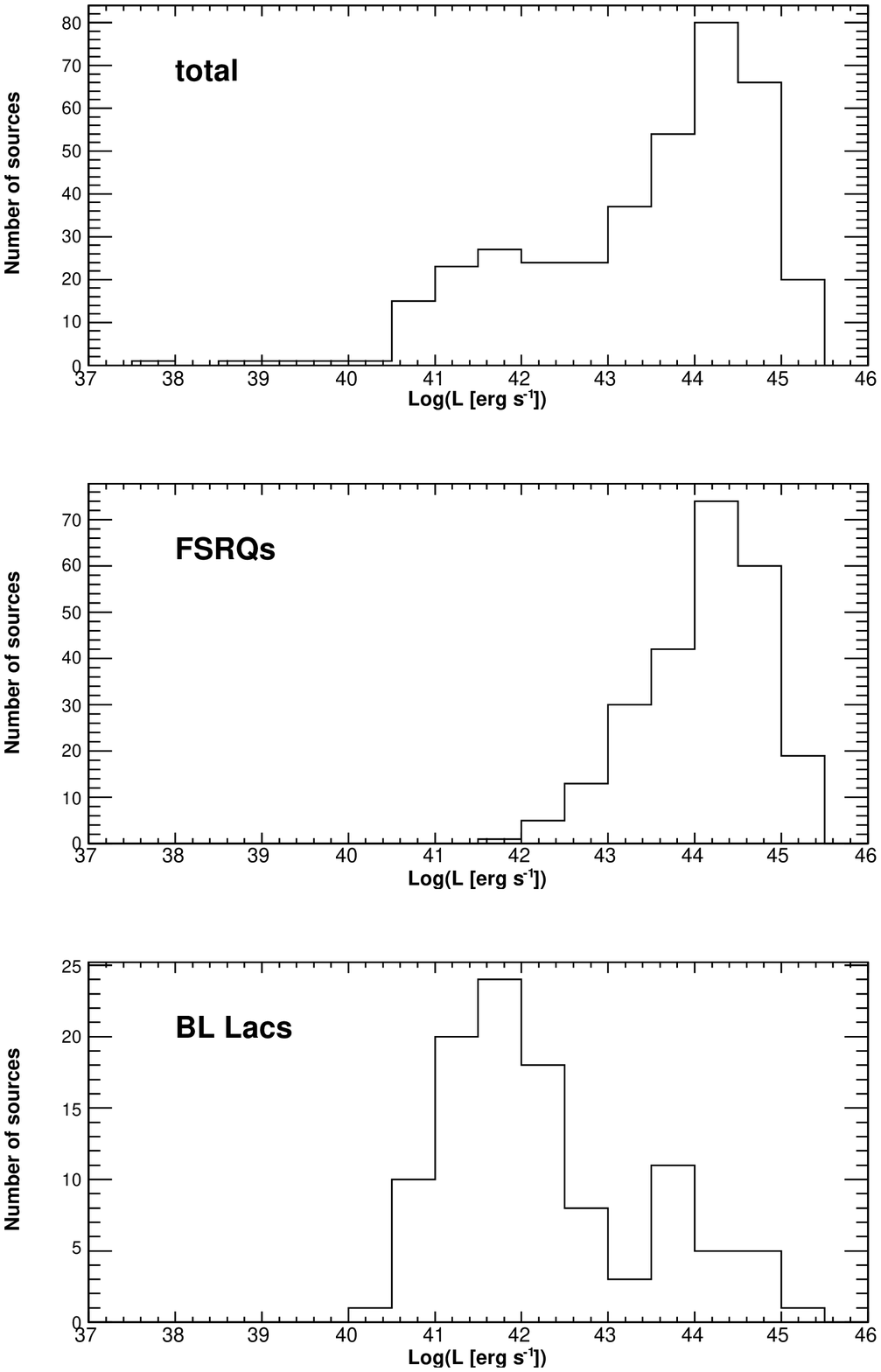}}}
\caption{Distributions of radio luminosities at 8.4~GHz for all sources in the clean sample (top), FSRQs (middle), and BL~Lacs (bottom).}
\label{fig:radiolum}
\end{figure}

In Figure~\ref{fig:radiolum}, we plot the corresponding radio luminosities,
calculated and $K$-corrected for the sources with measured redshifts.  The
distributions for FSRQ and BL~Lacs are clearly different, with the FSRQs
concentrated at higher radio luminosities
($\log$~$(L_{r,\,\mathrm{FSRQ}}$~[erg~s$^{-1}$]$)=44.1\pm0.7$~erg~s$^{-1}$) and
the BL~Lacs distributed over a broad range
($\log$~$L_{r,\,\mathrm{BLL}}$~[erg~s$^{-1}$]$)=42.3\pm1.1$~erg~s$^{-1}$).  This is
similar to the distribution found in the LBAS, and the only sources with radio
luminosities below $L_r = 10^{40}$~erg~s$^{-1}$ are a few nearby AGNs (Cen~A,
NGC~253, and NGC~4945).

\begin{figure}
\plotone{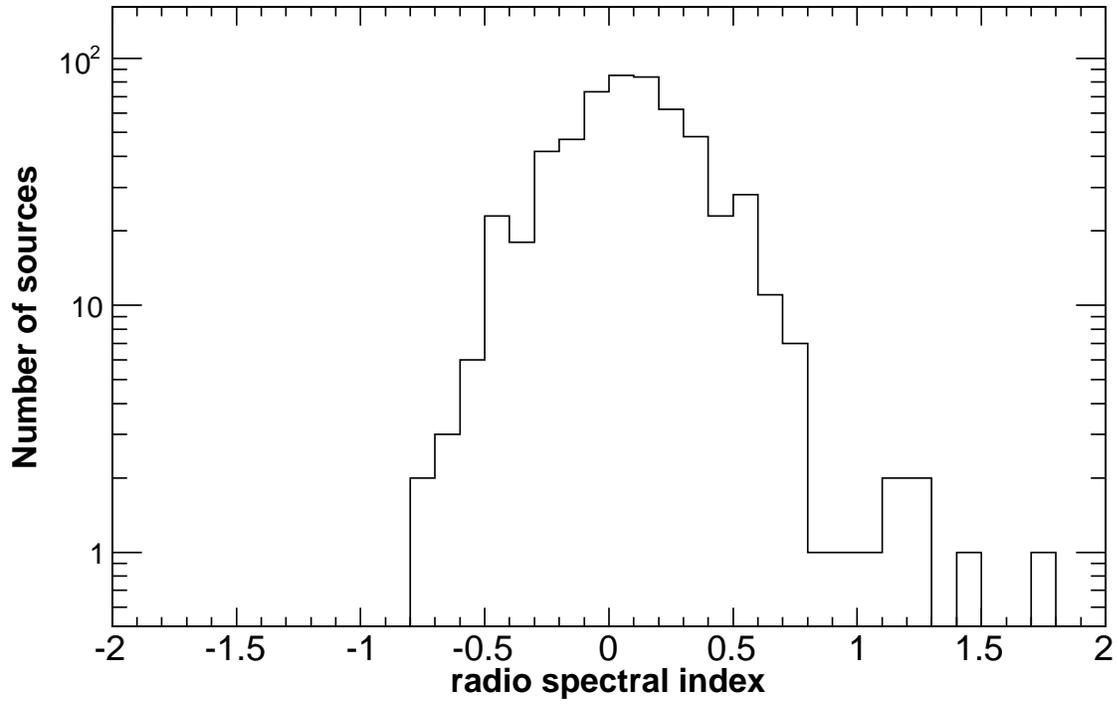}
\caption{Distribution of radio spectral indices between 1.4~GHz (0.8~GHz for southern sources) and 8.4~GHz for sources in the clean sample.}
\label{fig:radiospecind}
\end{figure}

Finally, in Figure~\ref{fig:radiospecind}, we plot the spectral index
distribution, calculated between either 1.4~GHz from NVSS for sources with
$\delta > -40\arcdeg$ or 0.84~GHz from the Sydney University Molonglo Sky
Survey \citep[SUMSS;][]{sumss} for sources with $\delta < -40\arcdeg$ and
8.4~GHz from CRATES (or other measurements from NED).  The overall distribution
is consistent with a flat spectral index ($\alpha=0.08 \pm 0.32$).  No
difference is found between FSRQs and BL~Lacs.  Interestingly, a tail of
sources with steeper spectral index is found, and it is mostly composed of
non-blazars, such as radio galaxies.  This suggests that \mbox{$\gamma$-ray}
sources can also be associated with radio sources that are less closely aligned
with the line of sight than blazars (see Section~\ref{sec:misaligned}).

Since the different classes have different properties in the \mbox{$\gamma$-ray} band,
the relationship between the radio emission and the high-energy emission merits
a deeper discussion.  A paper dedicated to this subject is in preparation, with
a focus on the correlation between radio and \mbox{$\gamma$-ray} fluxes.

\section{Discussion}

Of the 1043 high-latitude ($|b| > 10\arcdeg$) \mbox{$\gamma$-ray} sources in
the 1FGL catalog, 671 are associated with 709 AGNs; these constitute the 1LAC.
The 663 high-confidence AGN sources detected by {\it Fermi} in 11~months of
data can be compared with the 66 high-confidence AGNs detected by EGRET at
$>5 \sigma$ significance over its lifetime \citep{3EGcatalog}.  We estimate
that a large fraction of the other 372 high-latitude sources are likely AGNs on
the basis of their spectral and variability properties; indeed, we have found
plausible AGN candidates for $\sim$100 of them (see Section~\ref{sec:affil}.
In addition, $\sim$20 known millisecond pulsars, and possibly a comparable
number of yet unidentified millisecond pulsars, are among the remaining
high-latitude 1FGL sources.

The 1LAC represents a $\sim$5-fold increase in the number of sources associated
with \mbox{$\gamma$-ray} blazars over previously published source lists (in
particular, the 106 high-confidence blazars in the LBAS catalog).  It is likely
that other blazars remain unidentified, as implied by the observed north-south
anisotropy of the associated sources, which reflects the incompleteness of the
counterpart AGN catalogs.  This is partly supported by the similarity of the
\mbox{$\gamma$-ray} properties, including flux and photon spectral index, of
the unidentified sources to those of the 1LAC blazars \citep{unID}.  There are
also indications from our early radio follow-up of the unidentified sources
that the LAT is detecting blazars that are fainter in the radio than those
appearing in the catalogs of flat-spectrum radio sources from which we draw
candidate associations.

The overall properties of the 1LAC sources are generally consistent with those
of the LBAS \citep{LBAS,Spe09,SEDpaper}, including observations of:

\begin{enumerate}

\item A small number of non-blazar sources.

The misaligned AGNs include six radio galaxies, of which only three (NGC~1275,
Cen~A, and M~87) were previously reported.  Five of the six radio galaxies are
low-power FR-I galaxies associated with LAT sources with very high
probabilities ($P > 98$\%).  The sixth, NGC~6251, is a moderately powerful
radio galaxy \citep{Perley84} and has a lower association probability.  The
misaligned FR-II radio galaxy 3C~111 is associated with a LAT source (see
Table~\ref{tbl-lowlat}), but it is not a member of the 1LAC since it resides at
low Galactic latitude.  In addition, two steep-spectrum radio quasars and five
NLRGs are among the other non-blazar sources, though many of these have
prominent radio cores and some lack extended radio emission.

\item Redshift distributions peaking at $z \approx 1$ for 1LAC FSRQs and at low
redshift for 1LAC BL~Lacs with known redshifts.

The maximum redshift of the high-confidence 1LAC FSRQs is $z=3.10$, larger than
the maximum redshift ($z=2.286$) for an EGRET blazar.  For comparison, the
maximum redshift in both CGRaBS and Roma-BZCAT is for an FSRQ with $z > 5$.
The redshift distribution for BL~Lacs in the 1LAC with known redshift displays
a broad low-redshift peak from the lowest known redshift (for Mkn~421) at
$z=0.030$ to $z \approx 0.4$.  The good agreement with the {\it WMAP} redshift
distribution for FSRQs, with a comparable number of sources, supports the idea
that similar populations are being sampled by {\it WMAP} and the LAT.  Both the
FSRQ and BL~Lac redshift distributions are compatible with the corresponding
distributions from LBAS; the K-S probabilities that they derive from the same
parent distributions are 0.99 and 0.89 for FSRQs and BL~Lacs respectively.

No strong evidence for a new population of misaligned FSRQs emerging at lower
redshifts is found.  Other than NGC~6251, just mentioned, notably absent are
detections, at least at the 1LAC significance level of $TS > 25$, of nearby
powerful FR-II radio galaxies such as Cyg~A and Pic~A, though the low-latitude
FR-II radio galaxy 3C~111 is detected by the LAT.

\item A high BL~Lac/FSRQ ratio, close to unity.

This ratio is even higher than that found for the LBAS sources, which comprise
42 BL~Lacs and 57 FSRQs.   It stands in sharp contrast to the population of 3EG
blazars, in which FSRQs outnumber BL~Lacs by a factor of 3.

\item A high HSP/LSP ratio among BL~Lacs.

The large HSP/LSP radio for BL~Lacs is a result of the fact that the sample is
significance-limited.  The LAT detects hard-spectrum sources at higher
significance than soft-spectrum sources with comparable photon number fluxes.
Moreover, the LAT is far more sensitive to multi-GeV photons than was EGRET,
which lost sensitivity above several GeV due to self-vetoing effects.
Consequently, HSP-BL~Lacs, with harder spectra than LSP-BL~Lacs, are more
prevalent in the 1LAC than in the 3EG catalog.

\item Little evidence for different variability properties for FSRQs and
BL~Lacs.

No significant differences in the variability of FSRQs and BL~Lacs are observed
for sources at comparable fluxes.  Because HSP blazars are generally observed
at low flux levels, however, the comparison of variability properties primarily
concerns LSP/ISP-BL~Lacs and FSRQs.

\item A fairly strong correlation between photon spectral index and blazar
class for the detected sources.

FSRQs, which are almost all LSP blazars, are found to be soft, while BL~Lacs,
whether of the LSP, ISP, or HSP type, represent on average a population of
harder-spectrum sources.  The average photon spectral index
$\langle \Gamma \rangle$ continuously shifts to lower values (i.e., harder
spectra) as the class varies from FSRQs ($\langle \Gamma \rangle = 2.48$) to
LSP-BL~Lacs ($\langle \Gamma \rangle = 2.28$), ISP-BL~Lacs
($\langle \Gamma \rangle = 2.13$), and HSP-BL~Lacs
($\langle \Gamma \rangle = 1.96$).

\end{enumerate}

Furthermore, as noted by \citet{Ghi09} and seen in Figure~\ref{fig:index_lum},
the \mbox{$\gamma$-ray} blazars detected with the {\it Fermi}-LAT separate into
hard-spectrum, low-luminosity sources, primarily consisting of BL~Lac objects,
and high-luminosity, soft-spectrum sources, primarily consisting of FSRQs and
LSP-BL~Lacs.  Compared with the LBAS sample, the blazar divide now displays
an extension of the sample toward low-luminosity objects in each of the blazar
subclasses.  In addition, we identify a third branch consisting of radio
galaxies, which are distinguished from the aligned blazars by their soft
spectra and very low \mbox{$\gamma$-ray} luminosities.

The finding that FSRQs in the 1LAC almost all have soft ($\Gamma \ga 2.2$)
\mbox{$\gamma$-ray} spectra suggests a connection between the presence of strong
emission lines and the nonthermal electron maximum Lorentz factor or spectral
shape.  It is consistent with scenarios \citep{Ghi98,Boe02} in which the strong
ambient radiation fields, as revealed by the presence of emission lines,
affects the acceleration and cooling of particles and controls the formation of
the blazar SED.  The luminosity dependence of the \mbox{$\gamma$-ray} spectral
hardness is also consistent with the blazar sequence \citep{Ghi98,Fos98,
GhiTav08}, in which the frequencies of the synchrotron and Compton peaks are
inversely correlated with the apparent isotropic luminosities of the blazars.
This behavior has been interpreted in terms of an evolution of FSRQs into
BL~Lac objects as the supply of matter fueling the accretion disk declines, the
surrounding external radiation fields become less intense, and the jet weakens
\citep{Boe02,Cav02,GhiTav08}.

Important caveats must, however, be considered before drawing firm conclusions
about the blazar sequence from the 1LAC sources.  First, identification of
the blazars used to construct the blazar sequence is biased by the flux limits
and frequency ranges of the underlying radio and \mbox{X-ray} samples
\citep{Gio99,Pad03}, and different conclusions about its validity can be drawn
depending on the catalogs used to examine the sequence \citep{Niep06}.  Second,
specific sources that do not conform to the sequence have been identified
\citep{caccianiga2,j0810}.  Third, the 1LAC is significance-limited rather than
flux-limited, which results in preferential detection of lower-flux,
harder-spectrum sources.  This means that there is a strong bias against
detecting weak, soft-spectrum \mbox{$\gamma$-ray} sources from low-luminosity
LSP blazars that might violate the sequence.  On the other hand, hard-spectrum
HSP sources should be more easily detected, but there are no high-luminosity
HSP sources detected in the LAT band (see Figure~\ref{fig:index_lum}), as all
high-luminosity objects have SEDs of the LSP type.  This may, however, reflect
the difficulty of measuring the redshifts of BL~Lacs, despite efforts even with
10~m-class telescopes \citep{Shaw10}.  Even for sources associated with AGNs
with high probabilities, a large fraction lack redshift measurements, including
58\% of the high-confidence BL~Lacs and $\sim$30 BL~Lacs with $\Gamma \la 1.9$.
Considering the recent results of \citet{Plot10}, who find that many BL~Lacs
might have redshifts larger than 2 (up to $z=4.9$), and the indication that
BL~Lacs with unknown redshifts may be found preferentially at higher redshift
(see Section~\ref{sec:z}), it is possible that a number of distant,
high-luminosity HSP-BL~Lacs are included in the 1LAC sample.  According to our
SED classification, $\sim$20 BL~Lacs with unknown redshifts are already
classified as HSPs.  If they were all at high redshift (e.g., $z=2$), and
excluding those already detected at very high energy (which indicates a low
redshift due to a lack of EBL absorption), $\sim$7 objects could have
\mbox{$\gamma$-ray} luminosities $L_\gamma \ga 10^{48}$~erg~s$^{-1}$.

The existence of such sources, though quite rare, would be contrary to
expectations from the blazar sequence \citep{Ghi98,Fos98,GhiTav08}.  In the
absence of redshift measurements of possible high-luminosity, hard-spectrum
sources, we can only conclude that the {\it Fermi} results seem compatible
with the sequence.  Identifying hard-spectrum blazar sources at high redshifts
is also important for studies of the cosmological evolution of the
extragalactic background light \citep{EBL}, which are hampered by their
absence.

Related to this, it is interesting to compare the redshift distribution of the
LAT blazars (in particular, the FSRQs) with that of the objects detected by BAT
on {\it Swift} \citep{ajello09}.  Despite a fourfold increase in exposure time
with respect to the LBAS sample and a consideration of \mbox{$\gamma$-ray} sources
with $TS > 25$ in the 1LAC compared with $TS > 100$ in LBAS, the number of
LAT-detected blazars still drops sharply at $z \sim 2.5$, and only two of the
high-confidence 1LAC blazars have $z > 3$ (keeping in mind biases from the lack
of redshift measurements).  This is very different from the BAT survey, in
which half of all FSRQs have $z > 2$, and the distribution extends to
$z \sim 4$.  This behavior may be indicative of a shift in the SED peak
frequencies toward lower values (i.e., a ``redder'' SED) for blazars at high
redshifts.  Indeed, the jets of the high-redshift BAT blazars are found to be
more powerful than those of the LAT blazars and are among the most powerful
known \citep{Ghi10}.  The peak of the inverse Compton flux for these objects,
estimated to be in the MeV or even sub-MeV range, is located closer to the BAT
band than to the LAT band, and the LAT instead samples the cutoff region of the
inverse Compton spectrum.

Comparison of {\it Fermi} results with pre-launch {\it GLAST} expectations
depends on how the 1LAC results are characterized.  Except for a few dozen
millisecond pulsars, nearly all of the 1043 high-latitude \mbox{$\gamma$-ray}
sources in the 1FGL are likely to be AGNs, but only 671 high-latitude 1FGL
sources are associated with AGNs.  Adopting a radio/\mbox{$\gamma$-ray}
correlation, \citet{ss96} and \citet{cm98} predicted many thousands of blazars
to a flux limit of $2 \times 10^{-9}$ \pflux{} (for $E > 100$~MeV) reached
after five years for a source with $\Gamma = 2.1$, as did \citet{nt06}, who
treated the problem using radio and \mbox{X-ray} luminosity functions.
\citet{mp00}, using a physical blazar model matched to the EGRET data,
predicted an increased fraction of BL~Lac objects at fainter flux levels, with
nearly 2000 BL~Lacs and 1000 FSRQs at the one-year flux limit of
$4 \times 10^{-9}$ \pflux{} (for $E > 100$~MeV).  \citet{Der07} predicted that
$\sim$800 FSRQs and $\sim$200 BL~Lacs would be detected after one year, also
employing a physical blazar model.  Assuming the validity of the blazar
sequence, \citet{it09} predicted $\sim$600-1200 blazars to a flux limit of
$2 \times 10^{-9}$ \pflux{} (for $E > 100$~MeV).  However, predictions based on
a minimum flux depend sensitively on the source spectral index, as shown by
Figure~\ref{fig:index_flux}, so more reliable predictions needed to take this
effect into account.

The populations of AGNs that can be studied jointly at GeV and at TeV energies
has increased dramatically beyond what could have been expected based on the
experience of the EGRET era.  There is now a large and increasing population of
GeV--TeV AGNs that can be best studied through observations with both
{\it Fermi} and the ground-based \mbox{$\gamma$-ray} observatories.  Since its
launch, the number of TeV AGNs detected also by {\it Fermi} has steadily
increased as its exposure on these sources grows.  Complementing this, the TeV
observatories have started to observe the more promising {\it Fermi} AGNs, and
a number of detections have resulted.  One promising feature of joint GeV--TeV
observations is the power to infer the redshifts of BL~Lacs from spectral
modeling, as in the case of PG~1553$+$113 \citep{pg1553}.

\section{Conclusions}

We have presented the First LAT AGN Catalog (1LAC), consisting of 671
high-latitude \mbox{$\gamma$-ray} sources with $TS > 25$ measured in the first
11~months of the {\it Fermi}-LAT sky survey.  Due to multiple associations,
these sources are associated with 709 AGNs.  Besides a growing number of
blazars and six radio galaxies, the extragalactic \mbox{$\gamma$-ray} sky now
includes a small number of steep-spectrum quasars and narrow-line radio
galaxies.  Confirming many of the LBAS results, we find a strong correlation
between the 0.1-100~GeV \mbox{$\gamma$-ray} spectral index and the blazar type,
whether defined in terms of optical emission line strength or peak frequency of
the synchrotron component of the SED.  The relationship between the
\mbox{$\gamma$-ray} spectral properties and the spectral luminosity is
complicated by selection biases for source identification and redshift
determination, but the {\it Fermi} results seem compatible with the blazar
sequence.

The First LAT AGN Catalog represents significant progress in terms of the
number of detected sources, the diversity of source types, and the accuracy of
measured \mbox{$\gamma$-ray} properties.  This sample is likely to evolve in the
future as more unidentified sources can be associated with AGNs thanks to
better counterpart catalogs and continued correlated variability studies with
more LAT observations.  Bearing in mind the instrumental limitations described
above, the 1LAC provides an essential foundation on which to build a much
better understanding of the population of \mbox{$\gamma$-ray}--emitting AGNs.

\acknowledgments

\section*{Acknowledgments}

The \textit{Fermi} LAT Collaboration acknowledges generous ongoing support from
a number of agencies and institutes that have supported both the development
and the operation of the LAT as well as scientific data analysis.  These
include the National Aeronautics and Space Administration and the Department of
Energy in the United States; the Commissariat \`a l'Energie Atomique and the
Centre National de la Recherche Scientifique / Institut National de Physique
Nucl\'eaire et de Physique des Particules in France; the Agenzia Spaziale
Italiana and the Istituto Nazionale di Fisica Nucleare in Italy; the Ministry
of Education, Culture, Sports, Science and Technology (MEXT), High Energy
Accelerator Research Organization (KEK) and Japan Aerospace Exploration Agency
(JAXA) in Japan; and the K.~A.~Wallenberg Foundation, the Swedish Research
Council and the Swedish National Space Board in Sweden.

Additional support for science analysis during the operations phase is
gratefully acknowledged from the Istituto Nazionale di Astrofisica in Italy and
the Centre National d'\'Etudes Spatiales in France.

This research has made us of the NASA/IPAC Extragalactic Database (NED) which
is operated by the Jet Propulsion Laboratory, California Institute of
Technology, under contract with the National Aeronautics and Space
Administration.  Part of this work is based on archival data, software, or
online services provided by the ASI Science Data Center (ASDC).

Some of the results presented in this paper are based on observations obtained
with the Hobby-Eberly Telescope (HET), a joint project of the University of
Texas at Austin, the Pennsylvania State University, Stanford University,
Ludwig-Maximilians-Universit{\"a}t M{\"u}nchen, and
Georg-August-Universit{\"a}t G{\"o}ttingen.  The HET is named in honor of its
principal benefactors, William P.\ Hobby and Robert E.\ Eberly.  The Marcario
Low Resolution Spectrograph (LRS) is named for Mike Marcario of High Lonesome
Optics, who fabricated several optics for the instrument but died before its
completion.  The LRS is a joint project of the Hobby-Eberly Telescope
partnership and the Instituto de Astronom{\'i}a de la Universidad Nacional
Aut{\'o}noma de M{\'e}xico.

This work is also partly based on optical spectroscopy performed at the
Telescopio Nazionale Galileo (TNG), La Palma, Canary Islands (proposal
AOT20/09B).  These observations confirm some of the redshifts and
classifications found with the HET.  We thank the HET and TNG personnel for
their assistance during the observing runs.

The data in this paper are based partly on observations obtained at the Hale
Telescope, Palomar Observatory, as part of a collaborative agreement between
the California Institute of Technology, its divisions Caltech Optical
Observatories and the Jet Propulsion Laboratory (operated for NASA), and
Cornell University.

Some of the data presented herein were obtained at the W.~M.~Keck Observatory,
which is operated as a scientific partnership among the California Institute of
Technology, the University of California, and the National Aeronautics and
Space Administration.  The Observatory was made possible by the generous
financial support of the W.~M.~Keck Foundation.

The authors wish to recognize and acknowledge the very significant cultural
role and reverence that the summit of Mauna Kea has always had within the
indigenous Hawaiian community.  We are most fortunate to have the opportunity
to conduct observations from this mountain.

Some of the data in this paper are based on observations made with ESO
telescopes at the La Silla Observatory under programs \mbox{083.B-0460(B)} and
\mbox{084.B-0711(B).}

The National Radio Astronomy Observatory is a facility of the National Science
Foundation operated under cooperative agreement by Associated Universities, Inc. 

Funding for the Sloan Digital Sky Survey (SDSS) and SDSS-II has been provided
by the Alfred P.\ Sloan Foundation, the Participating Institutions, the
National Science Foundation, the U.~S.\ Department of Energy, the National
Aeronautics and Space Administration, the Japanese Monbukagakusho, the Max
Planck Society, and the Higher Education Funding Council for England.  The SDSS
Web site is http://www.sdss.org/.

The SDSS is managed by the Astrophysical Research Consortium (ARC) for the
Participating Institutions.  The Participating Institutions are the American
Museum of Natural History, Astrophysical Institute Potsdam, University of
Basel, University of Cambridge, Case Western Reserve University, The University
of Chicago, Drexel University, Fermilab, the Institute for Advanced Study, the
Japan Participation Group, The Johns Hopkins University, the Joint Institute
for Nuclear Astrophysics, the Kavli Institute for Particle Astrophysics and
Cosmology, the Korean Scientist Group, the Chinese Academy of Sciences
(LAMOST), Los Alamos National Laboratory, the Max-Planck-Institute for
Astronomy (MPIA), the Max-Planck-Institute for Astrophysics (MPA), New Mexico
State University, Ohio State University, University of Pittsburgh, University
of Portsmouth, Princeton University, the United States Naval Observatory, and
the University of Washington.

{\it Facilities:} \facility{{\it Fermi} LAT}.

\bibliography{biblio.bib}

\clearpage

\clearpage

\clearpage

\clearpage

\begin{deluxetable}{ccccccccccccccccccc}
\tablecolumns{19}
\rotate
\setlength{\tabcolsep}{0.0425in}
\tablewidth{0pc}
\tabletypesize{\tiny}
\tablecaption{First LAT AGN Catalog}
\tablehead{
\colhead{1FGL name}&
\colhead{Associated AGN}&
\colhead{RA\tablenotemark{a}}&
\colhead{DEC\tablenotemark{a}}&
\colhead{Ang.\ sep.\tablenotemark{b}}&
\colhead{$\theta_{95}$\tablenotemark{c}}&
\colhead{Assoc.\ prob.}&
\colhead{Opt.\ class}&
\colhead{SED class}&
\colhead{$z$}&
\colhead{$F_{35}$\tablenotemark{d}}&
\colhead{$\Delta F_{35}$\tablenotemark{d}}&
\colhead{$\Gamma$\tablenotemark{d}}&
\colhead{$\Delta \Gamma$\tablenotemark{d}}&
\colhead{$\sigma$\tablenotemark{d}}&
\colhead{Var.?\tablenotemark{d}}&
\colhead{Curv.?\tablenotemark{d}}&
\colhead{Note\tablenotemark{e}}&
\colhead{Clean?\tablenotemark{f}}}
\startdata
1FGL~J0000.9$-$0745&CRATES~J0001$-$0746&\phn0.32512&\phn$-$7.77417&0.089&0.153&0.96&BLL&LSP&\nodata&1.0&0.0&2.41&0.20&\phn5.6&N&N&S&Y\\
1FGL~J0004.7$-$4737&PKS~0002$-$478&\phn1.14867&$-$47.60517&0.032&0.153&0.99&FSRQ&LSP&0.880&0.8&0.3&2.56&0.17&\phn6.6&N&N&S&Y\\
1FGL~J0005.7$+$3815&B2~0003$+$38A&\phn1.48825&\phs38.33755&0.088&0.200&0.99&FSRQ&LSP&0.229&0.6&0.3&2.86&0.13&\phn8.4&N&N&S&Y\\
1FGL~J0008.3$+$1452&RX~J0008.0$+$1450&\phn2.02345&\phs14.83982&0.072&0.143&0.70&AGN&\nodata&0.045&0.8&0.2&2.00&0.21&\phn4.7&N&N&S&N\\
1FGL~J0008.9$+$0635&CRATES~J0009$+$0628&\phn2.26638&\phn\phs6.47256&0.119&0.117&0.93&BLL&LSP&\nodata&0.8&0.0&2.28&0.22&\phn5.0&N&N&S&Y\\
1FGL~J0011.1$+$0050&CGRaBS~J0011$+$0057&\phn2.87667&\phn\phs0.96439&0.153&0.420&0.96&FSRQ&LSP&1.492&0.6&0.2&2.51&0.15&\phn5.8&N&N&S&Y\\
1FGL~J0013.1$-$3952&PKS~0010$-$401&\phn3.24962&$-$39.90717&0.044&0.104&1.00&BLL&\nodata&\nodata&0.5&0.3&2.09&0.22&\phn5.0&N&N&S&Y\\
1FGL~J0013.7$-$5022&BZB~J0014$-$5022&\phn3.54675&$-$50.37575&0.075&0.135&1.00&BLL&HSP&\nodata&0.6&0.2&2.23&0.22&\phn4.4&N&N&S&Y\\
1FGL~J0017.4$-$0510&CGRaBS~J0017$-$0512&\phn4.39925&\phn$-$5.21158&0.040&0.079&1.00&FSRQ&LSP&0.227&1.5&0.3&2.60&0.07&20.2&Y&Y&S&Y\\
1FGL~J0017.7$-$0019&PKS~0013$-$00&\phn4.04621&\phn$-$0.25347&0.390&0.315&0.57&FSRQ&LSP&1.574&0.5&0.2&2.88&0.18&\phn5.1&N&N&S&N\\
1FGL~J0018.6$+$2945&BZB~J0018$+$2947&\phn4.61562&\phs29.79178&0.044&0.062&1.00&BLL&HSP&\nodata&0.8&0.0&1.48&0.35&\phn6.0&N&N&S&Y\\
1FGL~J0019.3$+$2017&PKS~0017$+$200&\phn4.90771&\phs20.36267&0.100&0.177&0.99&BLL&LSP&\nodata&0.7&0.2&2.38&0.15&\phn5.9&N&N&S&Y\\
1FGL~J0021.7$-$2556&CRATES~J0021$-$2550&\phn5.38563&$-$25.84703&0.110&0.116&0.86&BLL&ISP&\nodata&0.9&0.2&1.96&0.17&\phn7.3&N&N&S&Y\\
1FGL~J0022.5$+$0607&PKS~0019$+$058&\phn5.63517&\phn\phs6.13450&0.009&0.090&1.00&BLL&LSP&\nodata&1.5&0.3&2.15&0.11&10.1&N&N&S&Y\\
1FGL~J0023.0$+$4453&B3~0020$+$446&\phn5.89767&\phs44.94328&0.116&0.120&0.96&FSRQ&\nodata&1.062&1.0&0.3&2.46&0.16&\phn7.1&N&N&S&Y\\
1FGL~J0029.9$-$4221&PKS~0027$-$426&\phn7.57283&$-$42.41292&0.079&0.134&0.99&FSRQ&LSP&0.495&0.7&0.2&2.39&0.17&\phn6.6&N&N&S&Y\\
1FGL~J0033.5$-$1921&RBS~76&\phn8.39292&$-$19.35944&0.006&0.053&1.00&BLL&HSP&0.610&2.8&0.4&1.89&0.08&17.5&N&N&S&Y\\
1FGL~J0035.1$+$1516&RX~J0035.2$+$1515&\phn8.81125&\phs15.25111&0.026&0.072&1.00&BLL&HSP&\nodata&1.5&0.3&1.64&0.14&10.9&N&N&S&Y\\
1FGL~J0038.4$-$2504&PKS~0035$-$252&\phn9.56137&$-$24.98394&0.100&0.128&0.98&FSRQ&LSP&1.196&1.0&0.3&2.45&0.13&\phn9.3&Y&N&S&Y\\
1FGL~J0041.9$+$2318&PKS~0039$+$230&10.51896&\phs23.33367&0.043&0.201&0.98&FSRQ&\nodata&1.426&0.8&0.3&2.52&0.17&\phn5.0&N&N&S&Y\\
1FGL~J0045.3$+$2127&BZB~J0045$+$2127&11.33042&\phs21.46113&0.009&0.029&1.00&BLL&HSP&\nodata&2.1&0.3&1.84&0.12&14.6&N&N&S&Y\\
1FGL~J0047.3$-$2512&NGC~253&11.88806&$-$25.28812&0.097&0.176&1.00&AGN&\nodata&0.001&0.7&0.2&2.15&0.17&\phn6.2&N&N&S&Y\\
1FGL~J0048.0$-$8412&PKS~0044$-$84&11.11192&$-$84.37781&0.191&0.229&0.96&FSRQ&\nodata&1.032&1.3&0.0&2.69&0.17&\phn5.6&N&N&S&Y\\
1FGL~J0048.0$-$8412&PKS~0044$-$84&11.11192&$-$84.37781&0.191&0.229&0.96&FSRQ&\nodata&1.032&1.3&0.0&2.69&0.17&\phn5.6&N&N&S&Y\\
1FGL~J0049.8$-$5738&PKS~0047$-$579&12.49775&$-$57.64067&0.019&0.070&1.00&FSRQ&LSP&1.797&0.6&0.2&2.42&0.16&\phn8.3&N&N&S&Y\\
1FGL~J0050.0$-$0446&PKS~0047$-$051&12.58971&\phn$-$4.87242&0.119&0.184&0.98&FSRQ&\nodata&0.920&0.7&0.3&2.34&0.18&\phn5.6&N&N&S&Y\\
1FGL~J0050.2$+$0235&PKS~0047$+$023&12.43017&\phn\phs2.61772&0.125&0.116&0.98&BLL&\nodata&\nodata&0.5&0.3&2.27&0.19&\phn5.6&N&N&S&Y\\
1FGL~J0050.6$-$0928&PKS~0048$-$09&12.67217&\phn$-$9.48478&0.008&0.064&1.00&BLL&ISP&\nodata&4.5&0.5&2.20&0.05&27.4&Y&N&S&Y\\
1FGL~J0051.1$-$0649&PKS~0048$-$071&12.78421&\phn$-$6.83394&0.016&0.105&1.00&FSRQ&LSP&1.975&1.9&0.3&2.36&0.09&12.8&Y&N&S&Y\\
1FGL~J0058.0$+$3314&CRATES~J0058$+$3311&14.63363&\phs33.18811&0.104&0.110&0.95&BLL&\nodata&1.371&1.7&0.3&2.33&0.11&10.5&Y&N&S&Y\\
1FGL~J0058.4$-$3235&PKS~0055$-$328&14.50925&$-$32.57286&0.078&0.132&0.99&BLL&\nodata&\nodata&0.8&0.2&2.31&0.15&\phn7.0&N&N&S&Y\\
1FGL~J0100.2$+$0747&CRATES~J0100$+$0745&15.08662&\phn\phs7.76428&0.048&0.047&0.97&Unknown&\nodata&\nodata&2.6&0.4&1.86&0.09&15.9&Y&N&S&Y\\
1FGL~J0102.2$+$4223&CRATES~J0102$+$4214&15.61313&\phs42.23861&0.154&0.192&0.93&FSRQ&\nodata&0.874&0.6&0.3&2.74&0.15&\phn7.8&N&N&S&Y\\
1FGL~J0104.4$-$2406&PKS~0102$-$245&16.24250&$-$24.27456&0.212&0.235&0.94&FSRQ&\nodata&1.747&1.0&0.0&2.28&0.20&\phn4.5&N&N&S&Y\\
\enddata

\tablecomments{The first page of the table is shown here.  The full table is available in machine-readable form in the electronic version.}

\tablenotetext{a}{J2000 coordinate of the associated AGN.}

\tablenotetext{b}{Angular separation, in degrees, between the 1FGL source position and the position of the associated AGN.}

\tablenotetext{c}{$\theta_{95} \equiv \sqrt{\theta_1 \theta_2}$, where $\theta_1$ and $\theta_2$ are the semimajor and
semiminor axes (in degrees) of the \mbox{$\gamma$-ray} 95\% confidence region.}

\tablenotetext{d}{\mbox{$\gamma$-ray} properties from the 1FGL catalog.  $\mathbi{F}_\mathbf{35}$:\ The photon flux in units of
$10^{-9}$ \pflux{} for 1-100 GeV.  $\mathbf{\Delta} \mathbi{F}_\mathbf{35}$:\ $1\sigma$ uncertainty on $F_{35}$ in the same units.
An entry of ``0.0'' indicates that the value of $F_{35}$ is an upper limit.
$\mathbf{\Gamma}$:\ Photon number spectral index.  $\mathbf{\Delta \Gamma}$:\ $1\sigma$ uncertainty on $\Gamma$.
$\mathbf{\sigma}$:\ Detection significance.  {\bf Var.?}:\ ``Y'' indicates a probability $< 1$\% that the source is steady.
{\bf Curv.?}:\ ``Y'' indicates a probability $< 1$\% that a simple power law is a good fit to the spectrum.}

\tablenotetext{e}{The codes in this column provide information on multiple associations for a given 1FGL source.
{\bf S}:\ the 1FGL source is associated with exactly one AGN.
{\bf MM}:\ the 1FGL source is associated with at least two AGNs with high confidence ($P > 0.80$).
{\bf Mm}:\ the 1FGL source is associated with at least one AGN with high confidence ($P > 0.80$) and at least
one AGN with lower confidence ($0.50 < P < 0.80$).
{\bf mm}:\ the 1FGL source is associated with at least two AGNs with lower confidence ($0.50 < P < 0.80$).}

\tablenotetext{f}{``Y'' indicates that the source is in the clean sample, defined in Section~\ref{sec:cat}.}

\end{deluxetable}

\begin{deluxetable}{cccccccccccccccccc}
\tablecolumns{18}
\rotate
\setlength{\tabcolsep}{0.055in}
\tablewidth{0pc}
\tabletypesize{\tiny}
\tablecaption{\label{tbl-lowlat}AGN associations for low-latitude ($|b| < 10\arcdeg$) 1FGL sources}
\tablehead{
\colhead{1FGL name}&
\colhead{Associated AGN}&
\colhead{RA\tablenotemark{a}}&
\colhead{DEC\tablenotemark{a}}&
\colhead{Ang.\ sep.\tablenotemark{b}}&
\colhead{$\theta_{95}$\tablenotemark{c}}&
\colhead{Assoc.\ prob.}&
\colhead{Opt.\ class}&
\colhead{SED class}&
\colhead{$z$}&
\colhead{$F_{35}$\tablenotemark{d}}&
\colhead{$\Delta F_{35}$\tablenotemark{d}}&
\colhead{$\Gamma$\tablenotemark{d}}&
\colhead{$\Delta \Gamma$\tablenotemark{d}}&
\colhead{$\sigma$\tablenotemark{d}}&
\colhead{Var.?\tablenotemark{d}}&
\colhead{Curv.?\tablenotemark{d}}&
\colhead{Note\tablenotemark{e}}}
\startdata

1FGL~J0035.9$+$5951&1ES~0033$+$595&\phn\phn8.96935&\phs59.83461&0.020&0.028&1.00&BLL&\nodata&\nodata&\phn3.2&0.5&1.95&0.20&13.0&N&N&S\\
1FGL~J0046.8$+$5658&VCS1~J0047$+$5657&\phn11.75179&\phs56.96178&0.023&0.115&0.99&Unknown&\nodata&\nodata&\phn2.3&0.4&2.27&0.17&\phn7.5&N&N&S\\
1FGL~J0102.8$+$5827&TXS~0059$+$581&\phn15.69068&\phs58.40309&0.064&0.079&0.99&FSRQ&\nodata&0.644&\phn4.2&0.5&2.38&0.13&16.3&Y&N&S\\
1FGL~J0110.0$+$6806&4C~$+$67.04&\phn17.55364&\phs68.09478&0.021&0.074&1.00&Unknown&\nodata&\nodata&\phn2.0&0.5&2.35&0.21&\phn8.5&N&N&S\\
1FGL~J0254.2$+$5107&TXS~0250$+$508&\phn43.49003&\phs51.04902&0.089&0.114&0.99&Unknown&\nodata&\nodata&\phn2.3&0.5&2.42&0.22&\phn9.3&N&N&S\\
1FGL~J0303.1$+$4711&4C~$+$47.08&\phn45.89684&\phs47.27119&0.109&0.149&1.00&BLL&\nodata&\nodata&\phn1.4&0.4&2.56&0.15&\phn6.8&N&N&S\\
1FGL~J0334.3$+$6536&RX~J0333.9$+$6537&\phn53.48641&\phs65.61561&0.047&0.094&0.99&Unknown&\nodata&\nodata&\phn1.3&0.5&2.19&0.22&\phn4.1&N&N&S\\
1FGL~J0419.0$+$3811&3C~111&\phn64.58866&\phs38.02661&0.214&0.255&0.87&AGN&\nodata&0.049&\phn1.5&0.5&2.61&0.22&\phn4.3&N&N&S\\
1FGL~J0423.8$+$4148&4C~$+$41.11&\phn65.98337&\phs41.83409&0.025&0.031&1.00&Unknown&\nodata&\nodata&\phn3.5&0.5&1.87&0.07&15.2&N&N&S\\
1FGL~J0521.7$+$2114&RX~J0521.7$+$2112&\phn80.44152&\phs21.21429&0.021&0.030&1.00&Unknown&\nodata&\nodata&\phn5.5&0.6&1.94&0.18&21.1&N&N&S\\
1FGL~J0533.0$+$4825&RX~J0533.2$+$4823&\phn83.31611&\phs48.38134&0.054&0.110&1.00&FSRQ&\nodata&1.162&\phn2.3&0.4&2.43&0.35&10.3&Y&N&S\\
1FGL~J0648.7$-$1740&TXS~0646$-$176&102.11874&$-$17.73484&0.092&0.107&0.99&FSRQ&\nodata&1.232&\phn2.5&0.5&2.47&0.15&10.8&Y&N&S\\
1FGL~J0650.6$-$1635&PKS~0648$-$16&102.60242&$-$16.62770&0.075&0.151&0.98&Unknown&\nodata&\nodata&\phn1.4&0.4&2.46&0.17&\phn5.8&N&N&S\\
1FGL~J0656.2$-$0321&OH~$-$090&104.04634&\phn$-$3.38522&0.034&0.077&1.00&Unknown&\nodata&\nodata&\phn3.9&0.6&2.59&0.11&16.1&Y&N&S\\
1FGL~J0702.2$-$1954&TXS~0700$-$197&105.67875&$-$19.85612&0.120&0.159&0.96&Unknown&\nodata&\nodata&\phn1.7&0.4&1.92&0.16&\phn4.2&N&N&S\\
1FGL~J0721.4$+$0401&RX~J0721.3$+$0406&110.34963&\phn\phs4.11228&0.083&0.125&0.98&Unknown&\nodata&\nodata&\phn0.8&0.3&2.68&0.17&\phn6.8&N&N&S\\
1FGL~J0725.9$-$0053&PKS~0723$-$008&111.46100&\phn$-$0.91571&0.043&0.077&1.00&BLL&\nodata&0.128&\phn0.6&0.3&2.30&0.08&\phn6.1&N&N&S\\
1FGL~J0730.3$-$1141&PKS~0727$-$11&112.57963&$-$11.68683&0.013&0.022&1.00&FSRQ&\nodata&1.591&20.7&1.0&2.33&0.14&67.6&Y&N&S\\
1FGL~J0754.4$-$1147&OI~$-$187&118.61024&$-$11.78804&0.009&0.068&1.00&Unknown&\nodata&\nodata&\phn1.9&0.4&2.10&0.13&10.5&N&N&S\\
1FGL~J0825.8$-$2230&PKS~0823$-$223&126.50655&$-$22.50756&0.036&0.038&1.00&BLL&\nodata&\nodata&\phn5.3&0.5&2.14&0.17&26.5&N&N&S\\
1FGL~J0825.9$-$3216&PKS~0823$-$321&126.46405&$-$32.30645&0.036&0.110&1.00&Unknown&\nodata&\nodata&\phn1.4&0.4&2.68&0.12&\phn8.8&Y&N&S\\
1FGL~J0827.9$-$3738&PKS~B0826$-$373&127.01992&$-$37.51841&0.134&0.194&0.96&Unknown&\nodata&\nodata&\phn1.7&0.6&2.68&0.17&\phn5.7&Y&N&S\\
1FGL~J0845.0$-$5459&PMN~J0845$-$5458&131.26034&$-$54.96904&0.020&0.101&0.99&Unknown&\nodata&\nodata&\phn1.9&0.4&2.24&0.17&\phn7.7&N&N&S\\
1FGL~J0849.6$-$3540&VCS2~J0849$-$3541&132.44010&$-$35.68369&0.023&0.152&0.98&Unknown&\nodata&\nodata&\phn1.9&0.4&2.40&0.16&\phn6.0&N&N&S\\
1FGL~J0905.1$-$5736&PKS~0903$-$57&136.22158&$-$57.58494&0.041&0.068&1.00&FSRQ&\nodata&0.695&\phn1.8&0.4&2.36&0.18&\phn7.5&Y&N&S\\
1FGL~J1103.9$-$5355&PKS~1101$-$536&165.96759&$-$53.95019&0.019&0.041&1.00&Unknown&\nodata&\nodata&\phn6.1&0.6&2.05&0.19&19.5&Y&N&S\\
1FGL~J1122.9$-$6415&PMN~J1123$-$6417&170.83090&$-$64.29339&0.059&0.088&0.99&Unknown&\nodata&\nodata&\phn2.4&0.0&2.48&0.05&\phn4.7&Y&N&S\\
1FGL~J1307.3$-$6701&PKS~1304$-$668&197.07240&$-$67.11812&0.129&0.152&0.92&Unknown&\nodata&\nodata&\phn1.4&0.5&2.57&0.09&\phn5.3&Y&N&S\\
1FGL~J1327.0$-$5257&PMN~J1326$-$5256&201.70512&$-$52.93990&0.040&0.061&1.00&Unknown&\nodata&\nodata&\phn5.2&0.6&2.33&0.11&19.2&Y&N&S\\
1FGL~J1329.2$-$5605&PMN~J1329$-$5608&202.25477&$-$56.13407&0.055&0.095&1.00&Unknown&\nodata&\nodata&\phn4.1&0.6&2.56&0.15&15.1&Y&N&S\\
1FGL~J1330.7$-$7006&PKS~1326$-$697&202.54615&$-$70.05363&0.072&0.114&0.99&Unknown&\nodata&\nodata&\phn1.5&0.4&2.46&0.09&\phn8.3&Y&N&S\\
1FGL~J1400.9$-$5559&PMN~J1400$-$5605&210.17407&$-$56.08210&0.092&0.122&0.99&Unknown&\nodata&\nodata&\phn2.4&0.5&2.62&0.15&\phn9.1&Y&N&S\\
1FGL~J1514.1$-$4745&PMN~J1514$-$4748&228.66677&$-$47.80829&0.106&0.124&0.98&Unknown&\nodata&\nodata&\phn1.2&0.4&2.29&0.20&\phn5.7&N&N&S\\
1FGL~J1603.8$-$4903&PMN~J1603$-$4904&240.96119&$-$49.06820&0.007&0.027&1.00&Unknown&\nodata&\nodata&13.4&1.1&2.12&0.14&26.8&N&Y&S\\
1FGL~J1604.7$-$4443&PMN~J1604$-$4441&241.12925&$-$44.69221&0.046&0.066&1.00&Unknown&\nodata&\nodata&\phn7.7&0.8&2.46&0.04&22.2&Y&N&S\\
\enddata

\tablecomments{The first page of the table is shown here.  The full table is available in machine-readable form in the electronic version.}

\tablenotetext{a}{J2000 coordinate of the associated AGN.}

\tablenotetext{b}{Angular separation, in degrees, between the 1FGL source position and the position of the associated AGN.}

\tablenotetext{c}{$\theta_{95} \equiv \sqrt{\theta_1 \theta_2}$, where $\theta_1$ and $\theta_2$ are the semimajor and
semiminor axes (in degrees) of the \mbox{$\gamma$-ray} 95\% confidence region.}

\tablenotetext{d}{\mbox{$\gamma$-ray} properties from the 1FGL catalog.  $\mathbi{F}_\mathbf{35}$:\ The photon flux in units of
$10^{-9}$ \pflux{} for 1-100 GeV.  $\mathbf{\Delta} \mathbi{F}_\mathbf{35}$:\ $1\sigma$ uncertainty on $F_{35}$ in the same units.
An entry of ``0.0'' indicates that the value of $F_{35}$ is an upper limit.
$\mathbf{\Gamma}$:\ Photon number spectral index.  $\mathbf{\Delta \Gamma}$:\ $1\sigma$ uncertainty on $\Gamma$.
$\mathbf{\sigma}$:\ Detection significance.  {\bf Var.?}:\ ``Y'' indicates a probability $< 1$\% that the source is steady.
{\bf Curv.?}:\ ``Y'' indicates a probability $< 1$\% that a simple power law is a good fit to the spectrum.}

\tablenotetext{e}{The codes in this column provide information on multiple associations for a given 1FGL source.
{\bf S}:\ the 1FGL source is associated with exactly one AGN.
{\bf MM}:\ the 1FGL source is associated with at least two AGNs with high confidence ($P > 0.80$).
{\bf Mm}:\ the 1FGL source is associated with at least one AGN with high confidence ($P > 0.80$) and at least
one AGN with lower confidence ($0.50 < P < 0.80$).
{\bf mm}:\ the 1FGL source is associated with at least two AGNs with lower confidence ($0.50 < P < 0.80$).}
\end{deluxetable}

\begin{deluxetable}{cccccccccccccccc}
\tablecolumns{16}
\rotate
\setlength{\tabcolsep}{0.075in}
\tablewidth{0pc}
\tabletypesize{\tiny}
\tablecaption{\label{tbl-extras}AGN ``affiliations'' for 1FGL sources}
\tablehead{
\colhead{1FGL name}&
\colhead{Associated AGN}&
\colhead{RA\tablenotemark{a}}&
\colhead{DEC\tablenotemark{a}}&
\colhead{Ang.\ sep.\tablenotemark{b}}&
\colhead{$\theta_{95}$\tablenotemark{c}}&
\colhead{Opt.\ class}&
\colhead{SED class}&
\colhead{$z$}&
\colhead{$F_{35}$\tablenotemark{d}}&
\colhead{$\Delta F_{35}$\tablenotemark{d}}&
\colhead{$\Gamma$\tablenotemark{d}}&
\colhead{$\Delta \Gamma$\tablenotemark{d}}&
\colhead{$\sigma$\tablenotemark{d}}&
\colhead{Var.?\tablenotemark{d}}&
\colhead{Curv.?\tablenotemark{d}}}
\startdata
1FGL~J0016.6$+$1706&CRATES~J0015$+$1700&\phn\phn3.91662&\phs17.01128&0.247&0.193&FSRQ&\nodata&1.716&0.5&0.3&2.57&0.20&\phn4.7&N&N\\
1FGL~J0022.2$-$1850&1RXS~J002209.2$-$185333&\phn\phn5.53800&$-$18.89248&0.054&0.111&Unknown&HSP&\nodata&1.1&0.3&1.56&0.14&\phn9.4&N&N\\
1FGL~J0024.6$+$0346&CLASS~J0024$+$0349&\phn\phn6.18841&\phn\phs3.81766&0.045&0.099&FSRQ&\nodata&0.545&0.9&0.3&2.33&0.12&\phn8.9&Y&N\\
1FGL~J0038.0$+$1236&FRBA~J0037$+$1238&\phn\phn9.46180&\phs12.63863&0.060&0.098&Unknown&\nodata&\nodata&0.7&0.3&2.30&0.18&\phn7.0&N&N\\
1FGL~J0043.6$+$3424&CLASS~J0043$+$3426&\phn10.95352&\phs34.44059&0.042&0.063&Unknown&\nodata&\nodata&1.7&0.3&2.09&0.10&12.4&N&N\\
1FGL~J0048.0$+$2232&CLASS~J0048$+$2235&\phn12.01092&\phs22.59006&0.050&0.089&FSRQ&\nodata&1.156&1.5&0.3&2.39&0.09&13.5&Y&N\\
1FGL~J0051.4$-$6242&RBS~119&\phn12.81952&$-$62.70120&0.025&0.067&Unknown&\nodata&\nodata&1.8&0.3&1.68&0.12&12.0&N&N\\
1FGL~J0054.9$-$2455&FRBA~J0054$-$2455&\phn13.69480&$-$24.92519&0.035&0.078&Unknown&HSP&\nodata&0.7&0.2&1.95&0.22&\phn5.2&N&N\\
1FGL~J0110.0$-$4023&RBS~158&\phn17.48483&$-$40.34794&0.051&0.079&Unknown&HSP&\nodata&0.8&0.0&1.34&0.32&\phn4.2&N&N\\
1FGL~J0115.7$+$0357&CLASS~J0115$+$0356&\phn18.91880&\phn\phs3.94536&0.021&0.076&BLL&\nodata&\nodata&1.4&0.3&2.07&0.15&\phn8.7&N&N\\
1FGL~J0124.6$-$0616&AT20G~J0124$-$0625&\phn21.21033&\phn$-$6.41722&0.146&0.163&BLL&\nodata&\nodata&0.6&0.3&2.20&0.17&\phn5.5&N&N\\
1FGL~J0134.4$+$2632&RX~J0134.4$+$2638&\phn23.61779&\phs26.64588&0.103&0.155&Unknown&HSP&\nodata&0.9&0.3&2.26&0.16&\phn6.6&N&N\\
1FGL~J0157.0$-$5259&RBS~259&\phn29.23839&$-$53.03285&0.049&0.100&Unknown&\nodata&\nodata&0.8&0.2&1.85&0.20&\phn7.6&N&N\\
1FGL~J0217.9$-$6630&CRATES~J0216$-$6636&\phn34.21187&$-$66.61169&0.153&0.108&BLL&\nodata&\nodata&0.9&0.3&1.94&0.17&\phn7.0&N&N\\
1FGL~J0226.3$+$0937&FRBA~J0226$+$0937&\phn36.55704&\phn\phs9.62398&0.034&0.133&Unknown&\nodata&\nodata&0.9&0.3&1.99&0.17&\phn6.4&N&N\\
1FGL~J0256.9$+$2920&FRBA~J0256$+$2924&\phn44.22623&\phs29.41500&0.066&0.105&AGN&\nodata&0.190&0.6&0.3&1.69&0.30&\phn4.2&N&N\\
1FGL~J0315.6$-$5109&AT20G~J0314$-$5104&\phn48.60733&$-$51.07547&0.218&0.314&BLL&\nodata&\nodata&0.7&0.3&2.60&0.20&\phn6.0&N&N\\
1FGL~J0318.1$+$0254&CLASS~J0317$+$0248&\phn49.44965&\phn\phs2.81183&0.135&0.165&FSRQ&\nodata&0.748&0.8&0.3&2.20&0.18&\phn4.1&N&N\\
1FGL~J0333.7$+$2919&FRBA~J0333$+$2916&\phn53.45420&\phs29.27542&0.057&0.044&Unknown&ISP&\nodata&1.0&0.3&1.57&0.21&\phn6.5&N&N\\
1FGL~J0342.2$+$3859&CLASS~J0342$+$3859&\phn55.56779&\phs38.98507&0.012&0.102&FSRQ&\nodata&0.945&1.1&0.0&2.17&0.29&\phn4.7&N&N\\
1FGL~J0401.3$-$3152&PKS~0400$-$319&\phn60.58863&$-$31.79053&0.232&0.155&FSRQ&\nodata&1.288&0.6&0.3&2.37&0.17&\phn6.0&N&N\\
1FGL~J0445.2$-$6008&AT20G~J0445$-$6015&\phn71.25667&$-$60.25006&0.105&0.169&AGN&\nodata&0.097&0.8&0.3&1.98&0.29&\phn4.6&N&N\\
1FGL~J0506.9$-$5435&RBS~621&\phn76.74127&$-$54.58422&0.014&0.058&Unknown&\nodata&\nodata&1.0&0.0&1.42&0.31&\phn6.7&N&N\\
1FGL~J0515.2$+$7355&CLASS~J0516$+$7351&\phn79.12993&\phs73.85240&0.115&0.167&BLL&\nodata&0.249&0.7&0.3&2.04&0.25&\phn5.0&N&N\\
1FGL~J0515.9$+$1528&CLASS~J0515$+$1527&\phn78.94732&\phs15.45461&0.032&0.069&BLL&\nodata&\nodata&1.2&0.4&2.01&0.18&\phn8.4&N&N\\
1FGL~J0517.6$+$0857&CLASS~J0517$+$0858&\phn79.41693&\phn\phs8.97661&0.025&0.109&FSRQ&\nodata&0.328&1.1&0.4&2.58&0.12&\phn8.5&N&N\\
1FGL~J0537.7$-$5717&PKS~0541$-$834&\phn84.45325&$-$57.30806&0.014&0.073&FSRQ&\nodata&\nodata&1.0&0.0&1.81&0.42&\phn6.1&N&N\\
1FGL~J0538.4$-$3910&1RXS~J053810.0$-$390839&\phn84.54304&$-$39.14562&0.059&0.077&Unknown&HSP&\nodata&1.3&0.3&2.16&0.12&\phn9.0&N&N\\
1FGL~J0541.9$-$0204&CRATES~J0541$-$0154&\phn85.47800&\phn$-$2.07800&0.011&0.075&Unknown&\nodata&\nodata&1.5&0.6&2.30&0.12&\phn6.0&N&N\\
1FGL~J0600.5$-$2006&CRATES~J0601$-$2004&\phn90.47012&$-$20.07922&0.303&0.260&FSRQ&\nodata&1.216&0.7&0.3&2.41&0.15&\phn6.2&N&N\\
1FGL~J0603.0$-$4012&1WGA~J0602.8$-$4018&\phn90.71162&$-$40.31417&0.109&0.143&Unknown&\nodata&\nodata&1.2&0.3&2.25&0.16&\phn5.6&N&N\\
1FGL~J0604.2$-$4817&1ES~0602$-$482&\phn91.03918&$-$48.29059&0.011&0.062&Unknown&\nodata&\nodata&1.1&0.3&2.12&0.16&\phn7.2&N&N\\
1FGL~J0605.1$+$0005&CLASS~J0604$+$0000&\phn91.24341&\phn\phs0.01204&0.100&0.129&Unknown&\nodata&\nodata&1.1&0.3&1.94&0.23&\phn5.4&N&N\\
1FGL~J0608.1$-$0630&CRATES~J0609$-$0615&\phn92.41654&\phn$-$6.25161&0.457&0.306&Unknown&\nodata&\nodata&1.8&0.5&2.54&0.10&\phn8.1&N&Y\\
1FGL~J0609.3$-$0244&NVSS~J060915$-$024754&\phn92.31252&\phn$-$2.79840&0.065&0.096&Unknown&HSP&\nodata&1.2&0.3&2.02&0.20&\phn5.9&N&N\\
1FGL~J0622.2$+$3751&CLASS~J0621$+$3750&\phn95.49016&\phs37.84916&0.058&0.131&Unknown&\nodata&\nodata&2.0&0.4&2.36&0.09&10.2&N&Y\\
1FGL~J0706.5$+$3744&CLASS~J0706$+$3744&106.63207&\phs37.74344&0.007&0.063&BLL&HSP&\nodata&1.1&0.3&2.01&0.15&\phn8.2&N&N\\
1FGL~J0707.3$+$7742&FRBA~J0706$+$7741&106.71367&\phs77.69349&0.027&0.088&Unknown&\nodata&\nodata&1.4&0.3&2.30&0.13&10.5&N&N\\
1FGL~J0723.6$+$2908&CLASS~J0723$+$2859&110.97850&\phs28.99163&0.165&0.215&Unknown&\nodata&\nodata&0.9&0.3&2.10&0.17&\phn5.8&Y&N\\
\enddata

\tablecomments{The first page of the table is shown here.  The full table is available in machine-readable form in the electronic version.}

\tablenotetext{a}{J2000 coordinate of the associated AGN.}

\tablenotetext{b}{Angular separation, in degrees, between the 1FGL source position and the position of the associated AGN.}

\tablenotetext{c}{$\theta_{95} \equiv \sqrt{\theta_1 \theta_2}$, where $\theta_1$ and $\theta_2$ are the semimajor and
semiminor axes (in degrees) of the \mbox{$\gamma$-ray} 95\% confidence region.}

\tablenotetext{d}{\mbox{$\gamma$-ray} properties from the 1FGL catalog.  $\mathbi{F}_\mathbf{35}$:\ The photon flux in units of
$10^{-9}$ \pflux{} for 1-100 GeV.  $\mathbf{\Delta} \mathbi{F}_\mathbf{35}$:\ $1\sigma$ uncertainty on $F_{35}$ in the same units.
An entry of ``0.0'' indicates that the value of $F_{35}$ is an upper limit.
$\mathbf{\Gamma}$:\ Photon number spectral index.  $\mathbf{\Delta \Gamma}$:\ $1\sigma$ uncertainty on $\Gamma$.
$\mathbf{\sigma}$:\ Detection significance.  {\bf Var.?}:\ ``Y'' indicates a probability $< 1$\% that the source is steady.
{\bf Curv.?}:\ ``Y'' indicates a probability $< 1$\% that a simple power law is a good fit to the spectrum.}

\end{deluxetable}

\begin{deluxetable}{lccc}
\tablecolumns{4}
\tabletypesize{\footnotesize}
\tablecaption{\label{tbl-census}Census of 1LAC sources}
\tablewidth{0pt}

\tablehead{
\colhead{}&
\multicolumn{3}{c}{Number of AGNs in:}\\[3pt]
\colhead{AGN type}&
\colhead{Entire 1LAC sample}&
\colhead{High-confidence sample\tablenotemark{a}}&
\colhead{Clean sample\tablenotemark{a}}
}
\startdata
{\bf All}&{\bf 709}&{\bf 663}&{\bf 599}\\
\\
{\bf FSRQ}&{\bf 296}&{\bf 281}&{\bf 248}\\
{\ldots}LSP&189&185&171\\
{\ldots}ISP&\phn\phn3&\phn\phn2&\phn\phn1\\
{\ldots}HSP&\phn\phn2&\phn\phn2&\phn\phn1\\
\\
{\bf BL Lac}&{\bf 300}&{\bf 291}&{\bf 275}\\
{\ldots}LSP&\phn69&\phn67&\phn62\\
{\ldots}ISP&\phn46&\phn44&\phn44\\
{\ldots}HSP&118&117&113\\
\\
{\bf Other AGN}&{\bf \phn41}&{\bf \phn30}&{\bf \phn26}\\
\\
{\bf Unknown}&{\bf \phn72}&{\bf \phn61}&{\bf \phn50}\\
\enddata

\tablenotetext{a}{See Section~\ref{sec:cat} for the definitions of these samples.}

\end{deluxetable}

\begin{deluxetable}{cccc}
\tabletypesize{\footnotesize}
\tablecaption{\label{tbl-tev}Positional coincidences of 1LAC sources with TeV sources}
\tablewidth{0pt}

\tablehead{
\colhead{LAT Source}&
\colhead{TeV Association}&
\colhead{RA\tablenotemark{a}}&
\colhead{DEC\tablenotemark{a}}
}

\startdata
1FGL~J0222.6$+$4302&3C 66A&\phn35.8000&\phs43.0117\\
1FGL~J0319.7$+$1847&RBS 0413&\phn49.9658&\phs18.7594\\
1FGL~J0416.8$+$0107&1ES 0414+009&\phn64.2184&\phs\phn1.0901\\
1FGL~J0449.5$-$4350&PKS 0447$-$437&\phn72.3529&$-$43.8358\\
1FGL~J0507.9$+$6738&1ES 0502+675&\phn76.9842&\phs67.6233\\
1FGL~J0521.7$+$2114\tablenotemark{b}&VER J0521+211\tablenotemark{b}&\phn80.4792&\phs21.1900\\
1FGL~J0710.6$+$5911&RGB J0710+591&107.6254&\phs59.1390\\
1FGL~J0721.9$+$7120&S5 0716+714&110.4725&\phs71.3433\\
1FGL~J0809.5$+$5219&1ES 0806+524&122.4550&\phs52.3161\\
1FGL~J1015.1$+$4927&1ES 1011+496&153.7671&\phs49.4336\\
1FGL~J1103.7$-$2329&1ES 1101$-$232&165.9083&$-$23.4919\\
1FGL~J1104.4$+$3812&Mkn 421&166.1138&\phs38.2089\\
1FGL~J1136.6$+$7009&Mkn 180&174.1100&\phs70.1575\\
1FGL~J1221.3$+$3008&1ES 1218+304&185.3413&\phs30.1769\\
1FGL~J1230.8$+$1223&M 87&187.7058&\phs12.3911\\
1FGL~J1221.5$+$2814&W Com&185.3821&\phs28.2331\\
1FGL~J1256.2$-$0547&3C 279&194.0463&\phn$-$5.7894\\
1FGL~J1325.6$-$4300&Cen A&201.3667&$-$43.0183\\
1FGL~J1426.9$+$2347&PKS 1424+240&216.7516&\phs23.8000\\
1FGL~J1428.7$+$4239&H 1426+428&217.1358&\phs42.6725\\
1FGL~J1555.7$+$1111&PG 1553+113&238.9292&\phs11.1900\\
1FGL~J1653.9$+$3945&Mkn 501&253.4675&\phs39.7600\\
1FGL~J2000.0$+$6508&1ES 1959+650&299.9996&\phs65.1486\\
1FGL~J2009.5$-$4849&PKS 2005$-$489&302.3721&$-$48.8219\\
1FGL~J2158.8$-$3013&PKS 2155$-$304&329.7196&$-$30.2217\\
1FGL~J2202.8$+$4216&BL~Lac&330.6804&\phs42.2778\\
1FGL~J2347.1$+$5142\tablenotemark{b}&1ES 2344+514\tablenotemark{b}&356.7700&\phs51.7050\\
1FGL~J2359.0$-$3035&H 2356$-$309&359.7875&$-$30.6228\\
\enddata

\tablenotetext{a}{J2000 coordinate, in degrees, from TeVCat ({\tt http://tevcat.uchicago.edu/}).}
\tablenotetext{b}{This source is at low Galactic latitude ($|b| < 10\arcdeg$) and is thus not formally
in the 1LAC but appears in Table \ref{tbl-lowlat}.}
\end{deluxetable}

\begin{deluxetable}{cccccccccc}
\setlength{\tabcolsep}{0.06in}
\rotate
\tabletypesize{\tiny}
\tablecaption{Positional coincidences of 1LAC sources with EGRET/{\it AGILE} sources
\label{tbl-fmrdetect}}
\tablewidth{0pt}

\tablehead{
\colhead{1FGL name}&
\colhead{3EG name}&
\colhead{EGR name}&
\colhead{GEV name}&
\colhead{1AGL name}&
\colhead{LAT flux\tablenotemark{a}}&
\colhead{LAT ph. index}&
\colhead{3EG flux\tablenotemark{a}}&
\colhead{EGRET ph. index}&
\colhead{Probable class}
}

\startdata
1FGL~J0204.5$+$1516&3EG J0204$+$1458&\phantom{c}EGR J0204$+$1505&&&\phn\phn1.79 $\pm$ 0.58&2.47 $\pm$ 0.18&\phn23.6 $\pm$ \phn5.6&2.23 $\pm$ 0.28&AGN\\
1FGL~J0210.6$-$5101&3EG J0210$-$5055&\phantom{c}EGR J0210$-$5058&GEV J0210$-$5053&&\phn14.59 $\pm$ 0.96&2.37 $\pm$ 0.04&\phn85.5 $\pm$ \phn4.5&1.99 $\pm$ 0.05&BL Lac\\
1FGL~J0222.6$+$4302&3EG J0222$+$4253&\phantom{c}EGR J0223$+$4300&GEV J0223$+$4254&&\phn21.45 $\pm$ 0.97&1.93 $\pm$ 0.02&\phn18.7 $\pm$ \phn2.9&2.01 $\pm$ 0.14&BL Lac\\
1FGL~J0238.6$+$1637&3EG J0237$+$1635&&GEV J0237$+$1648&&\phn43.37 $\pm$ 1.07&2.14 $\pm$ 0.02&\phn65.1 $\pm$ \phn8.8&1.85 $\pm$ 0.12&BL Lac\\
1FGL~J0339.2$-$0143&3EG J0340$-$0201&\phantom{c}EGR J0338$-$0203&&&\phn\phn4.00 $\pm$ 0.65&2.50 $\pm$ 0.10&118.8 $\pm$ 22.0&1.84 $\pm$ 0.25&FSRQ\\
1FGL~J0416.5$-$1851&3EG J0412$-$1853&\phantom{c}EGR J0413$-$1851&&&\phn\phn3.59 $\pm$ 0.58&2.37 $\pm$ 0.10&\phn49.5 $\pm$ 16.1&3.25 $\pm$ 0.68&FSRQ\\
1FGL~J0423.2$-$0118&3EG J0422$-$0102&&&&\phn13.99 $\pm$ 0.90&2.42 $\pm$ 0.04&\phn50.2 $\pm$ 10.4&2.44 $\pm$ 0.19&FSRQ\\
1FGL~J0433.5$+$2905&3EG J0433$+$2908&\phantom{c}EGR J0433$+$2906&GEV J0433$+$2907&&\phn\phn6.57 $\pm$ 0.83&2.13 $\pm$ 0.06&\phn22.0 $\pm$ \phn2.8&1.90 $\pm$ 0.10&BL Lac\\
1FGL~J0442.7$-$0019&3EG J0442$-$0033&\phantom{c}EGR J0442$-$0027&GEV J0441$-$0044&&\phn17.27 $\pm$ 0.91&2.44 $\pm$ 0.04&\phn79.0 $\pm$ 10.1&2.37 $\pm$ 0.18&FSRQ\\
1FGL~J0448.6$+$1118&3EG J0450$+$1105&&&&\phn\phn5.90 $\pm$ 0.88&2.51 $\pm$ 0.09&109.5 $\pm$ 19.4&2.27 $\pm$ 0.16&BL Lac\\
1FGL~J0455.6$-$4618&3EG J0458$-$4635&&&&\phn\phn6.15 $\pm$ 0.83&2.57 $\pm$ 0.09&\phn\phn7.7 $\pm$ \phn2.1&2.75 $\pm$ 0.35&FSRQ\\
1FGL~J0457.0$-$2325&3EG J0456$-$2338&\phantom{c}EGR J0456$-$2334&&&\phn51.14 $\pm$ 1.07&2.21 $\pm$ 0.02&\phn14.7 $\pm$ \phn4.2&3.14 $\pm$ 0.47&FSRQ\\
1FGL~J0501.0$-$0200&3EG J0500$-$0159&&&&\phn\phn3.74 $\pm$ 0.70&2.50 $\pm$ 0.11&\phn11.2 $\pm$ \phn2.3&2.45 $\pm$ 0.27&FSRQ\\
1FGL~J0509.3$+$0540&&\phantom{c}EGR J0509$+$0550&GEV J0508$+$0540&&\phn\phn5.77 $\pm$ 0.73&2.16 $\pm$ 0.06&\nodata&\nodata&BL Lac\\
1FGL~J0516.7$-$6207&3EG J0512$-$6150&&&&\phn\phn5.76 $\pm$ 0.95&2.28 $\pm$ 0.09&\phn\phn7.2 $\pm$ \phn1.7&2.40 $\pm$ 0.26&XXX\\
1FGL~J0531.0$+$1331&3EG J0530$+$1323&\phantom{c}EGR J0530$+$1331&GEV J0530$+$1340&&\phn16.85 $\pm$ 1.28&2.64 $\pm$ 0.06&\phn93.5 $\pm$ \phn3.6&2.46 $\pm$ 0.04&FSRQ\\
1FGL~J0538.8$-$4404&3EG J0540$-$4402&\phantom{c}EGR J0540$-$4358&GEV J0540$-$4359&1AGL J0538$-$4424&\phn37.77 $\pm$ 1.06&2.27 $\pm$ 0.02&\phn25.3 $\pm$ \phn3.1&2.41 $\pm$ 0.12&BL Lac\\
1FGL~J0540.9$-$0547&3EG J0542$-$0655&&&&\phn\phn3.73 $\pm$ 1.01&2.37 $\pm$ 0.12&\phn66.5 $\pm$ 19.5&\nodata&FSRQ\\
1FGL~J0721.9$+$7120&3EG J0721$+$7120&\phantom{c}EGR J0723$+$7134&GEV J0719$+$7133&1AGL J0722$+$7125&\phn17.26 $\pm$ 0.75&2.15 $\pm$ 0.03&\phn17.8 $\pm$ \phn2.0&2.19 $\pm$ 0.11&BL Lac\\
1FGL~J0738.2$+$1741&3EG J0737$+$1721&\phantom{c}EGR J0737$+$1720&&&\phn\phn4.60 $\pm$ 0.48&2.02 $\pm$ 0.06&\phn16.4 $\pm$ \phn3.3&2.60 $\pm$ 0.28&BL Lac\\
1FGL~J0742.2$+$5443&3EG J0743$+$5447&\phantom{c}EGR J0743$+$5438&&&\phn\phn5.22 $\pm$ 0.75&2.45 $\pm$ 0.09&\phn30.3 $\pm$ \phn5.0&2.03 $\pm$ 0.20&FSRQ\\
1FGL~J0830.5$+$2407&3EG J0829$+$2413&\phantom{c}EGR J0829$+$2415&&&\phn\phn8.35 $\pm$ 0.84&2.79 $\pm$ 0.09&\phn24.9 $\pm$ \phn3.9&2.42 $\pm$ 0.21&FSRQ\\
1FGL~J0831.6$+$0429&&\phantom{c}EGR J0829$+$0510&&&\phn\phn7.35 $\pm$ 0.76&2.50 $\pm$ 0.07&\nodata&\nodata&BL Lac\\
1FGL~J0842.2$+$7054&3EG J0845$+$7049&&&&\phn\phn8.06 $\pm$ 1.04&2.98 $\pm$ 0.12&\phn10.2 $\pm$ \phn1.8&2.62 $\pm$ 0.16&FSRQ\\
1FGL~J0850.0$-$1213&3EG J0852$-$1216&\phantom{c}EGR J0852$-$1224&&&\phn\phn5.02 $\pm$ 0.64&2.27 $\pm$ 0.07&\phn44.4 $\pm$ 11.6&1.58 $\pm$ 0.58&FSRQ\\
1FGL~J0854.8$+$2006&3EG J0853$+$1941&\phantom{c}EGR J0853$+$2015&&&\phn\phn7.03 $\pm$ 0.82&2.38 $\pm$ 0.07&\phn10.6 $\pm$ \phn3.0&2.03 $\pm$ 0.35&BL Lac\\
1FGL~J0957.7$+$5523&&\phantom{c}EGR J0957$+$5513&GEV J0956$+$5508&&\phn11.39 $\pm$ 0.62&2.05 $\pm$ 0.03&\nodata&\nodata&FSRQ\\
1FGL~J1000.1$+$6539&3EG J0958$+$6533&\phantom{c}EGR J0956$+$6524&&&\phn\phn2.59 $\pm$ 0.67&2.51 $\pm$ 0.16&\phn15.4 $\pm$ \phn3.0&2.08 $\pm$ 0.24&BL Lac\\
1FGL~J1104.4$+$3812&3EG J1104$+$3809&\phantom{c}EGR J1104$+$3813&GEV J1104$+$3809&1AGL J1104$+$3754&\phn16.93 $\pm$ 0.58&1.81 $\pm$ 0.02&\phn13.9 $\pm$ \phn1.8&1.57 $\pm$ 0.15&BL Lac\\
1FGL~J1133.1$+$0033&3EG J1133$+$0033&&&&\phn\phn2.71 $\pm$ 0.53&2.18 $\pm$ 0.11&\phn10.6 $\pm$ \phn3.0&2.73 $\pm$ 0.63&BL Lac\\
1FGL~J1159.4$+$2914&3EG J1200$+$2847&&GEV J1201$+$2906&&\phn12.24 $\pm$ 0.74&2.37 $\pm$ 0.04&\phn50.9 $\pm$ 11.9&1.98 $\pm$ 0.22&FSRQ\\
1FGL~J1221.5$+$2814&&&&1AGL J1222$+$2851&\phn\phn7.82 $\pm$ 0.61&2.06 $\pm$ 0.04&\nodata&\nodata&BL Lac\\
1FGL~J1224.7$+$2121&3EG J1224$+$2118&&&&\phn\phn8.07 $\pm$ 0.75&2.55 $\pm$ 0.07&\phn13.9 $\pm$ \phn1.8&2.28 $\pm$ 0.13&FSRQ\\
1FGL~J1225.8$+$4336&3EG J1227$+$4302&&&&\phn\phn2.82 $\pm$ 0.69&2.81 $\pm$ 0.18&\phn21.7 $\pm$ \phn7.1&\nodata&XXX\\
1FGL~J1229.1$+$0203&3EG J1229$+$0210&\phantom{c}EGR J1229$+$0203&&1AGL J1228$+$0142&\phn55.30 $\pm$ 1.48&2.75 $\pm$ 0.03&\phn15.4 $\pm$ \phn1.8&2.58 $\pm$ 0.09&FSRQ\\
1FGL~J1239.5$+$0443&&\phantom{c}EGR J1237$+$0434&&1AGL J1238$+$0406&\phn\phn8.41 $\pm$ 0.75&2.35 $\pm$ 0.06&\nodata&\nodata&FSRQ\\
1FGL~J1256.2$-$0547&3EG J1255$-$0549&\phantom{c}EGR J1256$-$0552&GEV J1256$-$0546&1AGL J1256$-$0549&\phn68.84 $\pm$ 1.37&2.32 $\pm$ 0.02&179.7 $\pm$ \phn6.7&1.96 $\pm$ 0.04&FSRQ\\
1FGL~J1258.7$-$2221&&\phantom{c}EGR J1259$-$2209&&&\phn\phn5.66 $\pm$ 0.80&2.39 $\pm$ 0.08&\nodata&\nodata&FSRQ\\
1FGL~J1321.1$+$2214&3EG J1323$+$2200&&&&\phn\phn1.62 $\pm$ 0.50&2.21 $\pm$ 0.15&\phn18.1 $\pm$ \phn4.0&1.86 $\pm$ 0.35&FSRQ\\
1FGL~J1325.6$-$4300&3EG J1324$-$4314&&&&\phn20.40 $\pm$ 1.47&2.71 $\pm$ 0.06&\phn13.6 $\pm$ \phn2.5&2.58 $\pm$ 0.26&AGN\\
1FGL~J1337.7$-$1255&&\phantom{c}EGR J1337$-$1310&&&\phn\phn7.21 $\pm$ 1.12&2.50 $\pm$ 0.09&\nodata&\nodata&FSRQ\\
1FGL~J1408.9$-$0751&3EG J1409$-$0745&\phantom{c}EGR J1409$-$0736&GEV J1409$-$0741&&\phn\phn5.21 $\pm$ 0.71&2.42 $\pm$ 0.08&\phn97.6 $\pm$ \phn9.1&2.29 $\pm$ 0.11&FSRQ\\
1FGL~J1422.7$+$3743&3EG J1424$+$3734&\phantom{c}EGR J1424$+$3730&&&\phn\phn2.36 $\pm$ 0.72&2.63 $\pm$ 0.19&\phn16.3 $\pm$ \phn4.9&3.25 $\pm$ 0.46&BL Lac\\
1FGL~J1428.2$-$4204&3EG J1429$-$4217&\phantom{c}EGR J1428$-$4240&&&\phn\phn6.70 $\pm$ 0.82&2.31 $\pm$ 0.07&\phn29.5 $\pm$ \phn5.3&2.13 $\pm$ 0.21&FSRQ\\
1FGL~J1457.5$-$3540&3EG J1500$-$3509&&&&\phn33.27 $\pm$ 1.18&2.27 $\pm$ 0.02&\phn10.9 $\pm$ \phn2.8&2.99 $\pm$ 0.37&FSRQ\\
1FGL~J1503.5$-$1544&3EG J1504$-$1537&\phantom{c}EGR J1504$-$1539&&&\phn\phn0.76 $\pm$ 0.38&1.74 $\pm$ 0.19&\phn33.2 $\pm$ 10.3&\nodata&BL Lac\\
1FGL~J1505.1$-$3435&3EG J1500$-$3509&&&&\phn\phn1.11 $\pm$ 0.54&2.02 $\pm$ 0.19&\phn10.9 $\pm$ \phn2.8&2.99 $\pm$ 0.37&XXX\\
1FGL~J1512.8$-$0906&3EG J1512$-$0849&\phantom{c}EGR J1512$-$0857&&1AGL J1511$-$0908&127.10 $\pm$ 1.85&2.41 $\pm$ 0.01&\phn18.0 $\pm$ \phn3.8&2.47 $\pm$ 0.21&FSRQ\\
1FGL~J1607.1$+$1552&3EG J1605$+$1553&\phantom{c}EGR J1607$+$1533&&&\phn\phn3.86 $\pm$ 0.59&2.25 $\pm$ 0.08&\phn42.0 $\pm$ 12.3&2.06 $\pm$ 0.41&AGN\\
1FGL~J1609.0$+$1031&3EG J1608$+$1055&\phantom{c}EGR J1608$+$1051&&&\phn\phn6.19 $\pm$ 0.87&2.72 $\pm$ 0.10&\phn34.9 $\pm$ \phn5.6&2.63 $\pm$ 0.24&FSRQ\\
1FGL~J1613.5$+$3411&3EG J1614$+$3424&&GEV J1613$+$3432&&\phn\phn1.02 $\pm$ 0.48&2.29 $\pm$ 0.22&\phn26.5 $\pm$ \phn4.0&2.42 $\pm$ 0.15&FSRQ\\
1FGL~J1625.7$-$2524&3EG J1626$-$2519&&GEV J1626$-$2502&&\phn10.53 $\pm$ 1.42&2.36 $\pm$ 0.06&\phn42.6 $\pm$ \phn6.6&2.21 $\pm$ 0.13&FSRQ\\
1FGL~J1626.2$-$2956&3EG J1625$-$2955&\phantom{c}EGR J1625$-$2958&GEV J1626$-$2955&&\phn\phn5.02 $\pm$ 0.94&2.36 $\pm$ 0.10&258.9 $\pm$ 15.3&2.07 $\pm$ 0.07&FSRQ\\
1FGL~J1635.0$+$3808&3EG J1635$+$3813&&GEV J1636$+$3812&&\phn19.21 $\pm$ 1.17&2.47 $\pm$ 0.04&107.5 $\pm$ \phn9.6&2.15 $\pm$ 0.09&FSRQ\\
1FGL~J1635.4$+$8228&3EG J1621$+$8203&&&&\phn\phn3.78 $\pm$ 0.71&2.50 $\pm$ 0.12&\phn10.4 $\pm$ \phn3.0&2.29 $\pm$ 0.49&AGN\\
1FGL~J1702.7$-$6217&3EG J1659$-$6251&&&&\phn\phn4.74 $\pm$ 1.04&2.54 $\pm$ 0.13&\phn47.0 $\pm$ 13.1&2.54 $\pm$ 0.37&FSRQ\\
1FGL~J1728.2$+$0431&3EG J1727$+$0429&\phantom{c}EGR J1727$+$0416&&&\phn\phn6.66 $\pm$ 1.05&2.65 $\pm$ 0.11&\phn17.9 $\pm$ \phn4.1&2.67 $\pm$ 0.26&FSRQ\\
1FGL~J1733.0$-$1308&3EG J1733$-$1313&&&&\phn\phn8.73 $\pm$ 1.12&2.34 $\pm$ 0.07&\phn36.1 $\pm$ \phn3.4&2.23 $\pm$ 0.10&FSRQ\\
1FGL~J1740.0$+$5209&3EG J1738$+$5203&\phantom{c}EGR J1740$+$5213&&&\phn17.20 $\pm$ 1.03&2.71 $\pm$ 0.05&\phn18.2 $\pm$ \phn3.5&2.42 $\pm$ 0.23&FSRQ\\
1FGL~J1849.3$+$6705&&&&1AGL J1846$+$6714&\phn24.24 $\pm$ 0.89&2.25 $\pm$ 0.03&\nodata&\nodata&FSRQ\\
1FGL~J1911.2$-$2007&3EG J1911$-$2000&\phantom{c}EGR J1912$-$2000&&&\phn12.37 $\pm$ 1.07&2.42 $\pm$ 0.05&\phn17.5 $\pm$ \phn2.7&2.39 $\pm$ 0.18&FSRQ\\
1FGL~J2006.6$-$2302&3EG J2006$-$2321&&&&\phn\phn7.16 $\pm$ 0.95&2.68 $\pm$ 0.10&\phn19.8 $\pm$ \phn4.4&2.33 $\pm$ 0.36&FSRQ\\
1FGL~J2009.5$-$4849&&&GEV J2009$-$4827&&\phn\phn3.94 $\pm$ 0.49&1.90 $\pm$ 0.06& \nodata&\nodata&BL Lac\\
1FGL~J2025.6$-$0735&3EG J2025$-$0744&&GEV J2024$-$0812&1AGL J2026$-$0732&\phn29.15 $\pm$ 1.18&2.35 $\pm$ 0.03&\phn74.5 $\pm$ 13.4&2.38 $\pm$ 0.17&FSRQ\\
1FGL~J2031.5$+$1219&&\phantom{c}EGR J2032$+$1226&&&\phn\phn4.06 $\pm$ 0.90&2.41 $\pm$ 0.12&\nodata&\nodata&BL Lac\\
1FGL~J2035.4$+$1100&3EG J2036$+$1132&&&&\phn\phn7.74 $\pm$ 1.11&2.68 $\pm$ 0.10&\phn13.3 $\pm$ \phn3.1&2.83 $\pm$ 0.26&FSRQ\\
1FGL~J2056.3$-$4714&3EG J2055$-$4716&&&&\phn17.00 $\pm$ 1.03&2.54 $\pm$ 0.05&\phn23.6 $\pm$ \phn6.0&2.04 $\pm$ 0.35&FSRQ\\
1FGL~J2158.8$-$3013&3EG J2158$-$3023&\phantom{c}EGR J2200$-$3015&&&\phn21.44 $\pm$ 0.70&1.91 $\pm$ 0.02&\phn30.4 $\pm$ \phn7.7&2.35 $\pm$ 0.26&BL Lac\\
1FGL~J2202.8$+$4216&3EG J2202$+$4217&\phantom{c}EGR J2204$+$4225&&&\phn16.81 $\pm$ 1.01&2.38 $\pm$ 0.04&\phn39.9 $\pm$ 11.6&2.60 $\pm$ 0.28&BL Lac\\
1FGL~J2212.1$+$2358&3EG J2209$+$2401&&&&\phn\phn1.19 $\pm$ 0.50&2.13 $\pm$ 0.19&\phn14.6 $\pm$ \phn4.2&2.48 $\pm$ 0.50&FSRQ\\
1FGL~J2212.9$+$0654&&EGRc J2215$+$0653&&&\phn\phn3.01 $\pm$ 0.61&2.33 $\pm$ 0.11&\nodata&\nodata&FSRQ\\
1FGL~J2232.5$+$1144&3EG J2232$+$1147&&&&\phn14.70 $\pm$ 0.97&2.56 $\pm$ 0.05&\phn19.2 $\pm$ \phn2.8&2.45 $\pm$ 0.14&FSRQ\\
1FGL~J2235.7$-$4817&&\phantom{c}EGR J2233$-$4812&&&\phn\phn2.02 $\pm$ 0.56&2.43 $\pm$ 0.17&\nodata&\nodata&FSRQ\\
1FGL~J2253.9$+$1608&3EG J2254$+$1601&\phantom{c}EGR J2253$+$1606&GEV J2253$+$1622&1AGL J2254$+$1602&136.81 $\pm$ 1.74&2.47 $\pm$ 0.01&\phn53.7 $\pm$ \phn4.0&2.21 $\pm$ 0.06&FSRQ\\
1FGL~J2258.0$-$2757&&\phantom{c}EGR J2258$-$2745&&&\phn\phn5.61 $\pm$ 0.85&2.67 $\pm$ 0.10&\nodata&\nodata&FSRQ\\
1FGL~J2323.5$-$0315&3EG J2321$-$0328&&&&\phn\phn6.05 $\pm$ 0.80&2.45 $\pm$ 0.08&\phn38.2 $\pm$ 10.1&\nodata&FSRQ\\
\enddata

\tablenotetext{a}{In units of $10^{-8}$ \pflux{}.}

\tablecomments{Associated sources from the 3rd EGRET \citep[3EG;][]{3EGcatalog}, Revised
EGRET \citep[EGR;][]{EGR}, High-energy EGRET \citep[GEV;][]{Lamb1997}, and
one-year {\it AGILE} \citep[1AGL;][]{AGILEcatalog} catalogs. The LAT and EGRET fluxes
and spectral indices are also provided.}

\end{deluxetable}

\begin{deluxetable}{cccccc}
\tabletypesize{\tiny}
\tablecaption{\label{tab:hardXray}Positional associations with hard X-ray sources}
\tablewidth{0pt}

\tablehead{
\colhead{LAT Source}&
\colhead{Hard X-ray Source}&
\colhead{RA\tablenotemark{a}}&
\colhead{DEC\tablenotemark{a}}&
\colhead{Type}
}
\startdata
1FGL~J0217.8$+$7353 & 1ES 0212+735 & \phn34.32489\phn & \phs73.822807 & FSRQ\\
1FGL~J0319.7$+$4130 & NGC 1275 & \phn49.950871 & \phs41.501099 & AGN\\
1FGL~J0325.0$+$3403 & B2 0321+33B & \phn51.18626\phn & \phs34.176857 & AGN\\
1FGL~J0334.2$+$3233 & NRAO 140 & \phn54.128605 & \phs32.305359 & FSRQ\\
1FGL~J0405.6$-$1309 & PKS 0403$-$13 & \phn61.39616\phn & $-$13.142807 & FSRQ\\
1FGL~J0522.8$-$3632 & PKS 0521$-$36 & \phn80.746483 & $-$36.458878 & BLL\\
1FGL~J0531.0$+$1331 & PKS 0528+134 & \phn82.754601 & \phs13.562002 & FSRQ\\
1FGL~J0538.8$-$4404 & PKS 0537$-$441 & \phn84.750519 & $-$44.104771 & BLL\\
1FGL~J0539.1$-$2847 & PKS 0537$-$286 & \phn84.951698 & $-$28.645336 & FSRQ\\
1FGL~J0636.1$-$7521 & PKS 0637$-$75 & \phn99.080544 & $-$75.244789 & FSRQ\\
1FGL~J0710.6$+$5911 & BZB J0710+5908 & 107.664192 & \phs59.14875\phn & BLL\\
1FGL~J0746.6$+$2548 & B2 0743+25 & 116.609474 & \phs25.811897 & FSRQ\\
1FGL~J0750.6$+$1235 & PKS 0748+126 & 117.666161 & \phs12.520621 & FSRQ\\
1FGL~J0806.2$+$6148 & CGRaBS J0805+6144 & 121.314117 & \phs61.73798\phn & FSRQ\\
1FGL~J0842.2$+$7054 & 4C +71.07 & 130.395554 & \phs70.889877 & FSRQ\\
1FGL~J0949.0$+$0021 & CGRaBS J0948+0022 & 147.228745 & \phs\phn0.350572 & FSRQ\\
1FGL~J0956.5$+$6938 & M 82 & 148.94429\phn & \phs69.694511 & AGN\\
1FGL~J1048.7$+$8054 & CGRaBS J1044+8054 & 161.060471 & \phs80.926788 & FSRQ\\
1FGL~J1103.7$-$2329 & CRATES J1103$-$2329 & 165.899918 & $-$23.477991 & BLL\\
1FGL~J1104.4$+$3812 & Mkn 421 & 166.116058 & \phs38.208893 & BLL\\
1FGL~J1130.2$-$1447 & PKS 1127-14 & 172.528671 & $-$14.813811 & FSRQ\\
1FGL~J1136.2$+$6739 & BZB J1136+6737 & 174.020081 & \phs67.657433 & BLL\\
1FGL~J1221.3$+$3008 & B2 1218+30 & 185.328064 & \phs30.148579 & BLL\\
1FGL~J1222.5$+$0415 & 4C +04.42 & 185.589111 & \phs\phn4.235839 & FSRQ\\
1FGL~J1224.7$+$2121 & 4C +21.35 & 186.224121 & \phs21.383167 & FSRQ\\
1FGL~J1229.1$+$0203 & 3C 273 & 187.27565\phn & \phs\phn2.043433 & FSRQ\\
1FGL~J1256.2$-$0547 & 3C 279 & 194.041\phn\phn\phn & \phn$-$5.803322 & FSRQ\\
1FGL~J1305.4$-$4928 & NGC 4945 & 196.353302 & $-$49.473759 & AGN\\
1FGL~J1307.0$-$4030 & ESO 323-G77 & 196.643127 & $-$40.422039 & AGN\\
1FGL~J1325.6$-$4300 & Cen A & 201.362106 & $-$43.026054 & AGN\\
1FGL~J1331.9$-$0506 & PKS 1329$-$049 & 203.010651 & \phn$-$5.174589 & FSRQ\\
1FGL~J1417.8$+$2541 & 2E 1415+2557 & 214.479584 & \phs25.733086 & BLL\\
1FGL~J1428.7$+$4239 & 1ES 1426+428 & 217.147217 & \phs42.666458 & BLL\\
1FGL~J1442.8$+$1158 & 1ES 1440+122 & 220.704819 & \phs12.045237 & BLL\\
1FGL~J1512.8$-$0906 & PKS 1510$-$08 & 228.205551 & \phn$-$9.086402 & FSRQ\\
1FGL~J1517.8$-$2423 & AP Lib & 229.457718 & $-$24.370974 & BLL\\
1FGL~J1555.7$+$1111 & PG 1553+113 & 238.886841 & \phs11.209181 & BLL\\
1FGL~J1626.2$-$2956 & PKS 1622$-$29 & 246.539917 & $-$29.818855 & FSRQ\\
1FGL~J1653.9$+$3945 & Mkn 501 & 253.446915 & \phs39.768932 & BLL\\
1FGL~J1829.8$+$4845 & 3C 380 & 277.403839 & \phs48.753803 & AGN\\
1FGL~J1925.2$-$2919 & PKS B1921$-$293 & 291.157959 & $-$29.235388 & FSRQ\\
1FGL~J1938.2$-$3957 & PKS 1933$-$400\tablenotemark{b} & 294.294\phn\phn\phn & $-$39.932\phn\phn\phn & FSRQ\\
1FGL~J2000.0$+$6508 & 1ES 1959+650 & 299.94632\phn & \phs65.150742 & BLL\\
1FGL~J2148.5$+$0654 & 4C +06.69 & 327.031952 & \phs\phn6.948918 & FSRQ\\
1FGL~J2202.8$+$4216 & BL Lac & 330.72821\phn & \phs42.273628 & BLL\\
1FGL~J2229.7$-$0832 & PKS 2227$-$08 & 337.407898 & \phn$-$8.519718 & FSRQ\\
1FGL~J2232.5$+$1144 & CTA 102 & 338.135529 & \phs11.73317\phn & FSRQ\\
1FGL~J2253.9$+$1608 & 3C 454.3 & 343.483856 & \phs16.153084 & FSRQ\\
1FGL~J2327.7$+$0943 & PKS 2325+093 & 351.892822 & \phs\phn9.633142 & FSRQ\\
1FGL~J2359.0$-$3035 & 1H 2351$-$315 & 359.779633 & $-$30.594545 & BLL\\
\enddata

\tablenotetext{a}{J2000 coordinate, in degrees, from the 54-month Palermo BAT catalog where available
or from the fourth IBIS catalog otherwise.}
\tablenotetext{b}{Hard X-ray source found only in the fourth IBIS catalog.}
\end{deluxetable}

\end{document}